\documentclass[aps,prc,twocolumn,showpacs,preprintnumbers,
               nofootinbib,float,superscriptaddress,longbibliography]{revtex4-1}
\usepackage{graphicx, fancybox}
\usepackage{amsmath,amssymb}
\usepackage{color}
\usepackage[dvipsnames]{xcolor}
\usepackage[colorlinks=true, pdfstartview=FitV, linkcolor=Orange, citecolor=Cyan, urlcolor=blue]{hyperref}
\usepackage{soul}
\usepackage{array}
\usepackage{url}
\usepackage[utf8]{inputenc}

\def\Glauber{\textsc{3d-glauber}}
\def\MUSIC{\textsc{music}}
\def\UrQMD{\textsc{urqmd}}
\def\iEBEMUSIC{\textsc{iebe-music}}

\def\dNdeta{$dN^\mathrm{ch}/d\eta$}
\def\snn{\sqrt{s_\mathrm{NN}}}

\usepackage[normalem]{ulem}
\graphicspath{{figs/}}   

\begin{document}

\title{Longitudinal dynamics and particle production in relativistic nuclear collisions}

\author{Chun Shen}
\email{chunshen@wayne.edu}
\affiliation{Department of Physics and Astronomy, Wayne State University, Detroit, Michigan, 48201, USA}
\affiliation{RIKEN BNL Research Center, Brookhaven National Laboratory, Upton, NY 11973, USA}

\author{Bj\"orn Schenke}
\email{bschenke@bnl.gov}
\affiliation{Physics Department, Brookhaven National Laboratory, Upton, NY 11973, USA}

\begin{abstract}
This work presents a three-dimensional dynamical initialization model for relativistic heavy-ion collisions, implementing local energy-momentum conservation and baryon charge fluctuations at string junctions. Constraining parameters using experimental data from p+p collisions at various collision energies, the model provides a very good description of the charged hadron and net proton rapidity distributions in Au+Au collisions from 7.7 to 200 GeV and Pb+Pb collisions at 8.77 and 17.3 GeV. We demonstrate the importance of fluctuations of baryon densities to string junctions for describing net-proton distributions at collision energies of 62.4 and 200 GeV. Including this improved baryon stopping description along with the requirement of strangeness neutrality also yields a good description of identified particle yields as functions of the collision energy above 7.7 GeV. We further study asymmetric p+Al and (p, d, $^3$He)+Au collisions at the top RHIC energy and p+Pb, Xe+Xe, and Pb+Pb collisions at LHC energies. We identify the produced particle rapidity distributions in asymmetric collision systems as particularly useful for constraining models of the early-time longitudinal dynamics. 
\end{abstract}

{\maketitle}

\section{Introduction}

The behavior of Quantum Chromodynamics (QCD) at finite temperature and density is of fundamental interest. Quantifying the phase structure of hot and dense QCD matter is one of the primary goals in relativistic nuclear physics. Extensive experimental heavy-ion programs are being conducted all over the world, such as the Large Hadron Collider (LHC) \cite{Evans:2008zzb}, Relativistic Heavy-Ion Collider (RHIC) \cite{STAR:2005gfr, PHENIX:2004vcz, PHOBOS:2004zne, BRAHMS:2004adc} with its Beam Energy Scan (BES) program \cite{Caines:2009yu, Mohanty:2011nm, Mitchell:2012mx, Odyniec:2015iaa}, and the NA61/SHINE experiment at the Super Proton Synchrotron (SPS) \cite{Gazdzicki:2008kk, Abgrall:2014xwa}. They cover a wide range of collision energy and provide us with a unique opportunity to quantify the phase diagram of strongly interacting matter and to study the thermodynamic and transport properties of QCD matter as a function of temperature and net baryon density \cite{Bzdak:2019pkr, Wu:2021xgu, An:2021wof}. 

To establish quantitative connections between the QCD phase structure and measurements from relativistic heavy-ion collisions over an extensive collision energy range, we need to model the entire dynamical evolution of the heavy-ion collisions event by event. Hybrid theoretical frameworks that combine relativistic viscous hydrodynamics and hadronic transport models have been developed to simulate the space-time evolution of heavy-ion collisions and have achieved a lot of successes in phenomenological studies over past decades \cite{Gale:2013da, Shen:2020mgh}. At collision energies below top RHIC energy, the simplifying assumption of boost-invariance, which is often employed when describing high energy collisions, is not justified and one has to model the full 3+1D dynamics \cite{Karpenko:2015xea,Shen:2017bsr,Schafer:2021csj}. Experimental data on particle rapidity distributions at varying collision energies and for different collision systems, in particular asymmetric ones, provide important constraints for models that include longitudinal 
dynamics.

The incoming heavy nuclei carry baryon and electric charges, which are conserved during the system's evolution. After the initial impact, the colliding nucleons lose their kinetic energy and lead to non-trivial distributions and correlations for these conserved charges as functions of collision energy. 
Along with the rapidity dependence of neutral and charged hadron production, the net baryon and electric charge rapidity distributions allow to trace the longitudinal dynamics in heavy-ion collisions, and provide important constraints on the initial baryon and electric charge distributions and their evolution. The final-state phase-space distributions of protons and neutrons are also crucial inputs for light-nuclei production, which is a sensitive probe for QCD critical fluctuations \cite{Sun:2017xrx,Sun:2018jhg,Oliinychenko:2020znl,Zhao:2020irc,Sun:2021dlz}.

The conserved charges are important to consider theoretically as the properties of the Quark-Gluon Plasma (QGP) are altered in an environment doped with net baryon and electric charges. As observed experimentally at RHIC Beam Energy Scan (BES) collision energies \cite{BRAHMS:2003wwg, BRAHMS:2009wlg, STAR:2008med, STAR:2017sal}, an increasing net baryon charge remains at mid-rapidity as the collision energy decreases, which is a prerequisite for scanning the QCD phase diagram in the plane of temperature and net baryon chemical potential.

In this work, we develop a dynamical 3D initial state model which parametrizes the energy loss processes during the nuclear impact stage. We will use particle rapidity distributions in p+p collisions to calibrate the model as a function of collision energy and extend our calculations to asymmetric light-heavy ion and symmetric heavy-ion collisions. This initial state model also correlates the initial stage energy loss with the baryon charge distribution in coordinate and momentum space. Using only a handful of effective parameters, we will focus on describing the longitudinal distribution of particle production for collision energies of a few GeV to the TeV scale for a variety of collision systems.

This paper is organized as follows: the next section describes several theoretical improvements over the \Glauber{} model first introduced in \cite{Shen:2017bsr}. This initial state model is dynamically connected with a viscous hydrodynamics + hadronic transport framework to simulate relativistic nuclear collisions at different collision energies, which will be discussed in Sec.~\ref{sec:hybridFramework}. The model parameters are calibrated with p+p collisions at different collision energies in Sec.~\ref{sec:calibration}. In the same section we apply the model to small asymmetric collision systems and compare event-by-event multiplicity distributions and pseudo-rapidity distributions to experimental data. Then we extend our model to study rapidity distributions of produced particles in heavy-ion collisions at SPS, RHIC, and LHC energies in Sec.~\ref{sec:results}. In particular, the initial state baryon stopping is constrained with the net proton rapidity distributions at different collision energies. We further study how well the system's chemistry is described as a function of collision energy by comparing identified particle ratios to experimental data. The paper concludes in Sec.~\ref{sec:conc}. 

\section{The improved 3D Monte-Carlo Glauber initial state model}\label{sec:model}

We discuss the \Glauber{} Monte-Carlo initial state model, first introduced in \cite{Shen:2017bsr}, with a focus on new developments and improvements to the model. 

\subsection{Subnucleonic structure}
In this work, nucleons are sampled from nuclear density distributions of the Woods-Saxon form \cite{PhysRev.95.577}, and individual nucleons are treated as collections of three valence quarks with positions and momenta that fluctuate from configuration to configuration.
The valence quarks' spatial positions are sampled from a 3D Gaussian distribution with a width $B = 4$\,GeV$^{-1}$,
\begin{equation}
    P(\vec{r}) \propto e^{-r^2/(2B^2)}.
\end{equation}
A detailed study of how observables depend on the parameter $B$ will be done in the future, when we plan to explore observables sensitive to anisotropic flow, which are expected to depend more strongly on $B$ than the multiplicity distributions studied in this work.

\subsubsection{Metropolis sampling of multiple valence quarks from the parton distribution function}

The energy deposited in the interaction region is determined by the energy lost by the valence quarks in every nucleon-nucleon collision. The first step is consequently the determination of the initial valence quark energy and momentum. We sample the longitudinal momentum fraction $\{x_i\}$ of the quark according to the proton's and neutron's valence parton distribution function (PDF), where $i$ labels the quark. Nuclear modifications (EPS09) are included when considering nucleons inside heavy nuclei \cite{Eskola:2009uj}. The parton's longitudinal momentum is $P^z_{q, i} = x_i P^z_N$ with the nucleon's momentum given by the beam rapidity $y_\mathrm{beam}$, $P^z_N = m_N \sinh(y_\mathrm{beam})$, where for the nucleon mass we use $m_{N}=0.938\,{\rm GeV}$. We impose
\begin{equation}
    \sum_{i=1}^3 P^z_{q, i} \leq P^z_N \Rightarrow \sum_{i=1}^3 x_i \leq 1
\end{equation}
so that the total momentum carried by the three valence quarks is smaller than or equal to the nucleon's momentum. We also impose a similar constraint on energy,
\begin{eqnarray}
    &&\sum_{i=1}^3 E_{q, i} \leq E_N \Rightarrow \nonumber \\
    &&\sum_{i=1}^3 \sqrt{\frac{m^2_\mathrm{parton}}{m_N^2 \cosh^2(y_\mathrm{beam})} + x_i^2 \tanh^2(y_\mathrm{beam})} \le 1.
\end{eqnarray}
We set the valence quark mass $m_\mathrm{parton} = 0.312$ GeV in our calculation. For $y_\mathrm{beam} \gg 1$, the energy constraint reduces to $\sum_{i=1}^3 x_i \leq 1$.

Here we develop a Metropolis algorithm to realize this constraint while keeping the single parton's $x$ distribution unchanged. Our sampling procedure is to first generate a large sample of $\{x_i\}$ from the PDF for $u$ and $d$ quarks. Then we randomly group them into triplets of $(uud)$ and $(udd)$ for protons and neutrons, respectively. At this stage, some triplets have a sum of $\{x_i\}$ larger than 1 and some have the sum smaller than 1. We define a score $s$ for each triplet as,
\begin{equation}
    s(\{x_i\}) = \left\{\begin{array}{cl}
        \sum_i x_i & \mbox{ if } \sum_i x_i \le 1  \\
        0 & \mbox{ if } \sum_i x_i > 1. 
    \end{array} \right.
\end{equation}
Then we randomly pick two triplets from the list and swap a pair of valence quarks with the same flavor. 
If the sum of the scores from the two triplets increases after the swap, we keep the change. Otherwise, we reject the swap.
By repeating this procedure many times, we can ensure all triplets have a sum smaller than 1. Since we do not throw out any samples, the single parton's $x$ distribution remains unchanged. We include the nuclear PDF modification when we sample valence quarks inside of nucleons of large nuclei, such as Au and Pb. 

\begin{figure}[ht!]
  \centering
  \includegraphics[width=0.9\linewidth]{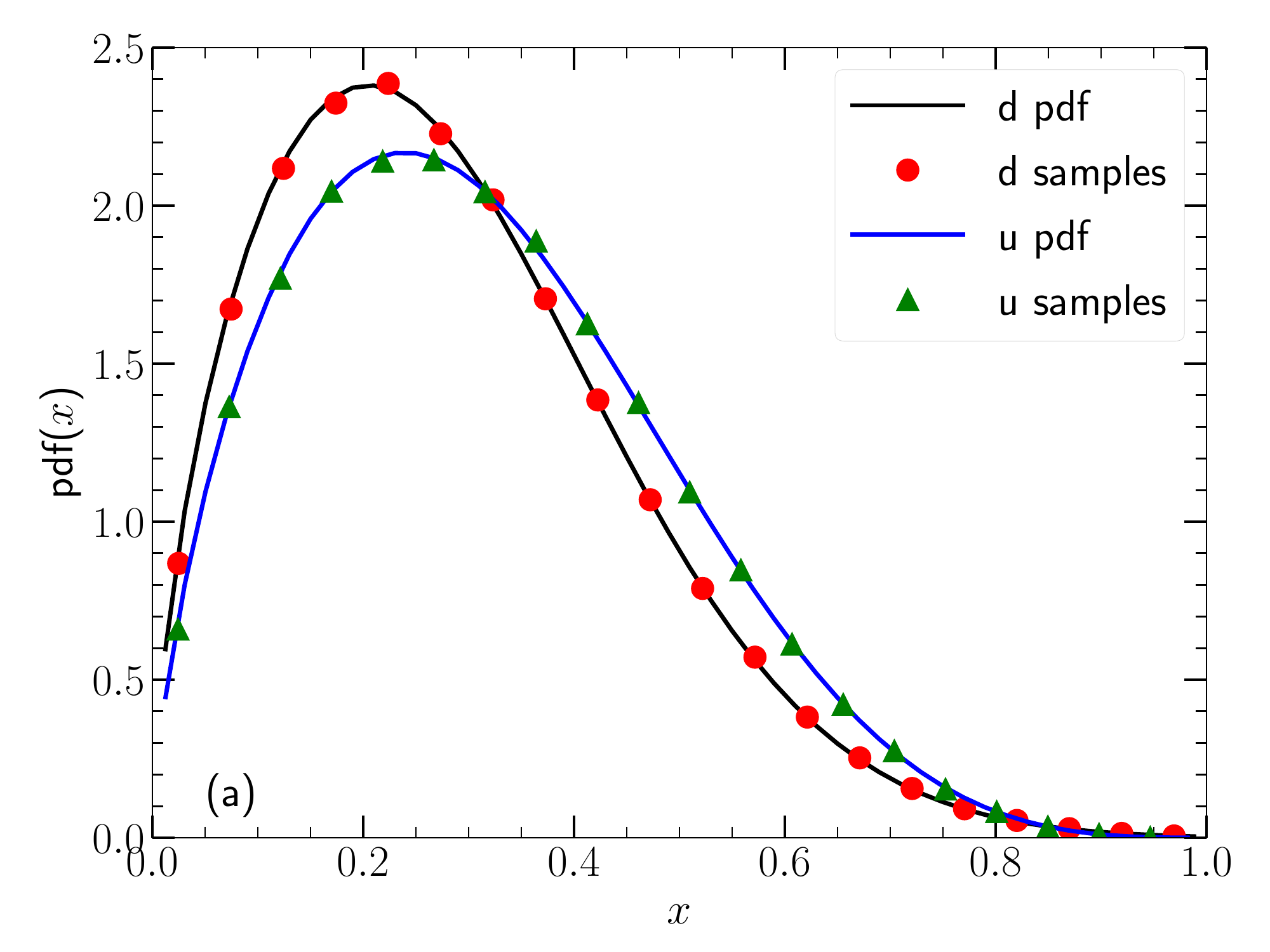}
  \includegraphics[width=0.9\linewidth]{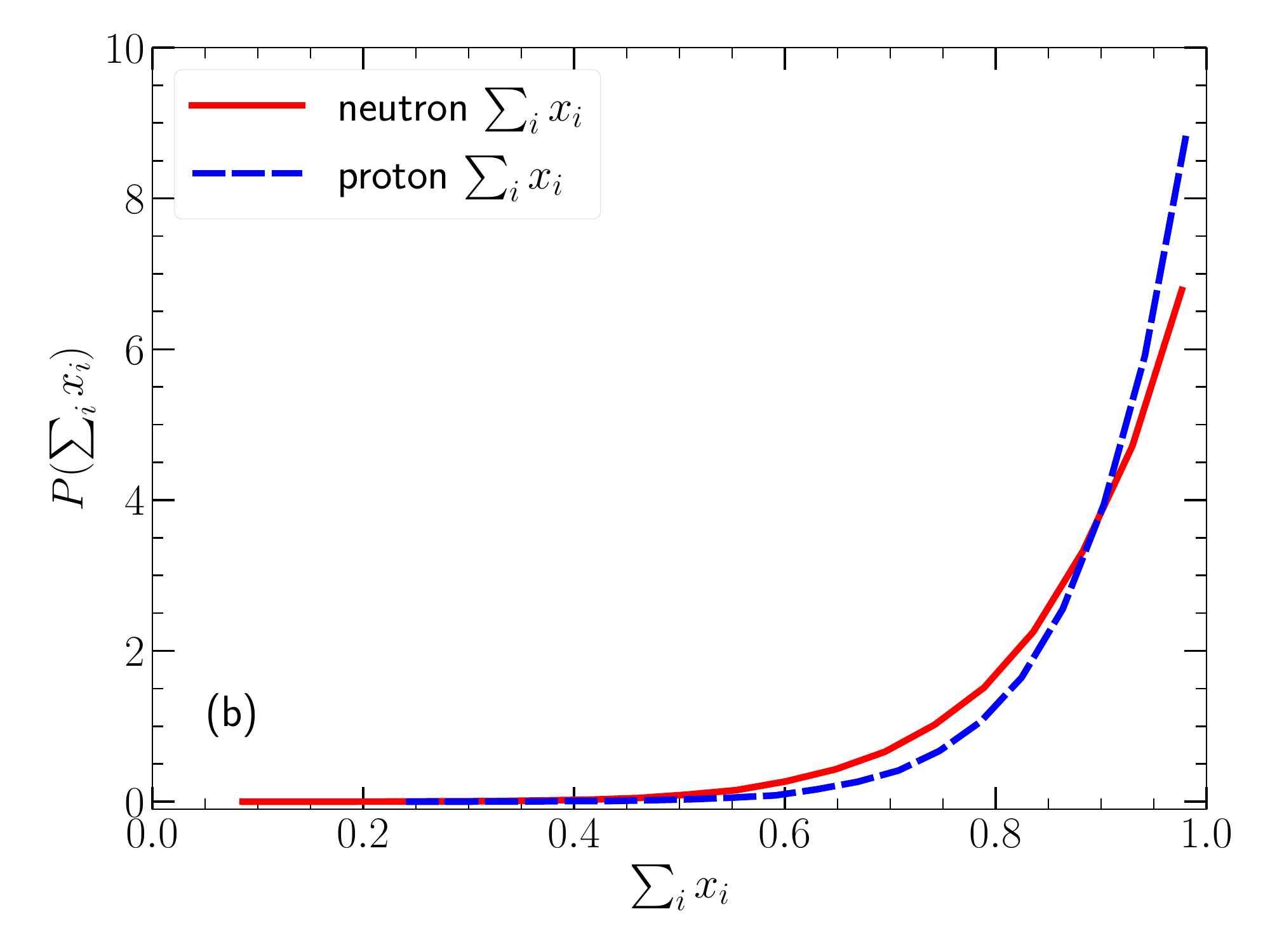}
  \caption{(Color online) Panel (a): The probability distributions of the sampled $u$ and $d$ valence quarks compared to their PDF  (CT10nnlo) \cite{Gao:2013xoa}. Panel (b): The probability distributions of the summed $x$ values of the three valence quarks inside protons and neutrons, respectively.}
  \label{fig:SamplePDF}
\end{figure}
%

Figure~\ref{fig:SamplePDF}a shows the $u$ and $d$ quarks' distributions from our Metropolis algorithm compared with the standard valence quark PDF (CT10nnlo) \cite{Gao:2013xoa}. The comparison  explicitly demonstrates that our algorithm does not modify the single parton distributions while constraining $\sum_i x_i \leq 1$ inside nucleons. Figure~\ref{fig:SamplePDF}b shows the probability distributions of the sum of three valence quarks' $x$ inside protons and neutrons. The distribution is more sharply peaked near one for protons than neutrons because protons carry two $u$ valence quarks whose mean $x$ is larger than that of the $d$ quark.

\subsubsection{Soft partonic cloud}

The three valence quarks do not carry all the energy and momentum of the nucleon, which is illustrated in Fig.~\ref{fig:SamplePDF}b. The remaining energy and momentum are distributed among the sea quarks and gluons. To consider the energy loss of these wee partons in this model, we treat them as a single soft cloud object inside the nucleon. Because the current string deceleration model assumes colliding partons having the same mass \cite{Shen:2017bsr}, we consider a soft partonic cloud has mass $m_\mathrm{parton}$. 
If the remaining energy inside the nucleon is larger than the valence quark mass $m_\mathrm{parton}$, we assign the following energy and momentum to the soft partonic cloud,
\begin{eqnarray}
    E_\mathrm{cloud} &=& E_N - \sum_{i=1}^3 E_{q, i} \equiv m_\mathrm{parton} \cosh(y_\mathrm{cloud}) \\
    P^z_\mathrm{cloud} &=& m_\mathrm{parton} \sinh(y_\mathrm{cloud}).
\end{eqnarray}
Otherwise, the remaining energy and momentum are attributed to the nucleon remnant which will be discussed below.
The energy loss of the soft partonic cloud will be treated the same way as the valence quarks during the nucleon-nucleon (NN) collisions.

\subsection{Parametrizing energy loss in nucleon-nucleon collisions}

We consider partons lose their energy and momentum through being decelerated by the longitudinal color fields from its colliding partner \cite{Li:2018ini, McLerran:2018avb}. It corresponds to a 1D deceleration model,
\begin{eqnarray}
    dP^\mu = - T^{\mu\nu} d\Sigma_\nu,
    \label{eq:StringDeceleration}
\end{eqnarray}
where $T^{\mu\nu} = \mathrm{diag}(\sigma, \sigma, \sigma, -\sigma)$ and $d\Sigma_\nu = (dz, 0, 0, -dt)$ \cite{Mishustin:2001ib, Bialas:2016epd, Shen:2017bsr, Shen:2017fnn}. The solution of Eq.~\eqref{eq:StringDeceleration} was discussed in details in Ref.~\cite{Shen:2017bsr}.

\subsubsection{Average rapidity loss}

Motivated by the baryon stopping extracted by the BRAHMS Collaboration \cite{BRAHMS:2009wlg}, we model the average rapidity loss function of the valence quark with an incoming rapidity $y_\mathrm{init}$ in the collision pair rest frame as,
\begin{equation}
    \langle y_\mathrm{loss} \rangle (y_\mathrm{init}) = A y_\mathrm{init}^{\alpha_2} [\tanh(y_\mathrm{init})]^{\alpha_1 - \alpha_2},
    \label{eq:ylossMean}
\end{equation}
We make sure that the choice of the three parameters $A, \alpha_1$, and $\alpha_2$ always give $\langle y_\mathrm{loss} \rangle < y_\mathrm{init}$. For small initial rapidity, $y_\mathrm{loss} \propto y_\mathrm{init}^{\alpha_1}$. And for large initial rapidity $\tanh(y_\mathrm{init}) \rightarrow 1$, so $y_\mathrm{loss} \propto y_\mathrm{init}^{\alpha_2}$.

\subsubsection{Multiplicity/Rapidity Loss Fluctuations}\label{subsec:yloss}

Given an initial incoming rapidity $y_\mathrm{init}$, we allow the amount of rapidity loss to fluctuate from one collision to another. We will see later that the rapidity loss fluctuation strongly affects the particle multiplicity distribution at mid-rapidity.
We introduce a logit-normal distribution whose mean agrees with our parametrization in Eq.~(\ref{eq:ylossMean}). The variance of the distribution, $\sigma$, is treated as a free parameter in our model. In contrast to the normal distribution, the logit-normal distribution allows us to impose lower and upper bounds on the possible values and can be written as,
\begin{equation}
    f(x, \mu, \sigma) = \frac{1}{\sigma \sqrt{2\pi}} \frac{1}{x(1 - x)} \exp\left(-\frac{(\mathrm{logit}(x) - \mu)^2}{2\sigma^2}\right),
    \label{eq:logitDis}
\end{equation}
where $\mu$ is the mean and $\sigma$ is the variance. The logit function is defined as ${\rm logit}(x) = \log(x/(1-x))$.
The samples $\{X_i\}$ from this distribution are bounded between 0 and 1. We can now map the upper bound to the incoming rapidity of the parton $y_\mathrm{init}$ and the mean $\mu = \langle y_\mathrm{loss} \rangle$.

In practice, we generate a random sample $X$ from a normal distribution with mean 0 and a given variance $\sigma$, $\mathcal{N}(0, \sigma)$. When computing
\begin{equation}
    Y = \frac{1}{1 + e^{-X}},
\end{equation}
the random number $Y$ lies in $(0, 1)$ and follows the logit-normal distribution. Finally, we use a second-order polynomial to map $(Y \rightarrow y_\mathrm{loss})$ requiring $(0 \rightarrow 0)$, $(\frac{1}{2} \rightarrow \langle y_\mathrm{loss} \rangle)$, and $(1 \rightarrow y_\mathrm{init})$. With this prescription, we sample $y_\mathrm{loss}$ for each individual valence quark and the soft cloud.

\subsubsection{Number of strings produced in NN collisions}\label{subsec:NstringFluct}
In individual nucleon-nucleon collisions, we allow for multiple partons to interact and lose energy. String-like energy-momentum sources are produced for each parton-parton collision. We allow individual partons to interact only once in one NN collision, while multiple strings can be produced from different pairs of partons in the NN collision.
The relative probability to produce more than one string $N_\mathrm{string} > 1$ is parameterized as
\begin{equation}
    P(N_\mathrm{string}) \propto e^{-(N_\mathrm{string} - 1)}.
    \label{eq:StringProb}
\end{equation}
Considering the three valence quarks and the soft partonic cloud, we allow one NN collision to produce a maximum of four strings. According to Eq.\,(\ref{eq:StringProb}), the average number of strings produced in one NN collision is $\langle N_\mathrm{string} \rangle \simeq 1.507$.

We will see in Sec.~\ref{sec:calibration} that the fluctuations in the number of string sources, resulting from sampling $N_\mathrm{string}$ from Eq.\,(\ref{eq:StringProb}), and the fluctuating amount of rapidity loss in each parton-parton collision contribute to the shape of the multiplicity distribution of produced particles.

In the situation when a projectile nucleon goes through multiple collisions with different nucleons from the target nucleus, we allow multiple strings to be attached to the same parton in a sequential order after all the partons in the projectile nucleon are connected with at least one string.

\subsection{Improved energy-momentum distribution of strings}

As the two nuclei collide with each other, we consider individual nucleon-nucleon collisions to be independent. The participant partons from the colliding nucleons lose energy and momentum during the impact according to the classical string deceleration model \cite{Bialas:2016epd, Shen:2017bsr}. We denote the initial and final rapidities of the projectile and target partons as $y^\mathrm{init}_{P/T}$ and $y^\mathrm{final}_{P/T}$, respectively. The energy and momentum lost during the collision are
\begin{eqnarray}
    E_\mathrm{loss} &=& m_\mathrm{parton}[\cosh(y^\mathrm{init}_{P}) + \cosh(y^\mathrm{init}_{T}) \nonumber \\
    && \qquad \qquad - \cosh(y^\mathrm{final}_P) - \cosh(y^\mathrm{final}_T)]\,, \label{eq:Eloss} \\
    P^z_\mathrm{loss} &=& m_\mathrm{parton}[\sinh(y^\mathrm{init}_P) + \sinh(y^\mathrm{init}_T) \nonumber \\
    && \qquad \qquad - \sinh(y^\mathrm{final}_P) - \sinh(y^\mathrm{final}_T)]. \label{eq:Pzloss}
\end{eqnarray}
These lost energy and momentum are distributed in strings in space-time, that serve as sources for the hydrodynamic fields. In this work we improve upon the Monte-Carlo \Glauber{} model introduced in \cite{Shen:2017bsr} by imposing energy-momentum conservation constraints on the string production. These global constraints enable us to make model predictions for particle rapidity distributions $dN_\mathrm{ch}/d\eta$ at different collision energies.

The energy-momentum source terms for the strings can be written as,
\begin{equation}
    J_\mathrm{string}^\mu(\vec{x}_\perp, \eta_s) = e_\mathrm{string}(\vec{x}_\perp, \eta_s) u^\mu_\mathrm{string}(\eta_s) 
\end{equation}
with
\begin{equation}
    u^\mu_\mathrm{string}(\eta_s) = (\cosh(y(\eta_s)), 0, 0, \sinh(y(\eta_s))).
\end{equation}
Here the local energy density inside a string is assumed to be constant in rapidity with half Gaussian shaped drop offs at both ends in the longitudinal direction modulated by a factorized transverse profile,
\begin{equation}
    e_\mathrm{string}(\vec{x}_\perp, \eta_s) = f_\perp(\vec{x}_\perp) e_\mathrm{string}(\eta_s),
\end{equation}
where the transverse profile is assumed to be a normalized 2D Gaussian,
\begin{equation}
    f_\perp(\vec{x}_{\perp}) = \frac{1}{2\pi \sigma_\perp^2}\exp \left(-\frac{(x_1 - x_{1, c})^2}{2 \sigma_\perp^2} - \frac{(x_2 - x_{2, c})^2}{2 \sigma_\perp^2} \right),
    \label{eq:transProf}
\end{equation}
where the center of the string sits at the binary collision point in the transverse plane, $x_{1(2), c} = (x^T_{1(2)} + x^P_{1(2)})/2$.
The longitudinal profile is parametrized as,
\begin{eqnarray}
    e_\mathrm{string}(\eta_s) &=& \mathcal{N}_e \exp \bigg[- \frac{(\vert \eta_s - \eta_{s, c} \vert - \Delta \eta_s)^2}{2\sigma_{\eta_s}^2} \nonumber \\
    && \qquad \qquad \times \theta(\vert \eta_s - \eta_{s, c} \vert - \Delta \eta_s) \bigg],
    \label{eq:StringEProf}
\end{eqnarray}
where $\mathcal{N}_e$ is a normalization factor, $\eta_{s,c}$ marks the center of the string, and $\sigma_{\eta_s}$ describes how fast the energy density drops at the string ends. The length of the string is characterized by $2 \Delta \eta_s$, which will be determined by the classical string deceleration model \cite{Shen:2017bsr, Bialas:2016epd}. Assuming a string extends from $\eta_{s,T}$ to $\eta_{s,P}$ (with $\eta_{s,P} > \eta_{s,T}$), the string center $\eta_{s, c} \equiv (\eta_{s,P} + \eta_{s,T})/2$, and $\Delta \eta_s \equiv (\eta_{s,P} - \eta_{s,T})/2$.

Since the rapidities of the decelerated partons at the string ends are $y^\mathrm{final}_T$ and $y^\mathrm{final}_P$, we assume the momentum rapidity profile inside the string is taking the following linear form,
\begin{eqnarray}
    y(\eta_s) &=& y^\mathrm{final}_T + \frac{y^\mathrm{final}_P - y^\mathrm{final}_T}{\eta_{s,P} - \eta_{s,T}} (\eta_s - \eta_{s,T}) \nonumber \\
    &=& y_\mathrm{CM} + \frac{\Delta y}{\Delta \eta_s} (\eta_s - \eta_{s,c}).
    \label{eq:StringRapProf}
\end{eqnarray}
Here, we defined the center-of-mass rapidity $y_\mathrm{CM} \equiv (y^\mathrm{final}_P + y^\mathrm{final}_T)/2$ and the local rest frame relative rapidity $\Delta y \equiv (y^\mathrm{final}_P - y^\mathrm{final}_T)/2$. One can show that $\tanh(y_\mathrm{CM}) = P^z_\mathrm{loss}/E_\mathrm{loss}$.

Based on Eqs.~(\ref{eq:StringEProf}) and (\ref{eq:StringRapProf}), we can determine the normalization factor $\mathcal{N}_e$ using the energy and net longitudinal momentum lost by the colliding partons in Eqs.~\eqref{eq:Eloss} and \eqref{eq:Pzloss}. The strings' energy and longitudinal momentum are,
\begin{eqnarray}
    E_\mathrm{loss} = E_\mathrm{string} &=& \int^{\eta_{s,P}}_{\eta_{s,T}} d\eta_s e(\eta_s) \cosh(y(\eta_s))\,,
\end{eqnarray}
and
\begin{eqnarray}
    P^z_\mathrm{loss} = P^z_\mathrm{string} &=& \int^{\eta_{s,P}}_{\eta_{s,T}} d\eta_s e(\eta_s) \sinh(y(\eta_s))\,.
    \label{eq:Pz}
\end{eqnarray}
By changing to the variable $\tilde{\eta}_s = \frac{\Delta y}{\Delta \eta_s}(\eta_s - \eta_{s,c})$, we can write the string's energy as,
\begin{equation}
    E_\mathrm{string} = \int^{\Delta y}_{- \Delta y} d\tilde{\eta}_s \frac{\Delta \eta_s}{\Delta y} e(\tilde{\eta}_s) \cosh (y_\mathrm{CM} + \tilde{\eta}_s).
\end{equation}
The local energy density profile in Eq.~(\ref{eq:StringEProf}) is symmetric with respect to $\tilde{\eta}_s \rightarrow -\tilde{\eta}_s$,
\begin{eqnarray}
    e_\mathrm{string}(\tilde{\eta}_s) &=& \mathcal{N}_e \exp \bigg[ - \frac{\Delta \eta_s^2}{2\sigma_{\eta_s}^2} \left(\left\vert \frac{\tilde{\eta}_s}{\Delta y} \right\vert - 1 \right)^2 \nonumber \\
    && \qquad \qquad \times \theta \left(\Delta\eta_s\left( \left\vert \frac{\tilde{\eta}_s}{\Delta y}  \right\vert -1 \right) \right) \bigg].
\end{eqnarray}
Therefore, the string's energy can be simplified as,
\begin{equation}
    E_\mathrm{string} = \cosh(y_\mathrm{CM}) \frac{\Delta \eta_s}{\Delta y} \int^{\Delta y}_{- \Delta y} d\tilde{\eta}_s e_\mathrm{string}(\tilde{\eta}_s) \cosh(\tilde{\eta}_s).
    \label{eq:StringE}
\end{equation}
Similarly, the string's longitudinal momentum can be written as
\begin{equation}
    P^z_{\mathrm{string}} = \sinh(y_\mathrm{CM}) \frac{\Delta \eta_s}{\Delta y} \int^{\Delta y}_{- \Delta y} d\tilde{\eta}_s e_\mathrm{string}(\tilde{\eta}_s) \cosh(\tilde{\eta}_s).
    \label{eq:StringPz}
\end{equation}
To fulfill Eqs. (\ref{eq:StringE}) and (\ref{eq:StringPz}), we must have 
\begin{equation}
    \frac{\Delta \eta_s}{\Delta y} \int^{\Delta y}_{- \Delta y} d\tilde{\eta}_s e(\tilde{\eta}_s) \cosh(\tilde{\eta}_s) = M_\mathrm{string},
    \label{eq:StringM}
\end{equation}
where $M_\mathrm{string}$ is the string's invariant mass, $M_\mathrm{string} = \sqrt{(E_\mathrm{string})^2 - (P^z_\mathrm{string})^2} = \sqrt{(E_\mathrm{loss})^2 - (P^z_\mathrm{loss})^2}$. From Eq.~(\ref{eq:StringM}), we can determine the normalization factor $\mathcal{N}_e$ in Eq.\,\eqref{eq:StringEProf}.

\subsection{Partial coherent NN scatterings in high energy heavy-ion collisions}

In heavy-ion collisions, an individual nucleon scatters multiple times as it travels through the other nucleus. The partons inside one nucleon could lose energy multiple times from different nucleon-nucleon collisions.
The interval between two scatterings shrinks as the collision energy increases. To take into account the coherence effect in multiple scatterings, we introduce a model parameter $\lambda_\mathrm{coh}$, which controls the probability for string production. The probability of producing one more string when the colliding pair of nucleons has already produced $N_\mathrm{string}$ strings is parametrized as,
\begin{equation}
    P(\lambda_\mathrm{coh}, N_\mathrm{string}) \propto (1 - \lambda_\mathrm{coh}) \exp (- \lambda_\mathrm{coh} N_\mathrm{string}).
\end{equation}
For $\lambda_\mathrm{coh} = 1$, no additional strings will be produced after the primary NN collision. While $\lambda_\mathrm{coh} = 0$, allows all binary collisions to produce strings.

\subsection{Collision Remnants}

After the last collision of each participant parton, they carry their final rapidity $y^{\rm final}_{P/T}$, meaning that they have finite energy and momentum left. The remnants of the wounded nucleons (partons in the nucleon that did not collide and produce a string) also carry finite energy-momentum. In this model, we deposit all of them as energy-momentum sources into the hydrodynamic fields.
In this subsection, we will discuss our treatment of the parton remnants at the string ends and the nucleon remnants, which are part of the participant nucleon but are not connected to strings.

\subsubsection{Parton remnants}

After a nucleon-nucleon collision, the energy and momentum that were not deposited inside the flux tube remain in the partons at the string ends, which have their final rapidities, $y^\mathrm{final}_T$, and $y^\mathrm{final}_P$. The energy-momentum carried by these quarks will also be deposited into the medium with a Gaussian spatial profile. 
The source term for the left-going target or right-going projectile parton remnant can be written as,
\begin{equation}
    J^\mu_\mathrm{parton}(\vec{x}_\perp, \eta_s) = e(\vec{x}_{\perp}, \eta_{s}) u^\mu_\mathrm{parton}.
    \label{eq:RemSourceShape}
\end{equation}
Here the velocity of the projectile/target parton is
\begin{equation}
    u^\mu_\mathrm{parton} = (\cosh(y^\mathrm{final}_{P/T}), 0, 0, \sinh(y^\mathrm{final}_{P/T})).
\end{equation}
The energy density profile is,
\begin{equation}
    e(\vec{x}_{\perp}, \eta_s) = m_\mathrm{parton} f_\perp(\vec{x}_{\perp}) f_\eta(\eta_s),
\end{equation}
where $m_\mathrm{parton}$ is the parton's mass and the spatial smearing function $f_\perp$ is defined in Eq.~(\ref{eq:transProf}).
In the longitudinal direction, the smearing function $f_\eta$ is a normalized Gaussian,
\begin{equation}
    f_\eta (\eta_s) = \frac{1}{\sqrt{2\pi}\sigma_{\eta_s}} \exp \left( -\frac{(\eta_s - \eta_{s, P/T})^2}{2 \sigma_{\eta_s}^2} \right).
    \label{eq:RemSourceEtaProf}
\end{equation}

\subsubsection{Nucleon remnants}

During individual nucleon-nucleon collisions, the participant valence quarks inside the nucleon will not carry all the energy and momentum of the incoming nucleon. Therefore, we introduce the nucleon remnant to take into account the remaining energy-momentum that does not contribute to the string production. 
To determine the energy and momentum of the nucleon remnant, we start with the original energy and momentum of the colliding nucleon and subtract the energies and momenta of each sampled valence quark from the nucleon, that is connected to a string.
After all the NN collisions are done, we will have the energy and momentum for the nucleon remnant, $P^\mu_\mathrm{rem}$. Physically, the nucleon remnants carry colors and will radiate and lose energy. To produce hydrodynamic source terms from the nucleon remnants, we model their energy loss according to the same string deceleration model as above, but with a reduced average rapidity loss, $\langle y^\mathrm{rem}_\mathrm{loss} \rangle = \alpha_\mathrm{rem} \langle y_\mathrm{loss} \rangle$. By fitting the charge hadron production at forward rapidities in p+p collisions, we find $\alpha_\mathrm{rem} = 0.5$ (see Fig.~\ref{fig:ppRemLoss} below).

The energy-momentum source term $J^\mu_\mathrm{rem}$ needs to be time-like to ensure that we can find a positive local energy density and flow velocity for the hydrodynamic fields.
If the $P^\mu_\mathrm{rem}$ is a time-like vector, we can decompose it into an invariant mass and a rapidity,
\begin{equation}
    M_\mathrm{rem} = \sqrt{(P^t_\mathrm{rem})^2 - (P^z_\mathrm{rem})^2}
\end{equation}
and
\begin{equation}
    y_\mathrm{rem} = \mathrm{arctanh} \left( \frac{P^z_\mathrm{rem}}{P^t_\mathrm{rem}} \right).
\end{equation}

If the remnant energy-momentum vector is a space-like vector, we need to regulate $P^\mu_\mathrm{rem}$ by keeping its energy component and using the beam rapidity $y_\mathrm{beam}$ to determine its invariant mass
\begin{equation}
    M_\mathrm{rem} = \frac{P^t_\mathrm{rem}}{\cosh(y_\mathrm{beam})}\,,
\end{equation}
and longitudinal momentum
\begin{equation}
    P^z_\mathrm{rem} =  M_\mathrm{rem} \sinh(y_\mathrm{beam}).
\end{equation}
This regulation scheme ensures the system's total energy remains correct but introduces small violations on the total longitudinal momentum conservation. While it is not the only scheme to regulate space-like vectors, preserving the collision system's energy when mapping the initial state to the hydrodynamic stage is preferred for studying global particle production in this work. The spatial profile of the nucleon remnant sources $J_\mathrm{rem}^\mu$ is chosen to be the same as those for the parton remnants in Eqs.~\eqref{eq:RemSourceShape}-\eqref{eq:RemSourceEtaProf}.

\subsection{Initial state baryon number fluctuations}

As an alternative to depositing the baryon number at the string ends, as done in \cite{Shen:2017bsr}, we introduce the string junction model \cite{Kharzeev:1996sq}, where the baryon charge of the string can fluctuate towards the center of the string, according to the following probability,
\begin{equation}
    P(y^B_{P/T}) = (1 - \lambda_B) y_{P/T} + \lambda_B \frac{e^{(y^B_{P/T} - (y_P + y_T)/2)/2}}{4 \sinh((y_P - y_T)/4)}\,.
    \label{eq:BaryonJunction}
\end{equation}
Here, the $y_P$ and $y_T$ are the rapidities of the two string ends.

The second term on the right-hand side of Eq.~\eqref{eq:BaryonJunction} is motivated by the single baryon production cross-section derived in Ref.~\cite{Kharzeev:1996sq}.
The spatial profile of a baryon charge is chosen as a 3D Gaussian in $\vec{x}_\perp$ and $\eta_s$. It is placed inside the string at the space-time rapidity $\eta_s$, which satisfies $y(\eta_s) = y^B_{P/T}$ according to Eq.~\eqref{eq:StringRapProf}.
We choose $\lambda_B = 0.2$ to reproduce the net proton rapidity distribution measured by the BRAHMS Collaboration at 62.4 and 200 GeV \cite{BRAHMS:2003wwg, BRAHMS:2009wlg}. The effects of string junction fluctuations decrease with collision energy. The same choice of the parameter can also reproduce the STAR measured net proton yields at mid-rapidity in the RHIC BES program \cite{STAR:2017ieb}.
At low collision energies, one would expect an additional double baryon production process to become important \cite{Kharzeev:1996sq}. We will include this contribution in future phenomenological studies.
These string junction fluctuations introduce non-trivial baryon transport at the initial state. They are important to understand the net baryon charge distributions measured in the RHIC BES program as shown in Fig.~\ref{fig:BaryonJunction}.

\section{Dynamical initialization in a hydrodynamic + hadronic transport hybrid framework}\label{sec:hybridFramework}

To study particle production in relativistic nuclear collisions, we integrate the improved Monte-Carlo \Glauber{} model discussed in the previous section with a hydrodynamics + hadronic transport hybrid framework. For center of mass energies around 10 GeV and when interested in the region away from midrapidity also for much higher collision energies, the finite extension of the collision overlap region requires us to interweave the 3D initial state with hydrodynamics in a dynamical setup, which was discussed in detail in Ref.~\cite{Shen:2017bsr}.

The produced strings and remnants source the hydrodynamic fields,
\begin{eqnarray}
    \partial_\mu T^{\mu \nu} &=& J^\nu \\
    \partial_\mu J_B^\mu &=& \rho_B,
\end{eqnarray}
with the $J^\mu$ being the energy-momentum source terms from the produced strings and parton and nucleon remnants discussed in the previous section. The system's energy-momentum tensor $T^{\mu\nu}$ and baryon current $J_B^\mu$ are evolved in full (3+1)D \cite{Shen:2017bsr} with a lattice-based equation of state at finite density, \textsc{neos-bqs} \cite{Monnai:2019hkn}. We employ the \MUSIC{} hydrodynamic framework \cite{Schenke:2010nt,Schenke:2011bn,Paquet:2015lta,Denicol:2018wdp} to perform numerical simulations.
For the spatial shape of the source terms, we choose the transverse smearing width $\sigma_\perp = 0.5$ fm in Eq.~\eqref{eq:transProf}. The value of the longitudinal smearing parameter $\sigma_{\eta_s}$ in Eq.~\eqref{eq:StringEProf} depends on the collision energy as follows,
\begin{equation}
    \sigma_{\eta_s} = \left\{ \begin{array}{cl}
        0.2, & \quad \sqrt{s_\mathrm{NN}} < 25\, \mathrm{GeV} \\
        0.5, & \quad \sqrt{s_\mathrm{NN}} \in [25, 100]\,\mathrm{GeV} \\
        0.6, & \quad \sqrt{s_\mathrm{NN}} > 100\,\mathrm{GeV}
    \end{array}. \right.
\end{equation}

During the hydrodynamic evolution, we use a specific shear viscosity $\eta T/(e + P) = 0.12$, which gives a reasonable description of the anisotropic flow coefficients in central and semi-peripheral Au+Au collisions. We neglect the bulk viscous effects and baryon diffusion in this work.

As the collision system evolves to low energy density, we convert fluid cells back to particles on a 3D hyper-surface across space-time according to the Cooper-Frye procedure \cite{Cooper:1974mv}.
The hyper-surface is constructed at a constant energy density $e_\mathrm{sw}$ during the hydrodynamic evolution using the Cornelius algorithm \cite{Huovinen:2012is}. At the beginning of the hydrodynamic evolution, we include an additional ``cold corona'' hyper-surface at a constant proper time for those fluid cells with local energy density $e < e_\mathrm{sw}$. We cut off the corona surface at a cutoff energy density $e_\mathrm{low} = 0.05$ GeV/fm$^3$, below which the particle production from the Cooper-Frye procedure is negligible. The effects of the cold corona on particle production will be discussed in Appendix~\ref{app:corona}.
The thermally emitted hadrons are fed into a hadronic transport model, \UrQMD{} \cite{Bass:1998ca, Bleicher:1999xi}, which performs further scatterings and decays.
The complete set of dynamical evolution models is integrated in the \iEBEMUSIC{} framework \cite{iEBEMUSIC}.

\section{Model calibrations with small systems}\label{sec:calibration}

In this section, we discuss the charged hadron production in small collision systems. Because proton+proton collisions do not involve multiple NN scatterings, we use the rapidity-dependent charged hadron production \dNdeta{} and the normalized particle multiplicity distributions in these collisions to calibrate the valence quarks' rapidity loss in Eq.\,\eqref{eq:ylossMean} along with the rapidity loss of the nucleon remnants.
For p+p collisions, we use a switching energy density $e_\mathrm{sw} = 0.25$\,GeV/fm$^{3}$ for the conversion hyper-surface. At zero net baryon density, this switching energy density corresponds to a temperature of 150 MeV.

\begin{figure}[ht!]
  \centering
  \includegraphics[width=0.95\linewidth]{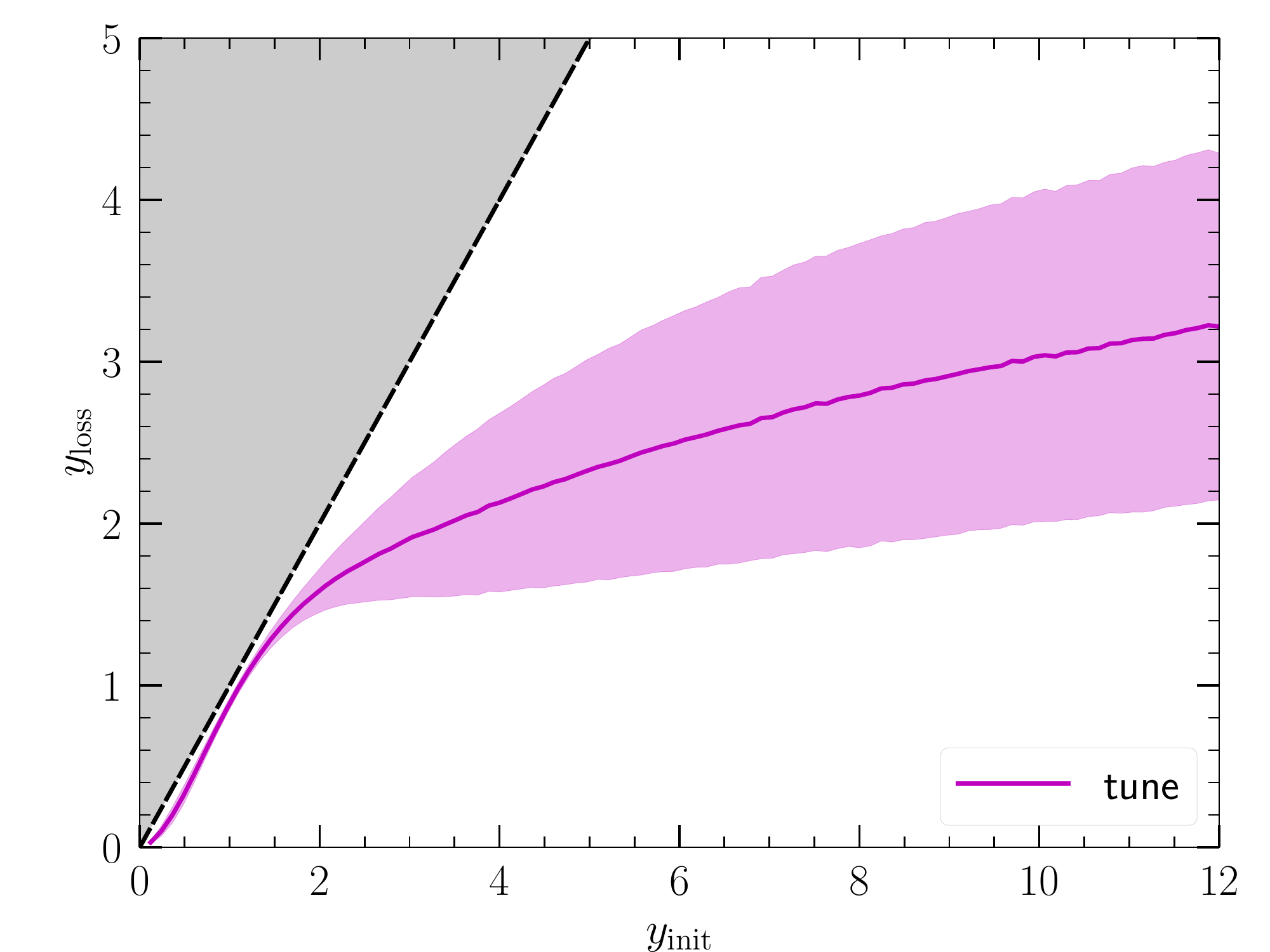}
  \caption{(Color online) The valence quark rapidity loss as a function of its initial rapidity in the collision pair rest frame. The shaded region indicates one standard deviation of the rapidity loss fluctuation. The gray area is excluded because $y_{\rm loss} \leq y_{\rm init}$.}
  \label{fig:yloss}
\end{figure}
%
Figure~\ref{fig:yloss} shows the mean and variance of the rapidity loss in the individual NN collisions as a function of the initial rapidity $y_\mathrm{init}$ calibrated using p+p collisions from 23.6 GeV to 200 GeV. For the parameters appearing in Eq.\,\eqref{eq:ylossMean} we find $A = 1.32$, $\alpha_1 = 1.8$, and $\alpha_2 = 0.35$.
The variance of the rapidity loss fluctuation in Eq.\,\eqref{eq:logitDis} is set to $\sigma = 0.6$ to describe the charged hadron multiplicity distribution p+p collisions at 200 GeV (see Fig.~\ref{fig:ppCalibration}b below). The variance is kept the same when extrapolated to LHC collision energies. Note that the mapping procedure for rapidity loss described in Sec.~\ref{subsec:yloss} automatically shrinks the magnitude of rapidity loss fluctuations as the mean rapidity loss is close to the initial rapidity for $y_\mathrm{init} \lesssim 2$.

\subsection{Small systems at RHIC}

We start our phenomenological discussion with minimum bias p+p collisions at collision energies relevant for the RHIC BES program.

\begin{figure}[ht!]
  \centering
  \includegraphics[width=0.9\linewidth]{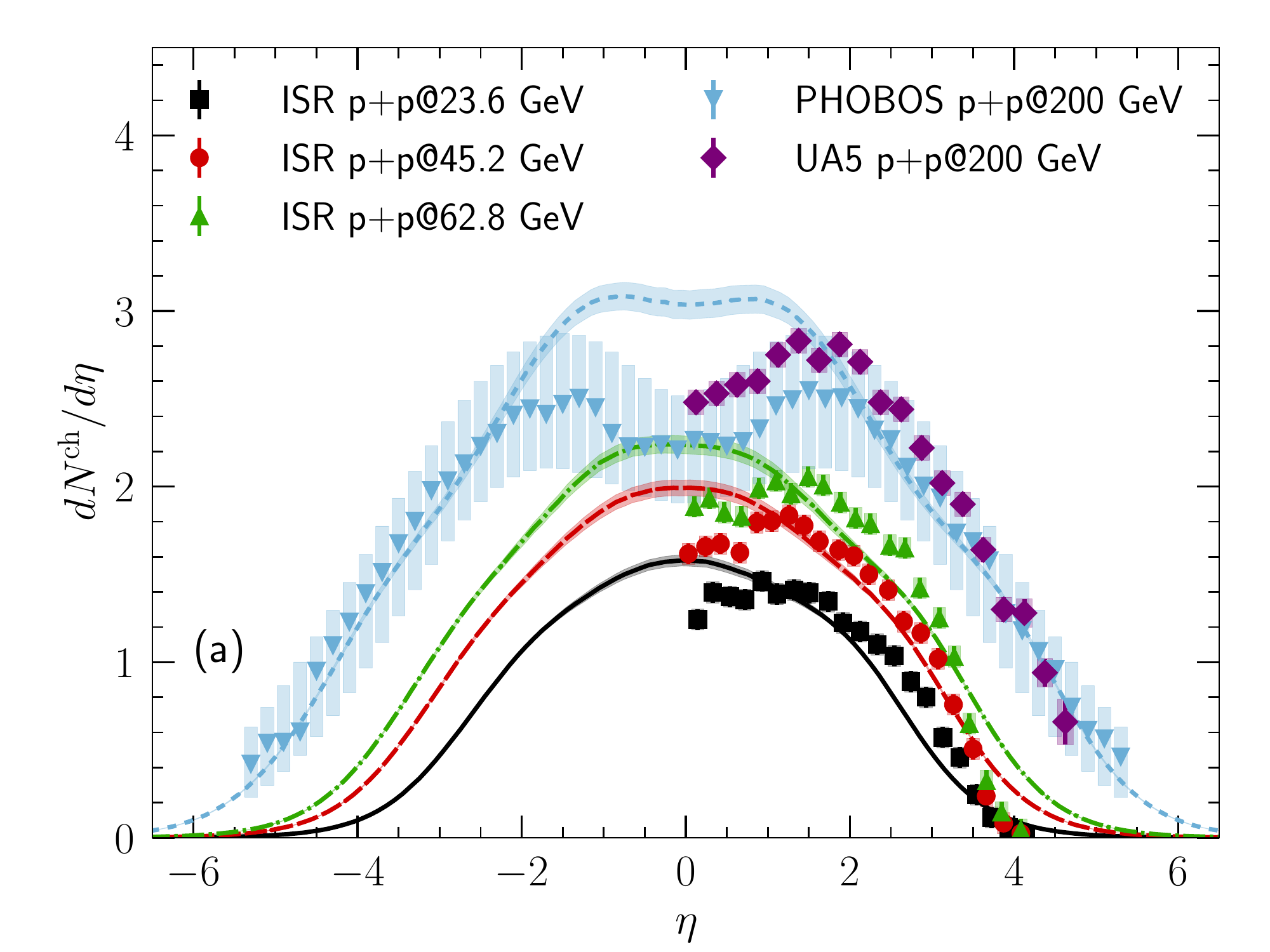}
  \includegraphics[width=0.9\linewidth]{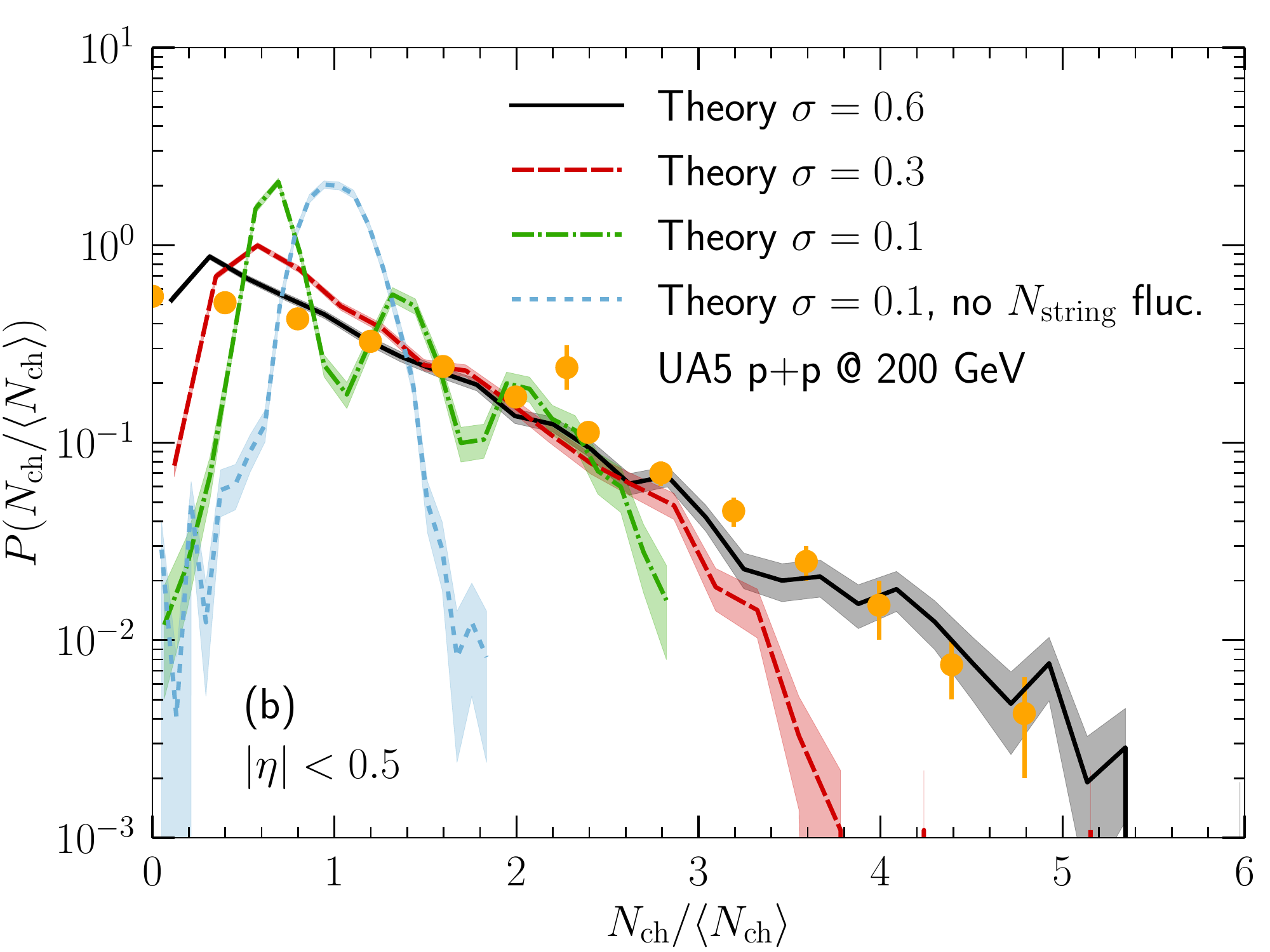}
  \caption{(Color online) Panel (a): Charged hadron pseudo-rapidity distributions in minimum bias p+p collisions from 23.6 GeV to 200 GeV center of mass energy compared with experimental data \cite{Aachen-CERN-Heidelberg-Munich:1977izz, UA5:1986yef, PHOBOS:2010eyu}. Panel (b): Event-by-event multiplicity distributions in p+p collisions at 200 GeV compared to experimental data from the UA5 Collaboration \cite{UA5:1988gup}.}
  \label{fig:ppCalibration}
\end{figure}

Figure~\ref{fig:ppCalibration}a shows our model calculation of the pseudo-rapidity distributions of charged hadrons compared to the experimental measurements in minimum bias p+p collisions from 23.6 GeV to 200 GeV. The average numbers of the produced charged particles are correlated with the amount of rapidity loss in individual NN collisions. Using the rapidity-loss parametrization shown in Fig.~\ref{fig:yloss}, our hybrid model can give a reasonable description of the particle production in minimum bias p+p collisions for $\vert \eta \vert > 2$, while it overestimates the mid-rapidity particle yield by about 10-20\% compared with the experimental measurements.

Figure~\ref{fig:ppCalibration}b shows the charged hadron multiplicity distribution in p+p collisions at 200 GeV. As stated above, in our model the number of charged hadrons produced at mid-rapidity is correlated with the amount of rapidity loss in individual NN collision. Therefore, the particle multiplicity fluctuations are directly related to the rapidity loss fluctuations. With a small variance of the rapidity loss fluctuations, $\sigma = 0.1$, the charged hadron multiplicity distribution at mid-rapidity has multiple peaks, resulting from the fluctuating number of contributing strings (see Sec.~\ref{subsec:NstringFluct}). These peaks in the charged hadron multiplicity distribution disappear as $\sigma$ increases to 0.3. We find that $\sigma = 0.6$ produces enough fluctuations at midrapidity to achieve good agreement with the UA5 data \cite{UA5:1988gup}. We also demonstrate that when turning off fluctuations of the number of strings in the case that  $\sigma = 0.1$, the distribution has only one peak and is significantly narrower than when allowing the number of strings to fluctuate.

\begin{figure}[ht!]
  \centering
  \includegraphics[width=0.9\linewidth]{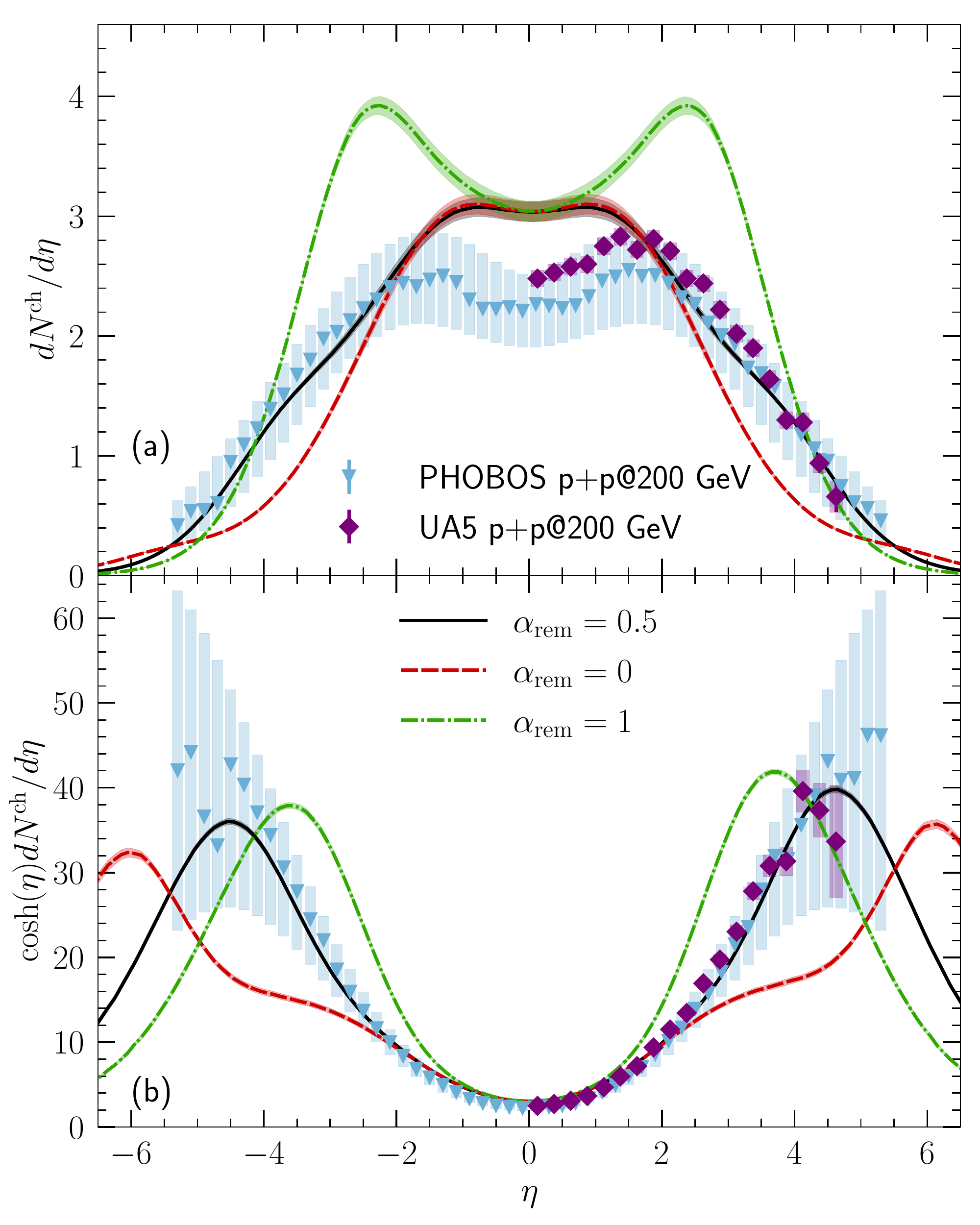}
  \caption{(Color online) Panel (a): Effects of nucleon remnants' energy loss on charged hadron pseudo-rapidity distributions in minimum-bias p+p collisions at 200 GeV. Panel (b): Estimation of the longitudinal energy distribution of charged hadrons.}
  \label{fig:ppRemLoss}
\end{figure}

In p+p collisions, it is instructive to understand how the 
nucleon remnant energy loss of the colliding nucleons affects the pseudo-rapidity distributions of produced hadrons.
Figure~\ref{fig:ppRemLoss}a shows that the amount of energy loss of the nucleon remnants has sizable effects on charged hadron production for $|\eta| > 2$ at 200 GeV. It is easier to understand how the particle production depends on the nucleon remnant energy loss fraction parameter $\alpha_\mathrm{rem}$ by looking at an approximation of the longitudinal energy distribution of charged hadrons in Figure~\ref{fig:ppRemLoss}b. Here the charged hadrons' longitudinal energy distribution is estimated by multiplying the charged hadron yields with a factor of $\cosh(\eta)$. Simulations without energy loss in the beam remnants $(\alpha_\mathrm{rem} = 0)$ underestimate the energy distribution for $|\eta| > 2$, while simulation results with $\alpha_\mathrm{rem} = 1$ overestimate the stopping power. 

After studying the particle production in p+p collisions with our hybrid framework, we extrapolate our calculations to the asymmetric light+heavy ion collisions at similar collision energies. 
We adopt the same model parameters as those in p+p collisions but use the switching energy density $e_\mathrm{sw}$ and the coherent parameter $\lambda_\mathrm{coh}$ which are tuned to match the identified particle yields in heavy-ion collisions. The values of $e_\mathrm{sw}$ and $\lambda_\mathrm{coh}$ at different collision energies are listed in Table~\ref{tab:ModelParams} below. The charged hadron results in p+p collisions are insensitive to these parameters.

\begin{figure}[ht!]
  \centering
  \includegraphics[width=1.0\linewidth]{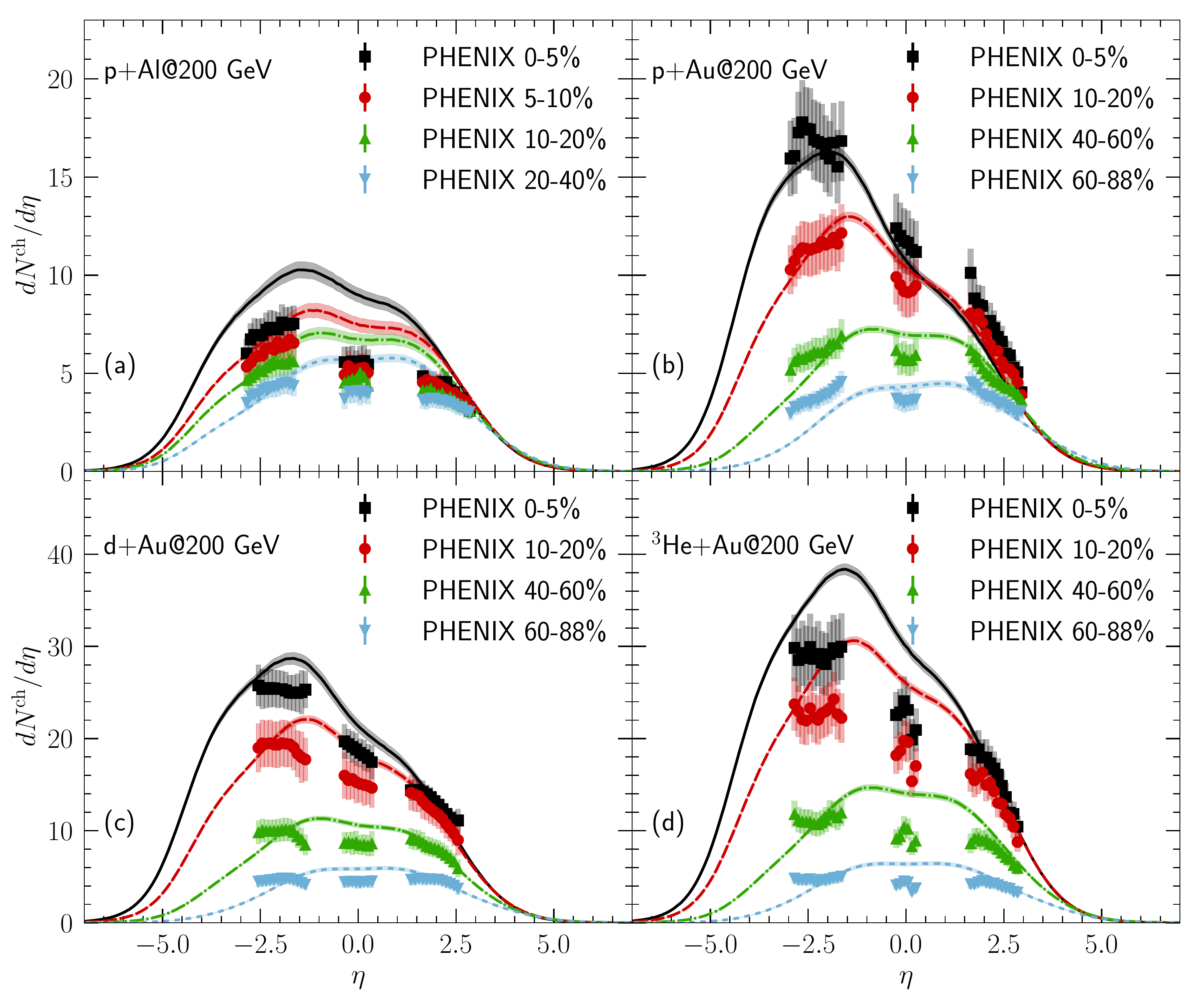}
  \caption{(Color online) Charged hadron pseudo-rapidity distributions for p+Al, p+Au, d+Au, and $^3$He+Au collisions at 200 GeV. Theoretical results in different centrality bins are compared with the experimental data from the PHENIX Collaboration \cite{PHENIX:2018hho}.}
  \label{fig:RHICSmallSystems}
\end{figure}

Figure~\ref{fig:RHICSmallSystems} shows the charged hadron pseudo-rapidity distributions in different centrality bins for p+Al, p+Au, d+Au, and $^3$He+Au collisions at 200 GeV. Our model predictions produce a reasonable description of the experimental data from the PHENIX Collaboration \cite{PHENIX:2018hho} for all four collision systems. The evolution of the asymmetric shape of \dNdeta{} from central to peripheral collisions is well captured.

We note that the full 3+1D simulations allow us to determine the collision centrality the same way as the PHENIX experiment, using the charged hadron multiplicity measured in the Au-going direction with pseudo-rapidity $-3.9 < \eta < -3.1$ \cite{PHENIX:2018hho}. The forward to the mid-rapidity correlation of particle production is crucial to reproduce the centrality dependence of charged particles in these asymmetric collision systems. Our hybrid model describes this correlation reasonably well in these asymmetric collisions. The particle yields at midrapidity in p+Al, d+Au, and $^3$He+Au collisions are overestimated, most significantly so in p+Al collisions. Fine-tuning of initial state energy loss around $y_\mathrm{init} \approx 5$ will likely help to improve the overall description. We leave such a calibration for future studies within a robust Bayesian framework.

\begin{figure}[ht!]
  \centering
  \includegraphics[width=0.95\linewidth]{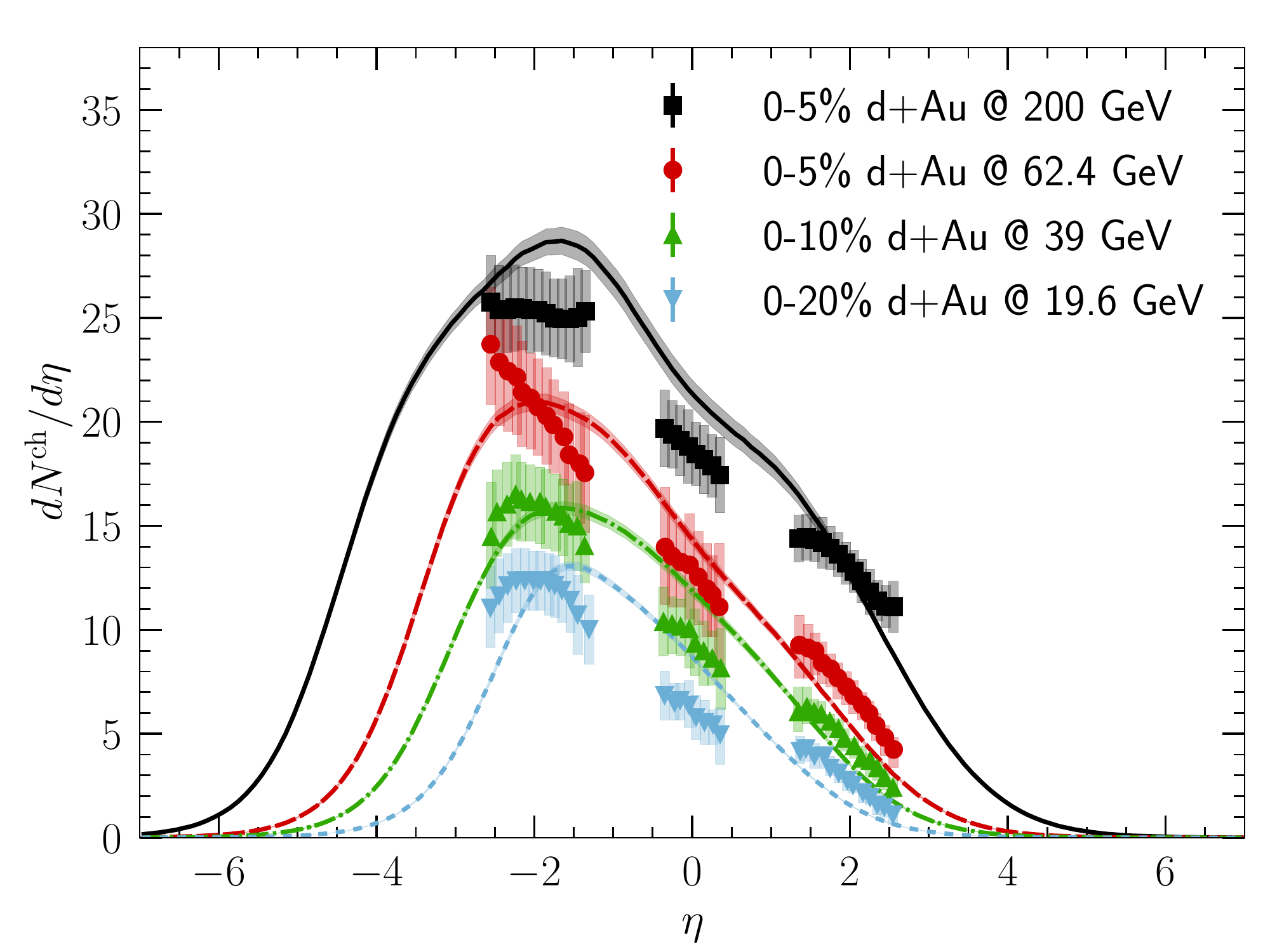}
  \caption{(Color online) Charged hadron pseudo-rapidity distributions for central d+Au collisions at 200, 62.4, 39, and 19.6 GeV. Theoretical results are compared with experimental data from the PHENIX Collaboration \cite{PHENIX:2017nae}.}
  \label{fig:RHICdAuBES}
\end{figure}

Figure~\ref{fig:RHICdAuBES} shows our model predictions for the collision energy dependence of the charged hadron pseudorapidity distribution in central d+Au collisions. We find a good description of the asymmetric charged hadron pseudo-rapidity distributions compared with the PHENIX data \cite{PHENIX:2017nae} from 200 GeV down to 19.6 GeV. The particle yields at mid-rapidity are about 10\% over-predicted. For $\snn < 40$ GeV, the theoretical \dNdeta{} curves peak at a slightly smaller $|\eta|$ compared with the PHENIX data on the Au-going side.

\begin{figure}[ht!]
  \centering
  \includegraphics[width=0.95\linewidth]{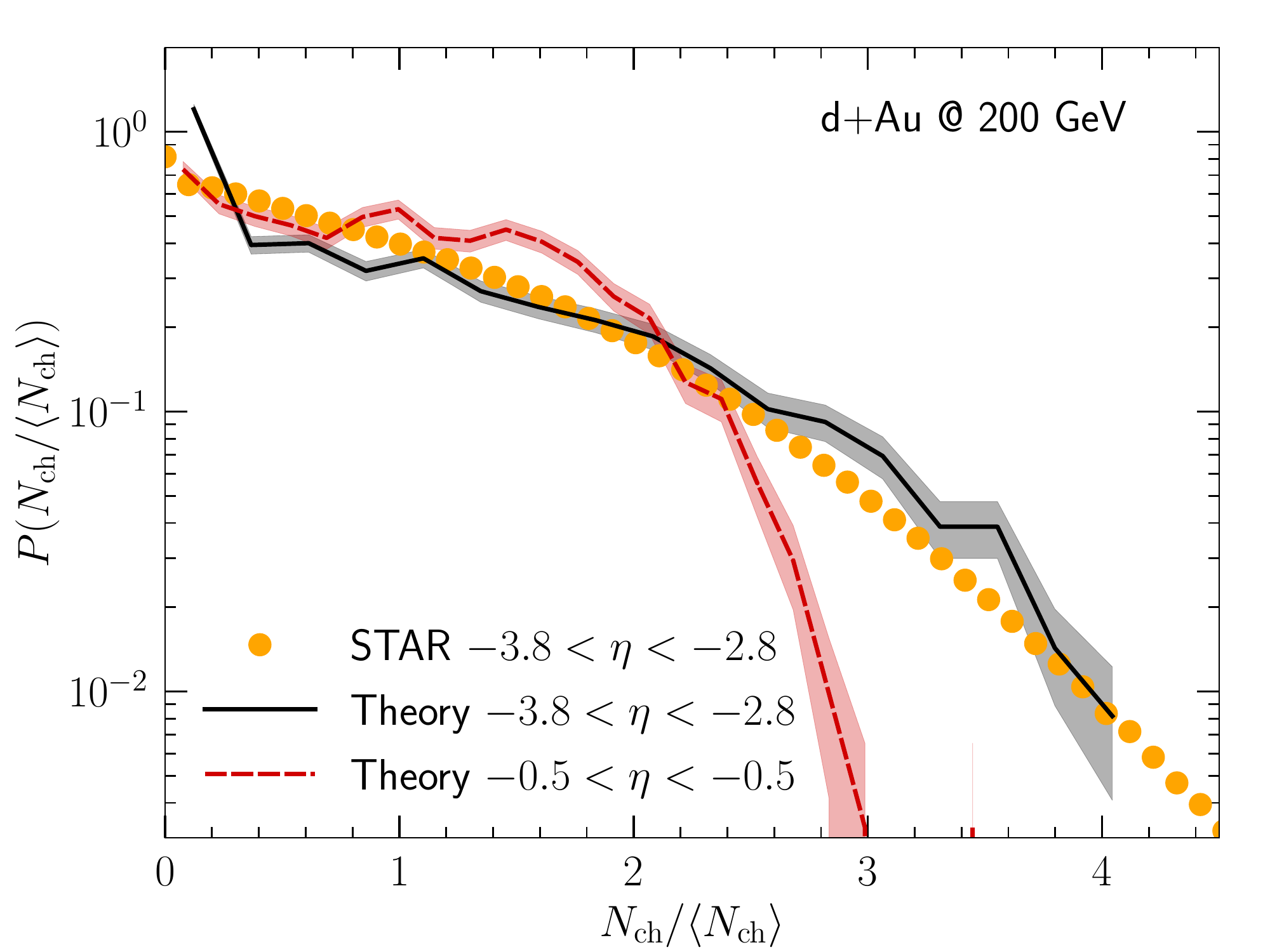}
  \caption{(Color online) Charged hadron multiplicity distribution in minimum bias d+Au collisions at 200 GeV compared with experimental data from the STAR Collaboration \cite{STAR:2008med}.}
  \label{fig:dAuMultDis}
\end{figure}

Figure~\ref{fig:dAuMultDis} shows the charged hadron multiplicity distribution in d+Au collisions at 200 GeV. Our model provides a good description of the charged hadron distribution in the Au-going rapidity region $-3.8 < \eta < -2.8$ measured by the STAR Collaboration \cite{STAR:2008med}. We also compare them with the normalized charged hadron multiplicity distribution at mid-rapidity. The mid-rapidity multiplicity distribution is narrower than that in the backward rapidity region and only extends to about three times the average. This shows that measurements of particle distributions in different rapidity windows will help to constrain the initial state longitudinal energy loss.

\subsection{Small systems at LHC}

We now further extrapolate our calculations to the higher LHC energies and present predictions for charged hadron production in p+p and p+Pb collisions. 

\begin{figure}[ht!]
  \centering
  \includegraphics[width=0.9\linewidth]{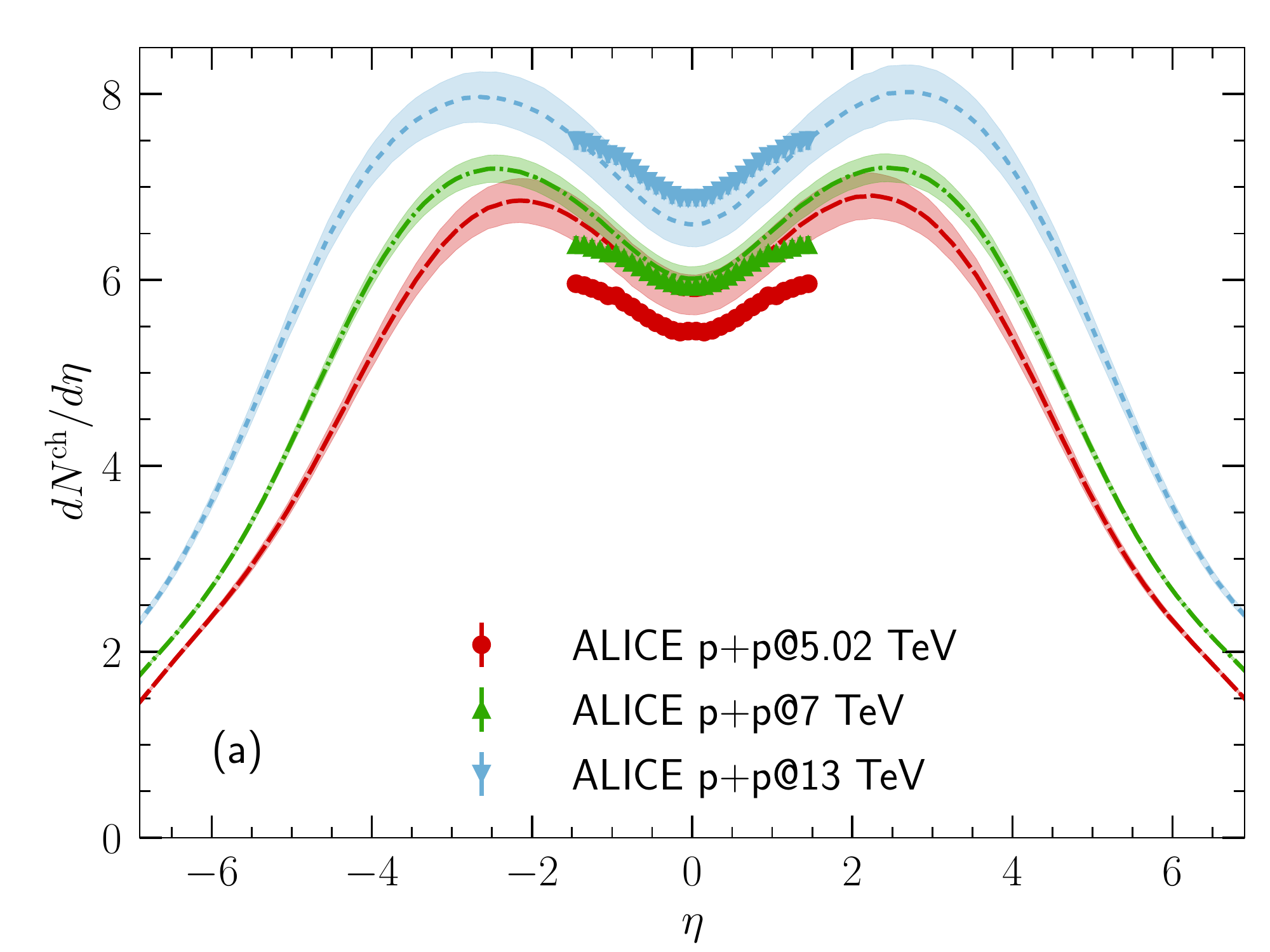}
  \includegraphics[width=0.9\linewidth]{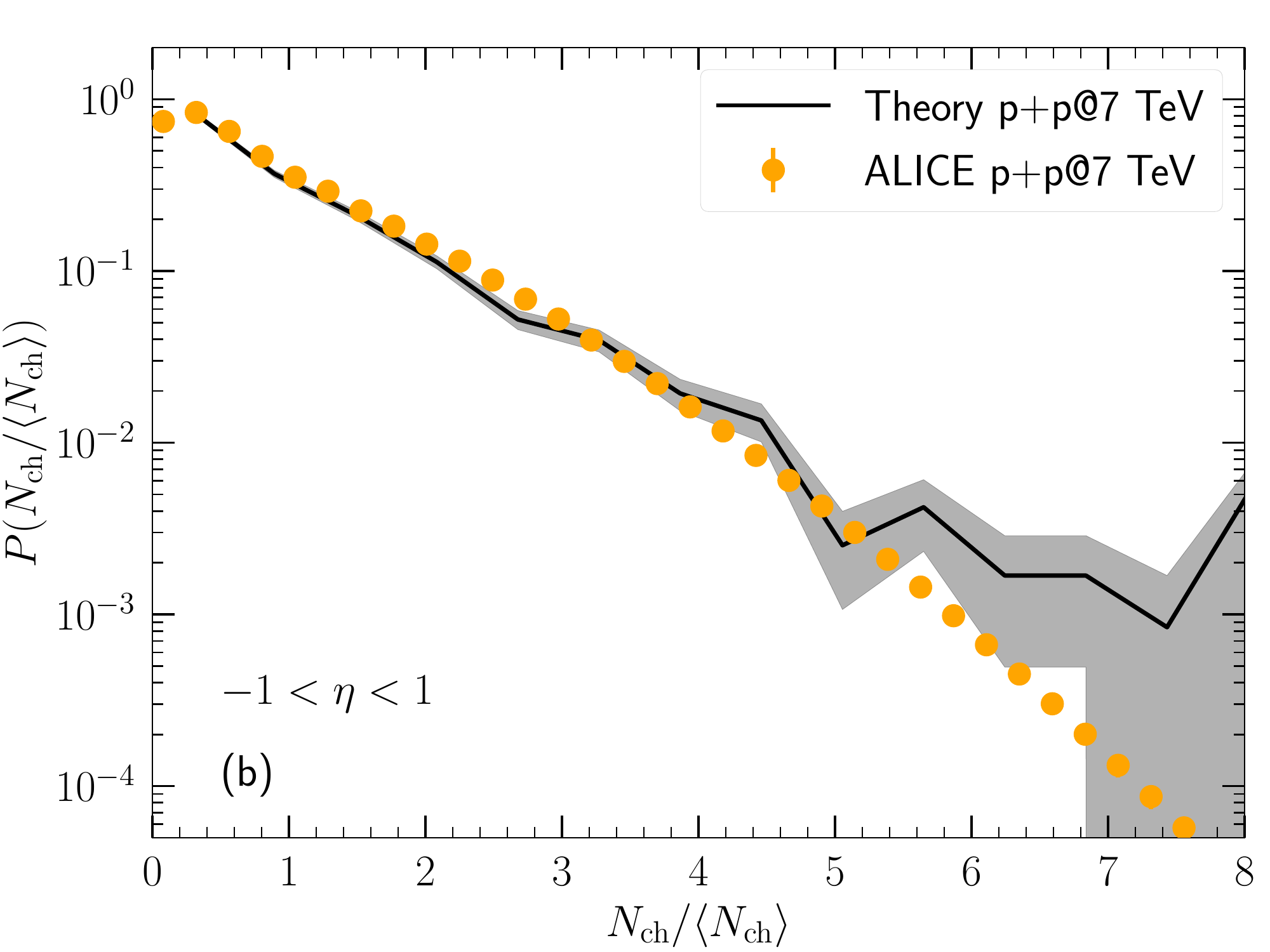}
  \caption{(Color online) Panel (a): Charged particle pseudo-rapidity distributions in minimum bias p+p collisions from 5.02 TeV to 13 TeV compared with experimental data from the ALICE Collaboration \cite{ALICE:2020swj}. Panel (b): Normalized multiplicity distribution for p+p collisions at 7 TeV compared with experimental data from the ALICE Collaboration \cite{ALICE:2015olq}. }
  \label{fig:ppCalibrationLHC}
\end{figure}

Figure~\ref{fig:ppCalibrationLHC}a shows the model to experimental data comparisons of the charged hadron pseudo-rapidity distributions in minimum bias p+p collisions for center of mass energies ranging from 5.02 TeV to 13 TeV. Using the rapidity loss parameterization shown in Fig.~\ref{fig:yloss}, our model gives a reasonable description of the mid-rapidity charged hadron production at 7 and 13 TeV, while overestimating the hadron yield by 10\% for 5.02 TeV. Figure~\ref{fig:ppCalibrationLHC}b shows the normalized multiplicity distribution at 7 TeV compared with the experimental data from the ALICE collaboration \cite{ALICE:2015olq}. Again, using the variance of rapidity fluctuations $\sigma = 0.6$, our model gives a good description of the data up to five times the average. The tail of the distribution is flatter than the data for $N_\mathrm{ch}/\langle N_\mathrm{ch} \rangle > 5$.

Moving to asymmetric p+Pb collisions at the LHC, because hydrodynamic simulations are performed in the local rest frame of nucleon-nucleon collisions, we need to apply a global rapidity boost of $\Delta y = 0.465$ towards the p-going direction \cite{ALICE:2014xsp, ALICE:2018wma, ATLAS:2015hkr} for all particles produced to the rest frame of the LHC detectors.

\begin{figure*}[ht!]
  \centering
  \includegraphics[width=0.6\linewidth]{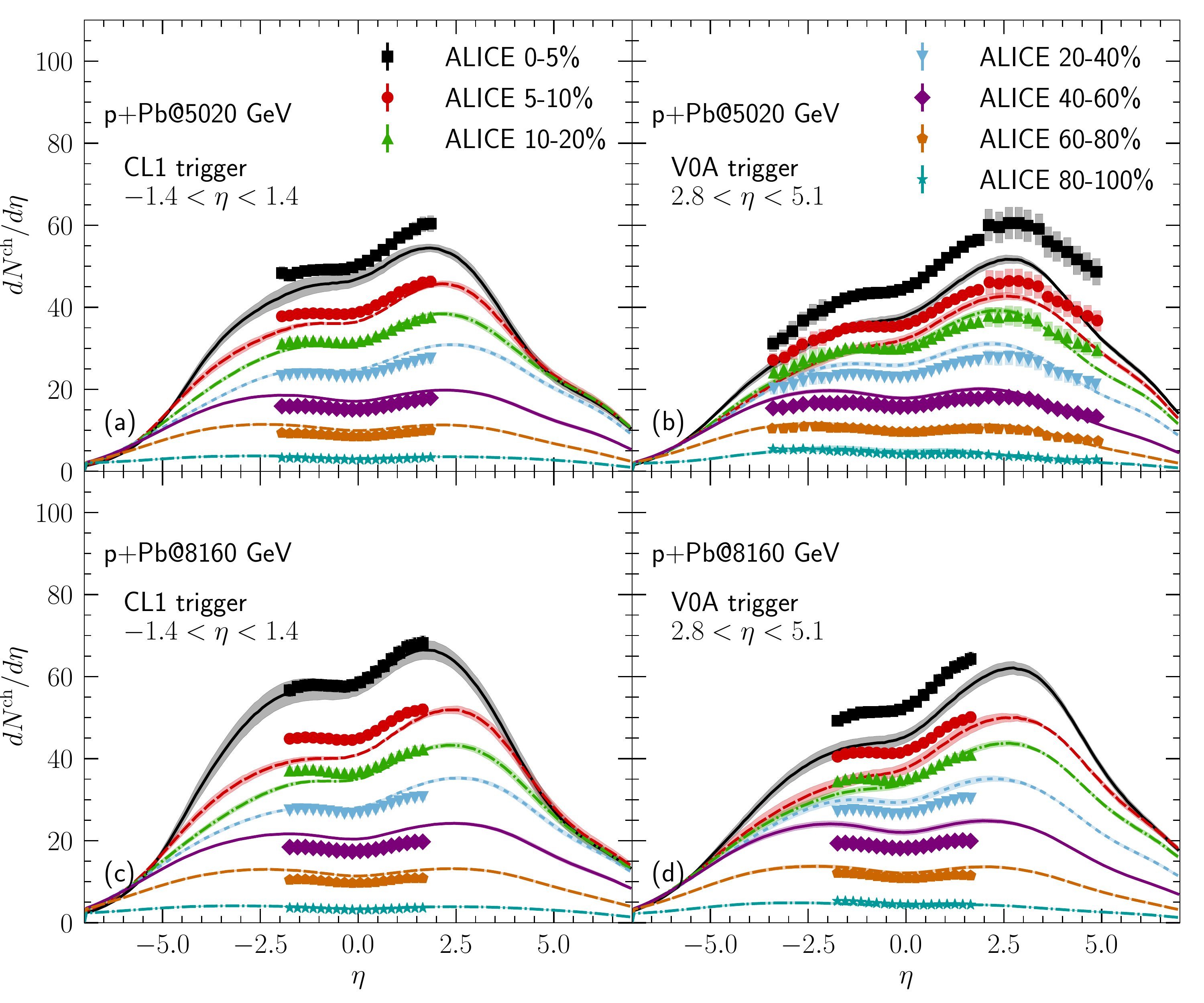}
  \caption{(Color online) Charged hadron pseudo-rapidity distributions in different centrality bins for p+Pb collisions at 5020 GeV (a, b) and 8160 GeV (c, d). Theoretical results with centrality defined by the CL1 (a, c) and V0A (b, d) triggers are compared with the ALICE measurements \cite{ALICE:2014xsp, ALICE:2018wma, Christensen:2017zqh}.}
  \label{fig:pPbLHC}
\end{figure*}

Figure~\ref{fig:pPbLHC} shows the charged hadron pseudo-rapidity distributions in p+Pb collisions at 5.02 and 8.16 TeV for different centrality bins. Similar to asymmetric collisions at RHIC, employing the correct criterion of centrality determination is crucial for a meaningful comparison. We compare the model results with the ALICE data for two different centrality triggers. The CL1 trigger uses the charged-particle multiplicity in the central region $\vert \eta \vert < 1.4$ to define the centrality classes. Figs.~\ref{fig:pPbLHC}a and c show that our predictions for \dNdeta$(\eta)$ are in good agreement with the ALICE measurements using the CL1 trigger. Within our model, the equivalent to the CL1 trigger selects the collision events with the most rapidity loss in 0-5\% p+Pb collisions. Because of the constraints from energy-momentum conservation, the particle production in the forward direction in these central collision events should be suppressed compared to peripheral collision events which have less rapidity loss. This anti-correlation between mid-rapidity and forward rapidity particle yields can indeed be seen in Figs.~\ref{fig:pPbLHC}a and c. The 0-5\% centrality class has almost the same or less charged hadrons at forward rapidity $\eta > 5$ (the Pb-going side) than the 5-10\% centrality bin.

In Figures~\ref{fig:pPbLHC}b and d, we define centrality bins using the charged hadron yields measured in the region $2.8 < \eta < 5.1$, corresponding to the V0A trigger. In this case, for central collisions the average charged hadron yield at mid-rapidity is smaller than for the CL1 trigger.

Our model gives a good description of the centrality dependence of p+Pb collisions with the central trigger but underestimates the charged particle yield by 10-15\% in 0-5\% central p+Pb collisions with the V0A trigger. This suggests that the correlation of particle multiplicities between central and forward rapidities is stronger in the data than in our model.

Although there is room for improvement in our (3+1)D model, we want to emphasize that the large difference between rapidity distributions obtained using different centrality selection methods indicates that it is crucial to perform the centrality selection the same way as the experiments, especially for asymmetric collision systems. This is only possible using the full (3+1)D dynamical modeling of relativistic nuclear collisions.

\section{Particle production in heavy-ion collisions}\label{sec:results}

In this section, we apply our calibrated framework to simulate heavy-ion collisions in the RHIC Beam Energy Scan and then extend the study to CERN SPS and LHC energies.

As we described in Sec.~\ref{sec:model}, we adjust the coherence parameter $\lambda_\mathrm{coh}$ for every collision energy so that the charged hadron multiplicity is reproduced in 0-5\% central collisions.
We also adjust the switching energy density to reproduce the $p/\pi$ ratio, as well as the smearing width $\sigma_{\eta_s}$ for every collision energy. Those parameters are listed in Table~\ref{tab:ModelParams} for different collision energies.

\begin{table}[]
    \centering
    \caption{Additional model parameters which are relevant for light+heavy and heavy-ion collisions at different collision energies.}
    \begin{tabular}{|c|c|c|}\hline
        $\sqrt{s_\mathrm{NN}}$\,[GeV] & $e_\mathrm{sw}$\,[GeV/fm$^3$] & $\lambda_\mathrm{coh}$ \\ \hline
        7.7 & 0.25 & 0.1 \\ \hline
        8.77 & 0.25 & 0.1 \\ \hline
        17.3 & 0.25 & 0.2 \\ \hline
        19.6 & 0.25 & 0.2 \\ \hline
        39 & 0.35 & 0.2 \\ \hline
        62.4 & 0.45 & 0.2 \\ \hline
        200 & 0.5 & 0.25 \\ \hline
        2760 & 0.2 & 0.35 \\ \hline
        5020 & 0.2 & 0.35 \\ \hline
        5440 & 0.2 & 0.35 \\ \hline
        8160 & 0.2 & 0.35 \\ \hline
    \end{tabular}
    \label{tab:ModelParams}
\end{table}

For heavy-ion collisions in the RHIC BES program, we perform full minimum bias simulations and determine the centrality based on the final charged hadron multiplicity at mid-rapidity, $\vert \eta \vert < 0.5$ \cite{STAR:2017sal}. In Appendix~\ref{app:centralityEst}, we will discuss a good initial-state estimator for collision events' centrality in symmetric heavy-ion collisions, which could be used to speed up simulations if one is interested in observables in a specific centrality bin.

\begin{figure}[ht!]
  \centering
  \includegraphics[width=1.0\linewidth]{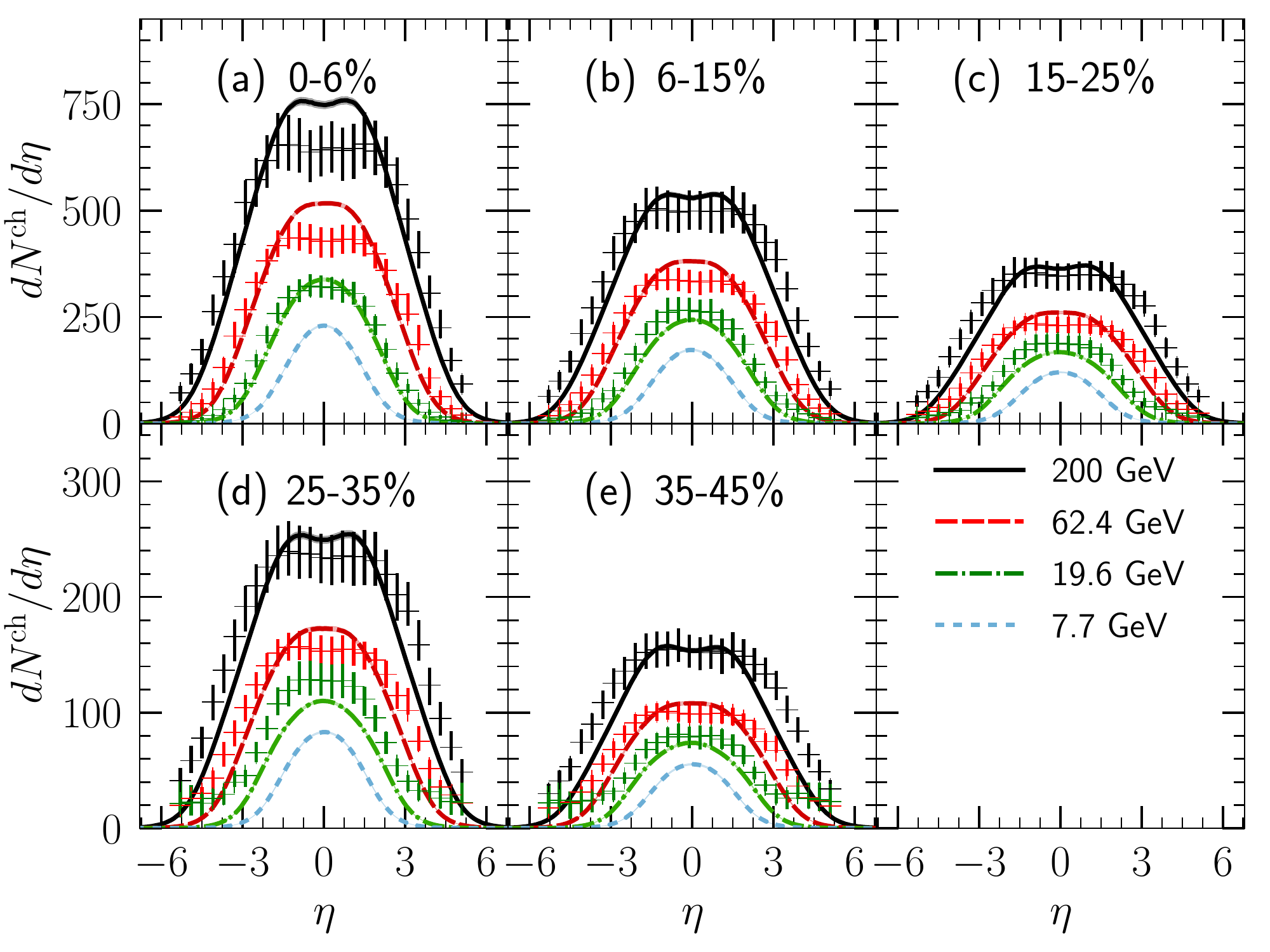}
  \caption{(Color online) The centrality dependence of charged hadron pseudo-rapidity distributions in Au+Au collisions at 19.6, 62.4, and 200 GeV. Theoretical calculations are compared with experimental data from the PHOBOS Collaboration \cite{PHOBOS:2005zhy}.}
  \label{fig:BES_dNchdeta}
\end{figure}
%
Figure~\ref{fig:BES_dNchdeta} shows the pseudo-rapidity distributions of charged hadrons in Au+Au collisions in five centrality bins from 200 GeV down to 7.7 GeV. Our model gives a good description of the experimental data from the PHOBOS Collaboration from central to semi-peripheral centralities. This figure together with Figs.~\ref{fig:ppCalibration}, \ref{fig:RHICSmallSystems}, and \ref{fig:RHICdAuBES} demonstrates that our hybrid framework can provide a consistent description of particle production from small p+p collisions to heavy-ion collisions at RHIC BES energies.

We find that the best fit values of the coherence parameter $\lambda_\mathrm{coh}$ in Table~\ref{tab:ModelParams} increase with the collision energy, which is consistent with the saturation picture in the high energy limit of the nucleus-nucleus collisions. The effective number of collisions reduces as collision energy increases.

Because net-baryon fluctuations could have the potential to reveal the existence and position of the QCD critical point in the phase diagram \cite{Vovchenko:2021kxx}, it is of crucial importance for the RHIC BES program to quantify and understand the energy loss of the baryon charges during the initial state as a function of collision energy.

\begin{figure}[ht!]
  \centering
  \includegraphics[width=1.0\linewidth]{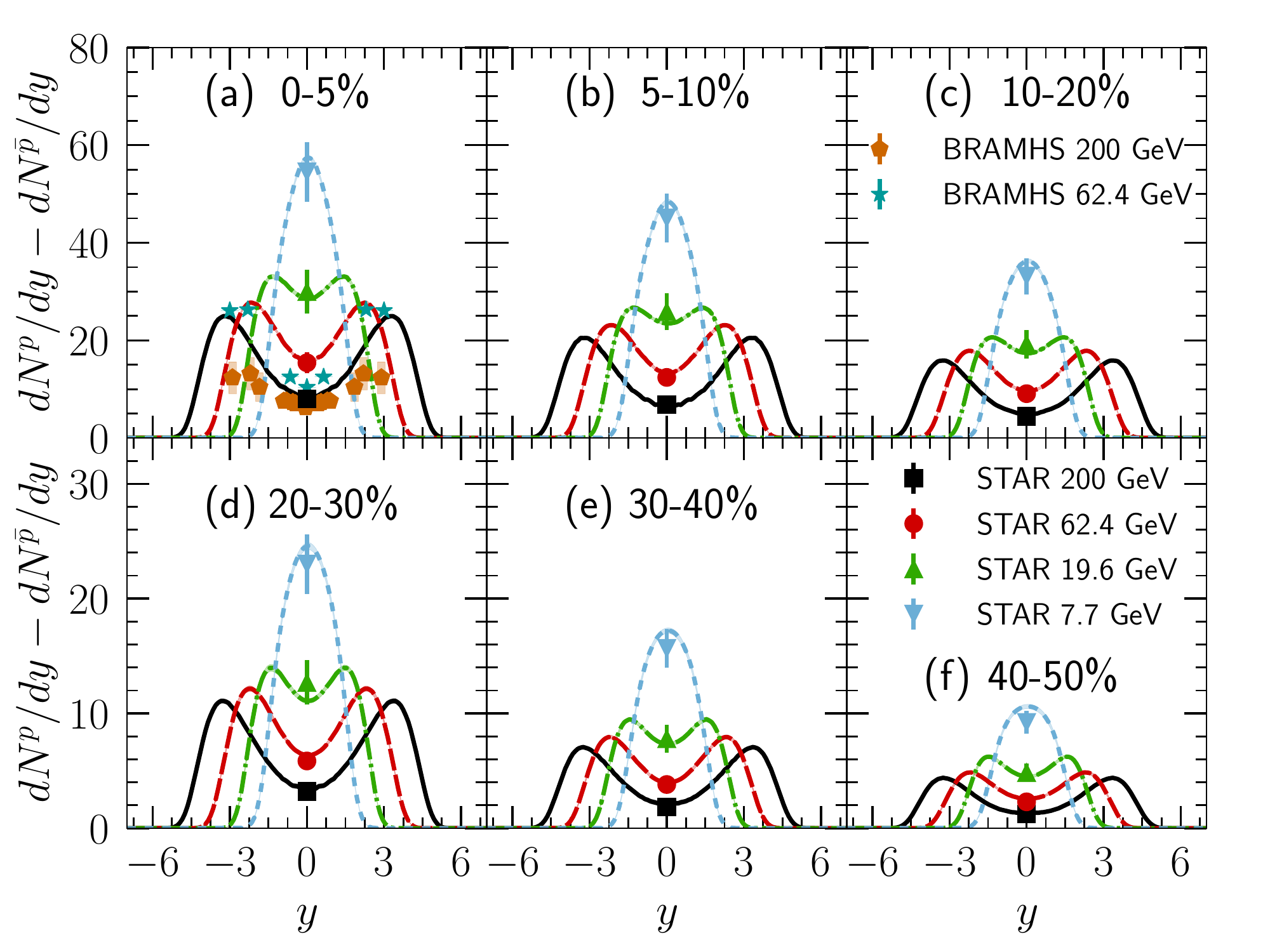}
  \caption{(Color online) The rapidity distributions of net protons for six centrality bins in Au+Au collisions at 7.7, 19.6, 62.4, and 200 GeV. Theoretical calculations are compared with the experimental data from the BRAMHS and STAR Collaborations \cite{BRAHMS:2003wwg, BRAHMS:2009wlg, STAR:2008med, STAR:2017sal}. Weak decays are included for both protons and anti-protons in the calculations.}
  \label{fig:BES_netProtondNdy}
\end{figure}
%
Experimentally, net-baryon distributions are hard to obtain, but net-proton distributions can be measured much more easily.
Figure~\ref{fig:BES_netProtondNdy} shows the net proton rapidity distribution from 7.7 GeV to 200 GeV. Our model predictions show remarkable agreement with the STAR and BRAMHS data. The baryon stopping in our hybrid framework is controlled by the energy loss of the colliding nucleon and the probability of baryon charge fluctuations according to the string junction model. Setting the parameter $\lambda_B = 0.2$, our hybrid framework provides a consistent description of the net proton rapidity distributions across centrality bins and collision energies.

\begin{figure}[ht!]
  \centering
  \includegraphics[width=0.95\linewidth]{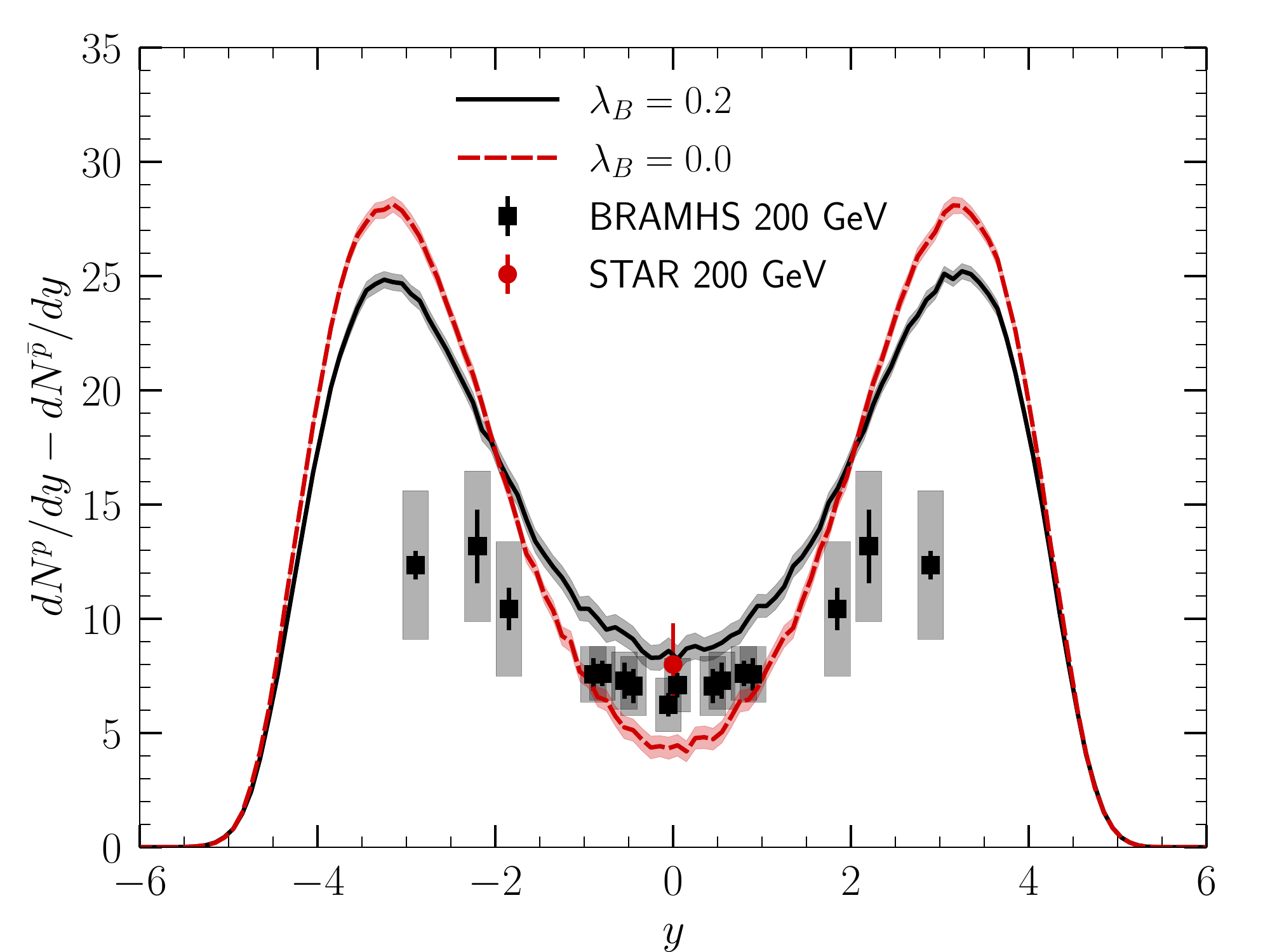}
  \caption{(Color online) The effects of baryon charge fluctuations to string junctions on the net proton rapidity distributions in Au+Au collisions at 200 GeV \cite{BRAHMS:2003wwg, STAR:2008med}.}
  \label{fig:BaryonJunction}
\end{figure}
%
Figure~\ref{fig:BaryonJunction} shows the effects of baryon charge fluctuations to string junctions on the net proton rapidity distributions. The non-zero $\lambda_B$ allows a finite probability for the initial baryon charges to fluctuate from the string ends to string junctions during the initial collision. At 200 GeV, there is a visible effect of baryon junction fluctuations to transport baryon charge from forward rapidity regions to mid-rapidity.

\begin{figure}[ht!]
  \centering
  \includegraphics[width=\linewidth]{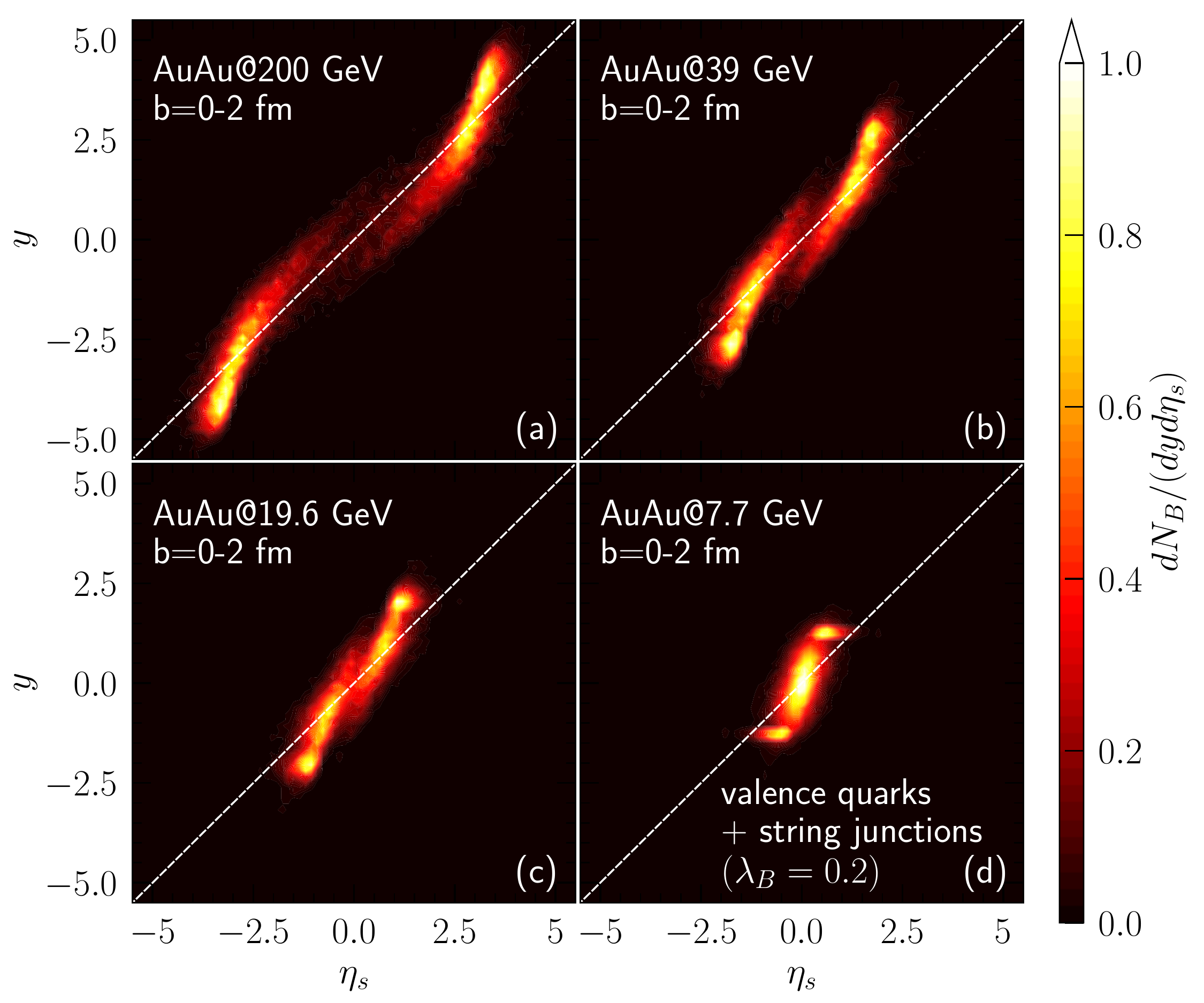}
  \caption{(Color online) The correlation of net baryon charge's space-time rapidity and momentum rapidity for central Au+Au collisions at $\snn = $ 200 GeV (a), 39 GeV (b), 19.6 GeV (c), and 7.7 GeV (d).}
  \label{fig:nBcorr}
\end{figure}
%
Figure~\ref{fig:nBcorr} shows the baryon charges' spatial-momentum correlations along the longitudinal direction after the initial state deceleration dynamics. The baseline $y = \eta_s$ represents the baryon's free-streaming limit at late times. In the forward rapidity regions, the baryon charge's rapidity is larger than its space-time rapidity for all collision energies. This structure is a consequence of the finite longitudinal overlapping region, as the baryon charges produced at later times (but at similar longitudinal position $z$) have a reduced space-time rapidity.

\begin{figure*}[ht!]
  \centering
  \includegraphics[width=0.7\linewidth]{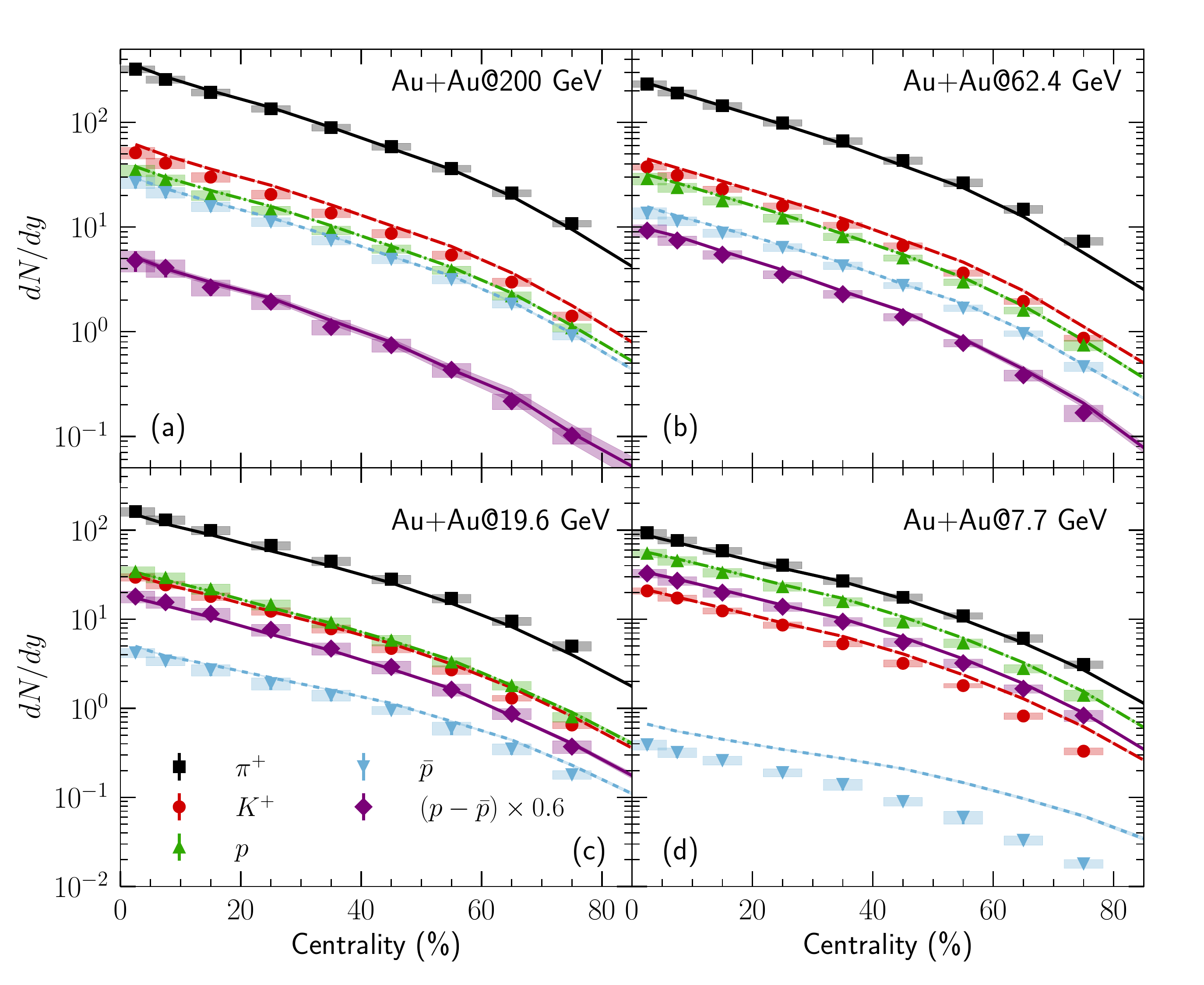}
  \caption{(Color online) The centrality dependence of identified particle yields in the mid-rapidity region of Au+Au collisions from 7.7 to 200 GeV compared with the experimental data from the STAR Collaboration \cite{STAR:2017sal}. Weak decays are included for protons and anti-protons in the calculations.}
  \label{fig:piddNdy}
\end{figure*}
Closer to the mid-rapidity region, the two blobs in Fig.~\ref{fig:nBcorr}a cross the $y=\eta_s$ baseline, meaning that the baryon charge's rapidity becomes smaller than its space-time rapidity. This qualitative change is introduced by the deceleration dynamics because the baryon charges move forward as they lose energy and momentum. The baryons that get stopped (with $y = 0$) sit at forward space-time rapidity. Also for $\snn = 39$ GeV, the baryon charges from the projectile and target nuclei form two distinct blobs. As the collision energy decreases more, these two distinct regions start to overlap. At $\snn = 7.7$ GeV, they are merged together and the $y < \eta_s$ correlation from the deceleration dynamics is mixed with the $y > \eta_s$ correlations from the collision geometry. The two blobs at $\vert y \vert \sim 2$ indicate a significant fraction of baryon charges is carried by the beam remnants.

Figure~\ref{fig:piddNdy} compares identified particle yields as functions of the collision centrality with the experimental data from the STAR Collaboration at mid-rapidity for collision energies from 7.7 to 200 GeV. The measured yields of light-flavor mesons, protons, and anti-protons are well reproduced by our hybrid framework, except for anti-protons at 7.7 GeV. We note that the centrality dependence of identified particle yields is a prediction of our model. 

The hadronic chemistry in our model is determined by the particlization energy density $e_\mathrm{sw}$ and the net baryon density distribution on this particlization hyper-surface. The latter is the result of convoluting the initial state baryon stopping with the propagation of the net baryon current in the hydrodynamic phase.
The constraints on strangeness neutrality $n_S = 0$ and net electric charge density $n_Q = 0.4 n_B$ in our employed equation of state play an essential role in generating differences between the yields of particles and their anti-partners \cite{Shen:2021nbe}.

\begin{figure}[ht!]
  \centering
  \includegraphics[width=0.9\linewidth]{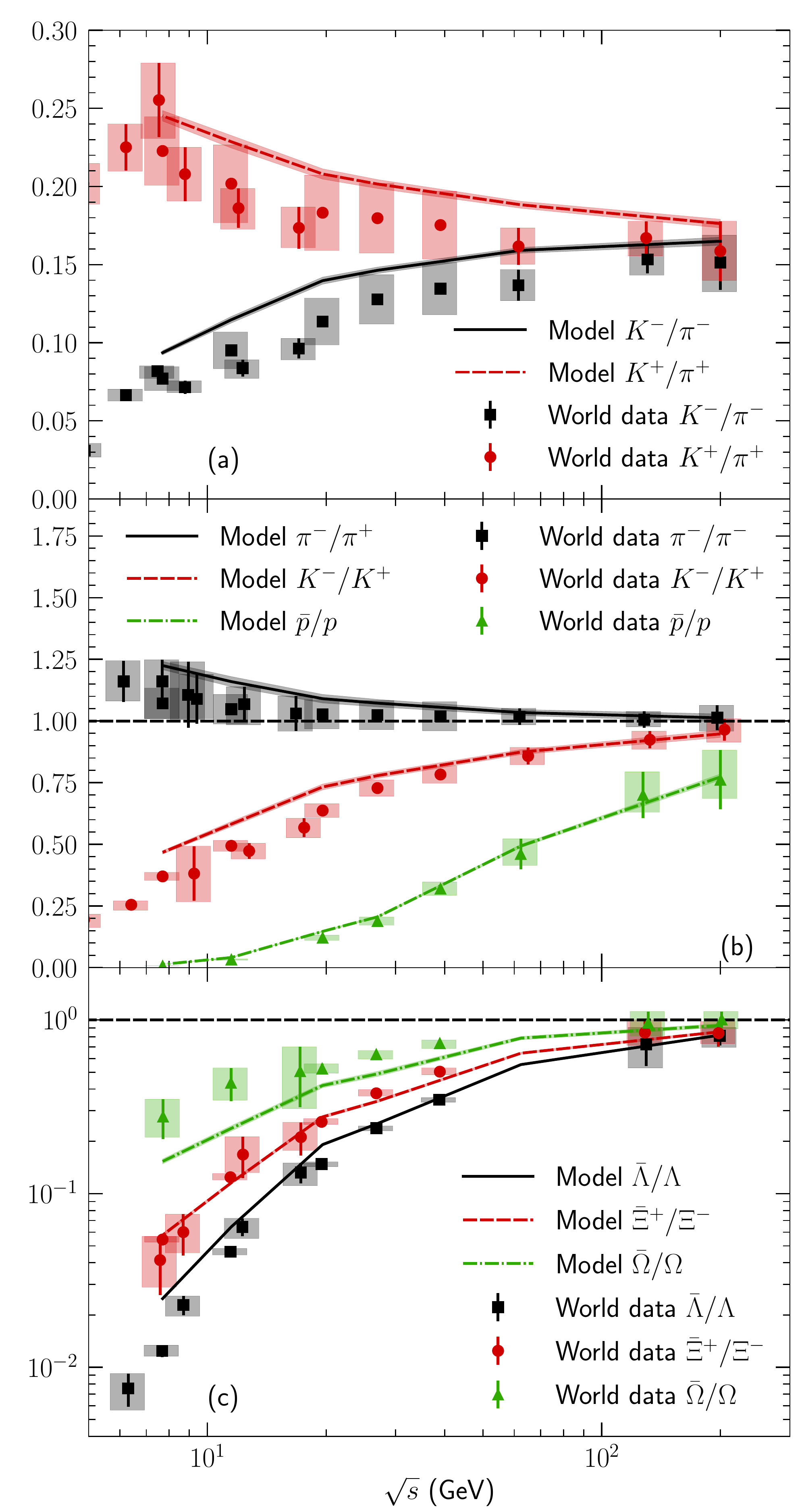}
  \caption{(Color online) Identified particle ratios as functions of the collision energy in central Au+Au collisions in the RHIC BES program compared with the world data \cite{E895:2001zms, NA49:2002pzu, NA49:2004irs, NA49:2008ysv, STAR:2017sal, STAR:2019bjj}. Weak decays are included for protons and anti-protons in the calculations but turned off when computing particle yields for multi-strange baryons.}
  \label{fig:pidRatios}
\end{figure}
%
Fig.~\ref{fig:pidRatios} shows various ratios of identified particles as functions of collision energy. Assuming a Grand Canonical Ensemble (GCE), these ratios are sensitive to how the averaged chemical potentials of conserved charges in the mid-rapidity fireball evolve as a function of the collision energy. Based on the quantum numbers carried by different species of hadrons, the ratio $K^+/\pi^+ \propto \exp(+\mu_S)$ and $K^-/\pi^- \propto \exp(-\mu_S)$. In our model, the strangeness chemical potential $\mu_S$ is related to the net baryon chemical potential $\mu_B$ via the strangeness neutrality condition, which requires $\mu_S \approx \mu_B/3$ \cite{Monnai:2019hkn, Monnai:2021kgu}. Figure~\ref{fig:pidRatios}a shows that the baryon stopping and strangeness neutrality constraint in our model can be enough to reproduce the collision energy dependence of the $K^+/\pi^+$ and $K^-/\pi^-$ ratios. Below 7.7 GeV, the ratio of $K^+/\pi^+$ stops increasing and starts to decrease as the collision energy goes down. The resulting simultaneous suppression of $K^+/\pi^+$ and $K^-/\pi^-$ ratios is usually interpreted as the canonical suppression of the strangeness production, and can not be reproduced in our grand canonical framework. 

Moving to the ratios of particles and anti-particles, the ratio $\pi^-/\pi^+ \propto \exp(-\mu_Q)$ carries information about the net electric charge chemical potential. The constraint $n_Q = 0.4 n_B$ for the Au nucleus determines how $\mu_Q$ is related to $\mu_B$ \cite{Monnai:2019hkn, Monnai:2021kgu}. The comparison of the $\pi^-/\pi^+$ ratio with the world data shown in Fig.~\ref{fig:pidRatios}b suggests that the baryon stopping and the $n_Q = 0.4 n_B$ constraint are enough to explain the mild increase of the $\pi^-/\pi^+$ ratio as the collision energy goes down. The ratios of $K^-/K^+ \propto \exp(-2(\mu_Q + \mu_S))$ contains a mixture of information on $\mu_Q$ and $\mu_S$. The ratio $\bar{p}/p \propto \exp(-2(\mu_Q + \mu_B))$ is dominated by the baryon stopping. 

Figure~\ref{fig:pidRatios}c shows the ratios of anti-particles to particles for strange baryons. The ratios of $\bar{\Lambda}/\Lambda$, $\bar{\Xi}^+/\Xi^-$, and $\bar{\Omega}/\Omega$ contain different weights of the strangeness chemical potential $\mu_S$. Our calculations reproduce the hierarchy of these ratios from 200 GeV down to 7.7 GeV well. The ratio of $\bar{\Omega}/\Omega$ is slightly underestimated.

\begin{figure}[ht!]
  \centering
  \includegraphics[width=1.0\linewidth]{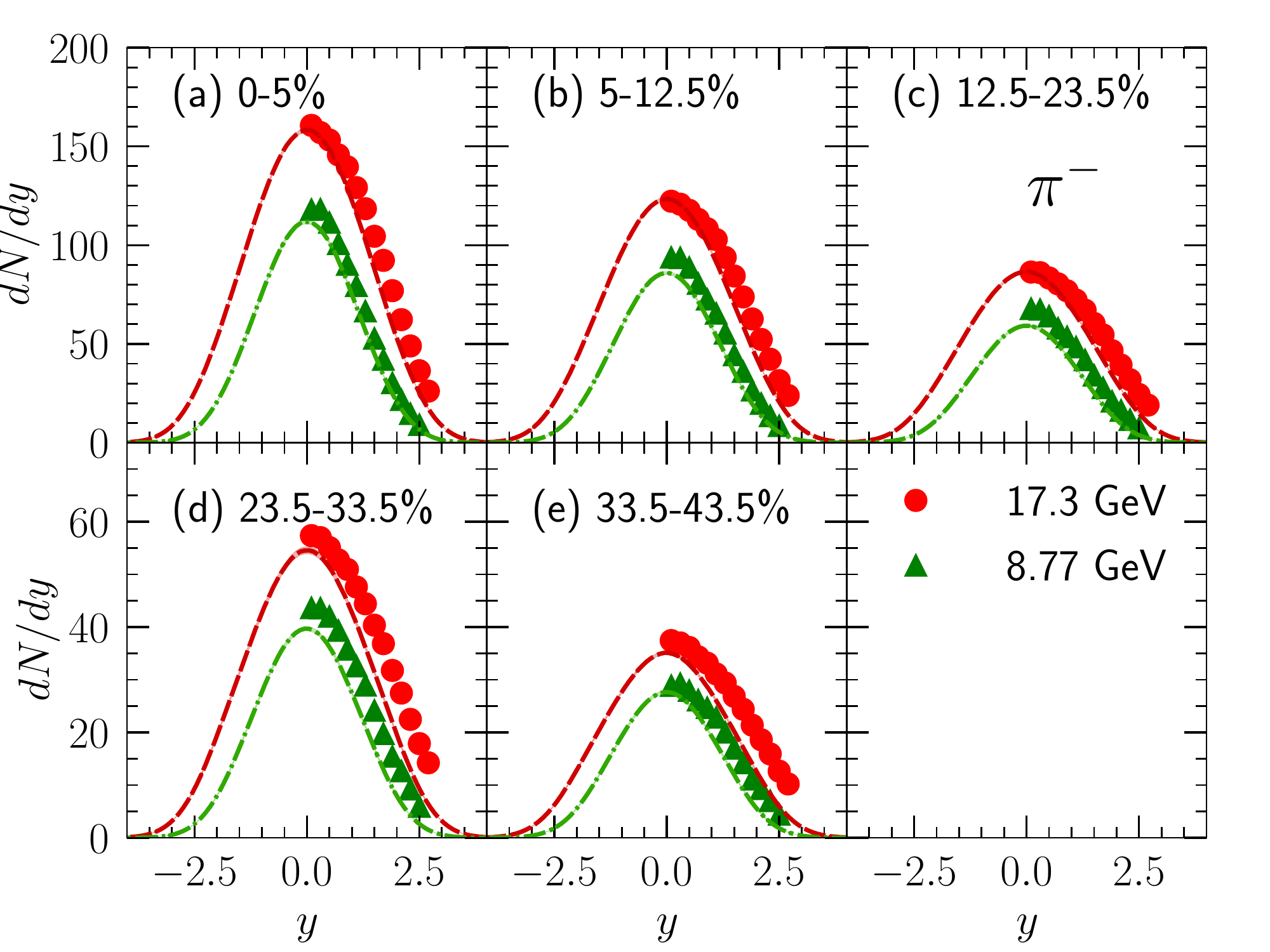}
  \includegraphics[width=1.0\linewidth]{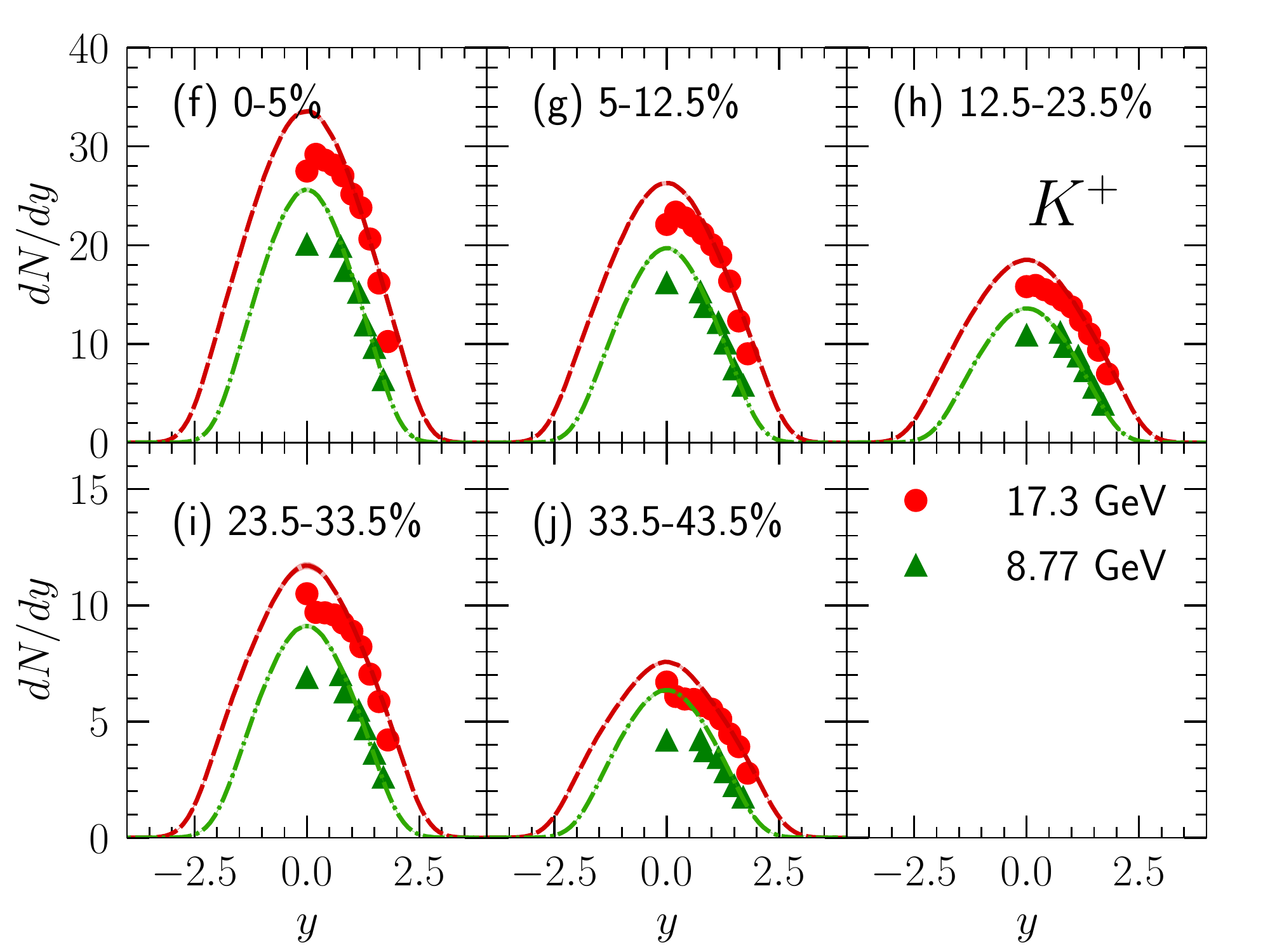}
  \includegraphics[width=1.0\linewidth]{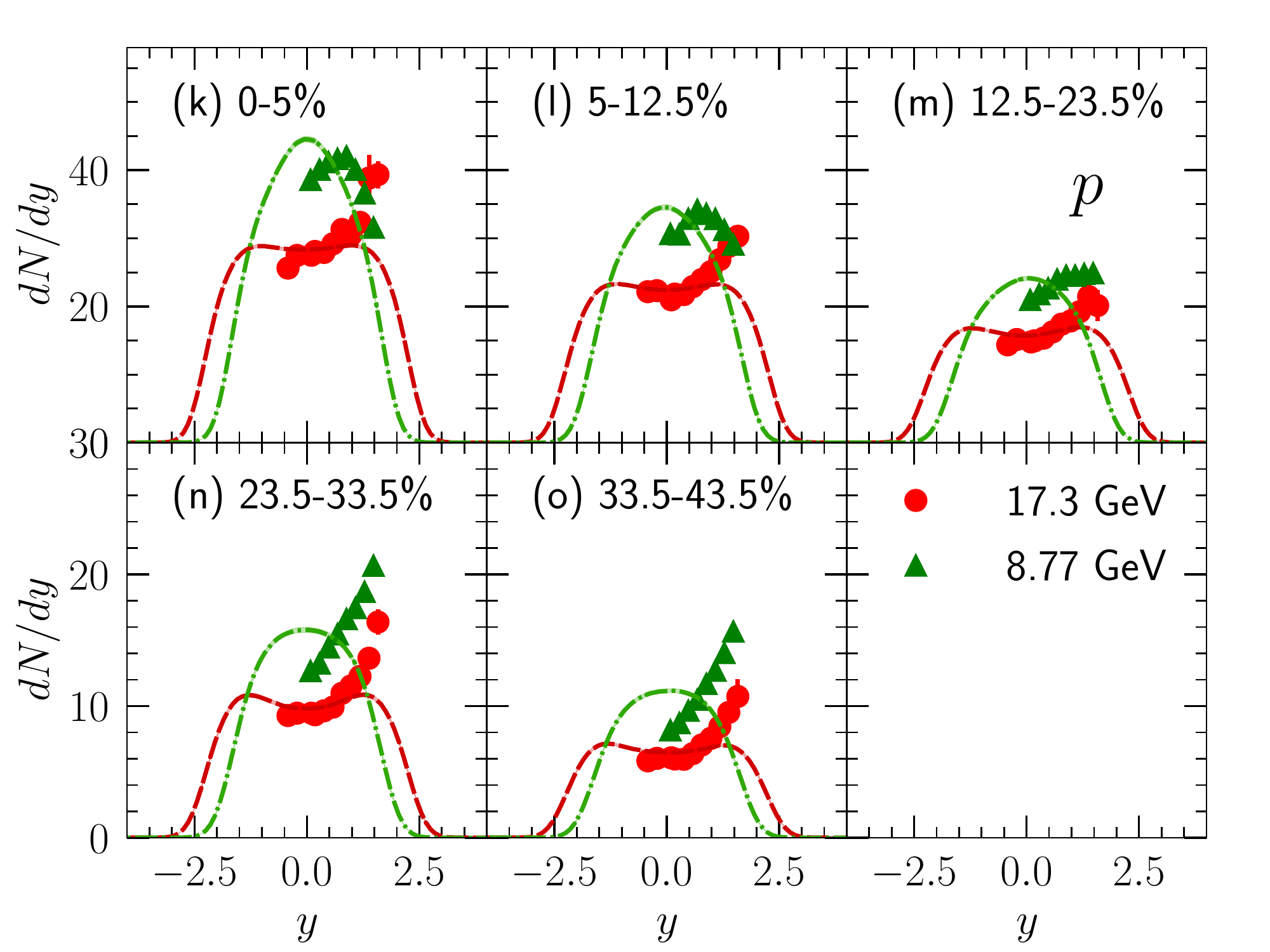}
  \caption{(Color online) Rapidity distributions of negative pions, positive kaons, and protons in Pb+Pb collisions at SPS collision energies. Theoretical results in different centrality bins are compared with the experimental data from the NA49 Collaboration \cite{NA49:2012rsi}.}
  \label{fig:SPS}
\end{figure}

We present results for identified particle production in Pb+Pb collisions at CERN SPS collision energies in Fig.~\ref{fig:SPS}. Compared to the available data from the RHIC BES program phase I, the measured rapidity-dependent particle yields $dN/dy$ at SPS energies provide valuable information. Our model can reasonably reproduce the $\pi^-$ and $K^+$ rapidity distributions from central to semi-peripheral collisions at 17.3 and 8.77 GeV center-of-mass energies.
Figures~\ref{fig:SPS}k-o further show the comparisons for the proton rapidity distributions at SPS energies.  Our model gives a good description of the proton yield near the mid-rapidity region for both collision energies. At 17.3 GeV, the plateau of proton yields in $\vert y \vert < 1$ is reproduced well by our model without any parameter tuning. 
At forward rapidity $y > 1$, the experimental data shows a monotonic increase of the proton yields, suggesting potential contamination from the spectator protons in the measurements. At $\snn = 8.77$\,GeV, the experimental data in the 0-5\% centrality bin shows that the proton yield peaks around $y = 1$, while our model calculations have most of the protons produced at $y = 0$. This difference suggests the initial state baryon stopping is somewhat overestimated at 8.77 GeV. For centralities beyond 20\%, the proton yield shows a monotonic increase with rapidity in the experimental data also at 8.77 GeV, which again suggests contamination from the spectators.

\begin{figure}[ht!]
  \centering
  \includegraphics[width=0.95\linewidth]{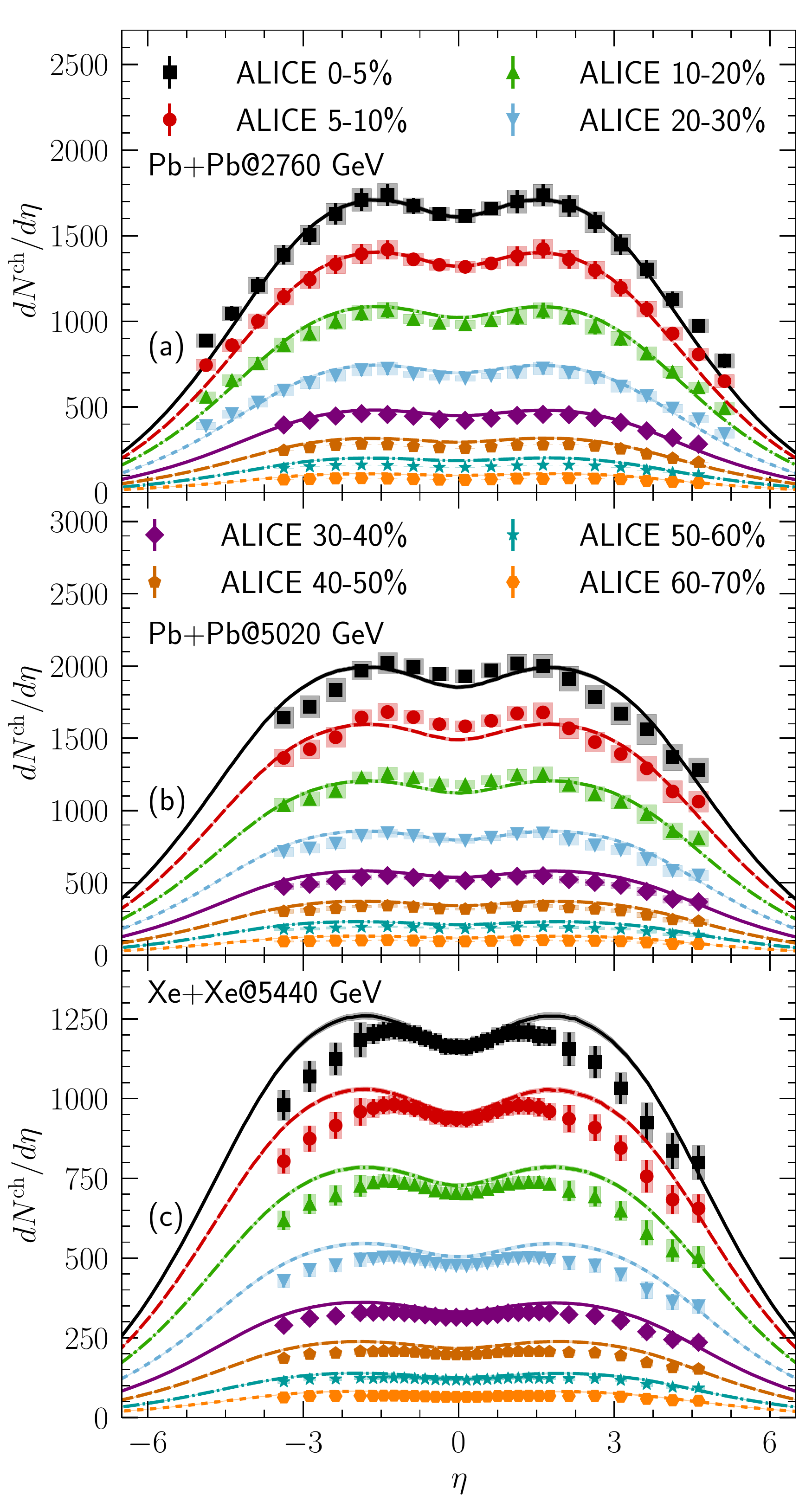}
  \caption{(Color online) Charged hadron pseudo-rapidity distributions in Pb+Pb (a,b) and Xe+Xe (c) collisions at LHC collision energies. Theoretical results in different centrality bins are compared with experimental data from the ALICE Collaboration \cite{ALICE:2013jfw, ALICE:2016fbt, ALICE:2018cpu}.}
  \label{fig:PbPbLHC}
\end{figure}
%
Last but not least, we extrapolate our model to predict the particle production in heavy-ion collisions at LHC energies. We determine collision centrality using the charged hadron multiplicity in the rapidity region of the ALICE V0A detector \cite{ALICE:2010khr}. Figure~\ref{fig:PbPbLHC} shows that our model gives excellent predictions for charged hadron pseudo-rapidity distributions $dN^\mathrm{ch}/d\eta (\eta)$ across centrality bins in Pb+Pb collisions at center of mass energies of 2.76 and 5.02 TeV.
Shifting to smaller collision systems, the charged hadron yields in $\vert \eta \vert < 1$ are still well reproduced for Xe+Xe collisions at 5.44 TeV. Meanwhile, the tails of $dN^\mathrm{ch}/d\eta (\eta)$ beyond $\vert \eta \vert > 2$ are slightly steeper than in the ALICE data.
For heavy-ion collisions, we have checked that the charged hadron rapidity distributions in different centrality bins have negligible dependence on the centrality triggers (V0A vs. CL1) in our model.

\section{Conclusions} \label{sec:conc}

In this work, we have developed a (3+1)D initial state model to study the longitudinal dynamics of particle production in various types of relativistic nuclear collisions. We include event-by-event fluctuations of the three-dimensional spatial energy-momentum and net baryon density distributions, which prove to be important for describing the collision energy, system size, and rapidity dependence of particle production.

We dynamically couple this new (3+1)D initial condition with a hydrodynamics + hadronic transport hybrid framework to simulate the entire dynamics of relativistic nuclear collisions at center of mass energies ranging from 7.7 GeV to 13 TeV. After calibrating the model with minimum bias p+p collisions at a few collision energies, we show that this hybrid theoretical framework can give quantitative predictions for particle production as a function of pseudo-rapidity or rapidity in asymmetric light+heavy and symmetric heavy-ion collisions across three orders of magnitude in collision energy. This unified description of particle production across system size and collision energy shows the effectiveness of our hybrid theoretical framework. Our work establishes the basis to study anisotropic flow and other collective multi-particle correlations in full three dimensions.

The presented hybrid framework enables us to study the correlation between particle production in mid-rapidity and forward regions. Reproducing this correlation is found to be crucial for the centrality determination in asymmetric light+heavy ion collisions. Missing these correlations, as (2+1)D boost-invariant simulations certainly do, would lead to the wrong subsets of events being sorted into a given centrality class. This would affect results for all other observables, including azimuthal momentum anisotropies, electromagnetic radiation \cite{Shen:2015qba, Shen:2016zpp, Gale:2021emg}, and high momentum probes \cite{Park:2016jap}. 

Our hybrid framework also provides a description of baryon stopping during the initial stage of the collision, which is particularly important for the RHIC BES and its studies of net-baryon fluctuations to determine the QCD critical point. We demonstrated that the baryon charge fluctuations in the string junction model are essential to reproduce the small but finite net proton numbers around midrapidity in 200 GeV collisions at RHIC. 

Comparing the computed identified particle yields with the RHIC BES and CERN SPS measurements, we show that the collision energy dependence of the identified particle ratios between 7.7 GeV and 200 GeV can be explained by a combination of using the grand canonical ensemble, baryon stopping, and the constraints of strangeness neutrality and $n_Q = 0.4 n_B$ in the nuclear equation of state at finite densities. To further extend the theoretical description of the hadronic chemistry in heavy-ion collisions below 7.7 GeV, we need to include canonical suppression of strangeness in our dynamical framework.

This theoretical framework paves the way to extracting the transport properties of hot nuclear matter at finite net baryon density and to determining critical behavior using the Bayesian inference method.

\acknowledgements
We thank Nicole Lewis, Akihiko Monnai, Scott Pratt, Sangwook Ryu, Prithwish Tribedy, Zhangbu Xu, and Wenbin Zhao for fruitful discussions. CS is supported in part by the U.S. Department of Energy (DOE) under award number DE-SC0021969 and DE-SC0013460 and in part by the National Science Foundation (NSF) under award number PHY-2012922. BPS is supported by the U.S. Department of Energy, Office of Science, Office of Nuclear Physics through Contract No. DE-SC0012704.
CS acknowledges a DOE Office of Science Early Career Award.
This research was done using resources provided by the Open Science Grid (OSG)~\cite{Pordes:2007zzb, Sfiligoi:2009cct}, which is supported by the National Science Foundation award \#2030508 and resources of the high-performance computing services at Wayne State University.
This work is in part supported by the U.S. Department of Energy, Office of Science, Office of Nuclear Physics, within the framework of the Beam Energy Scan Theory (BEST) Topical Collaboration.

\appendix
\section{Effects of cold corona on particle production}\label{app:corona}

In our hybrid simulations, we convert fluid cells into particles on a hypersurface with constant energy density $e_\mathrm{sw}$. 
Fluid cells that have an energy density smaller than $e_\mathrm{sw}$ from the beginning need to be treated separately. In this appendix, we study the effect of including contributions from this cold corona on final observables.
For heavy-ion collisions at high energy, we expect a negligible contribution to the final-state particle production from these cold fluid cells. However, if the collision system size is small or the collision energy is low, these cold corona fluid cells could have a sizable contribution to the total particle yield.

\begin{figure}[ht!]
  \centering
  \includegraphics[width=\linewidth]{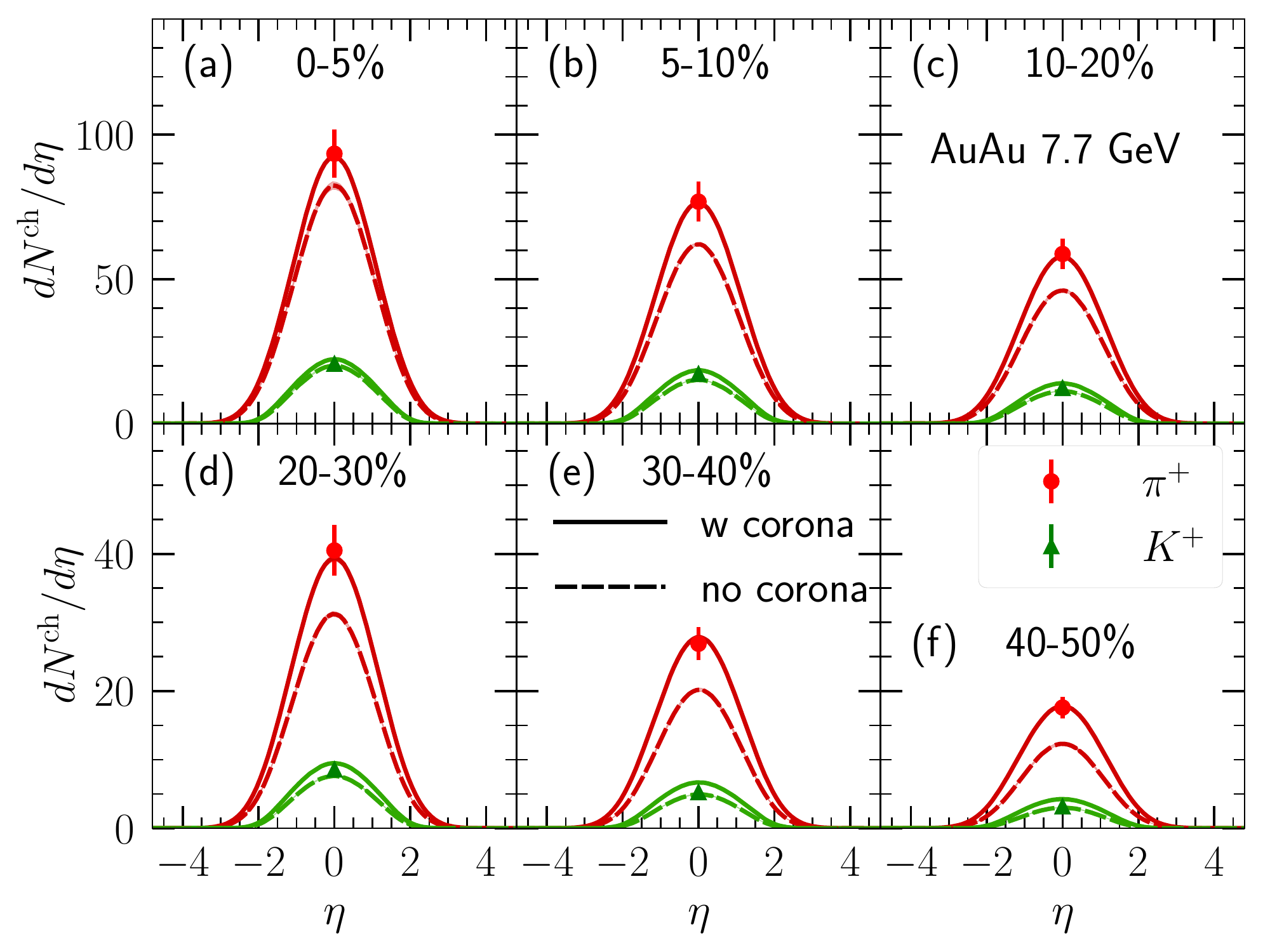}
  \includegraphics[width=0.95\linewidth]{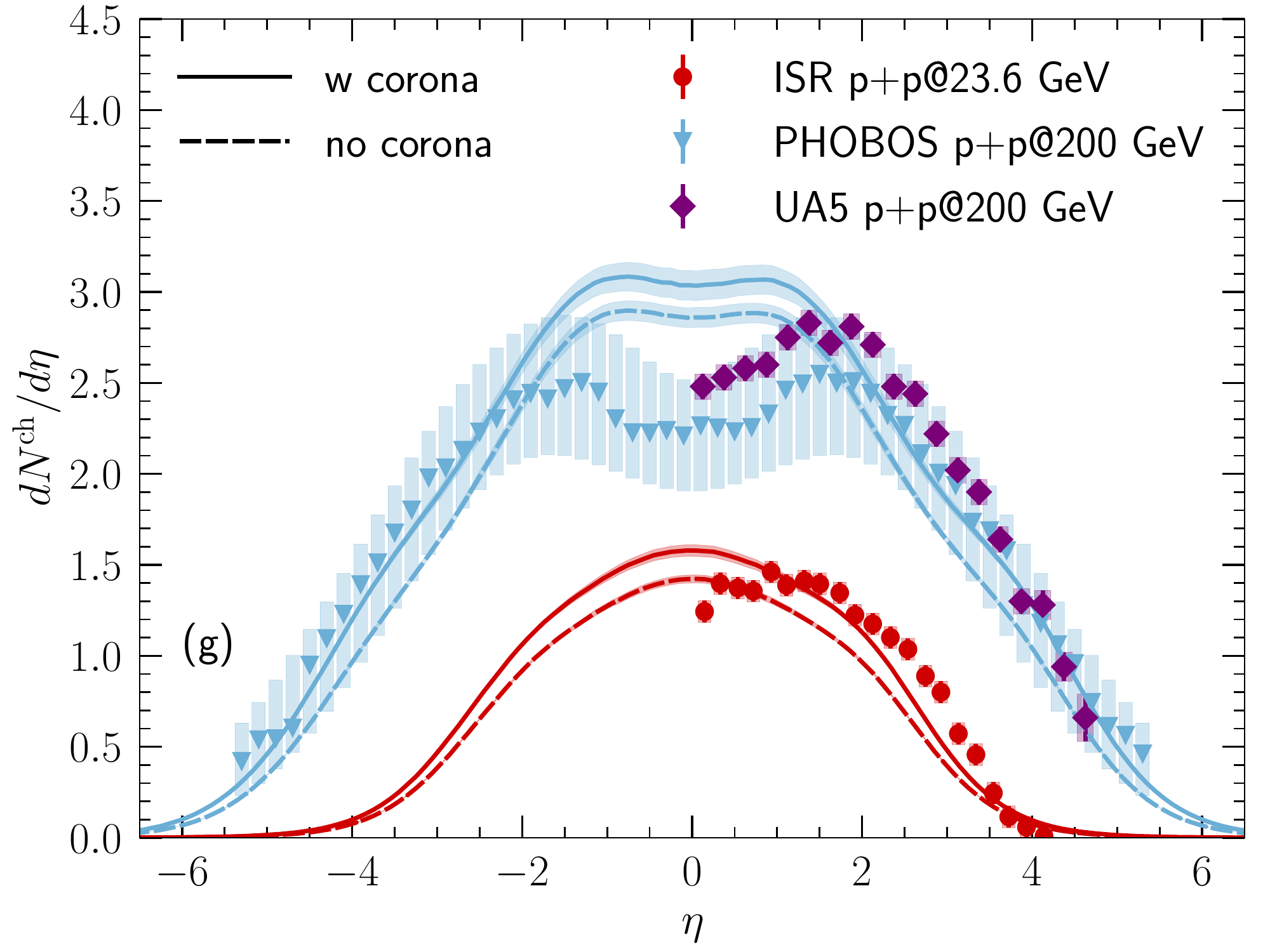}
  \caption{(Color online) The effects of cold corona cells on particle production in Au+Au collisions at 7.7 GeV (panels (a)-(f)) and minimum bias p+p collisions (panel (g).}
  \label{fig:Corona}
\end{figure}
%
As discussed in Sec.~\ref{sec:hybridFramework}, we estimate the particle production from these cold corona fluid cells with the Cooper-Frye prescription using their local temperature and chemical potentials.

Figures~\ref{fig:Corona}a-f show that the corona contributes significantly to the pion and kaon yields in Au+Au collisions at $\snn = 7.7$\,GeV. The corona's relative contribution increases from 15\% to 30\% from central to peripheral centrality bins. We checked that the relative abundance of different hadron species does not change noticeably when including particles emitted from the cold corona.

Figure~\ref{fig:Corona}g shows that cold corona cells give 5-10\% contributions to the mid-rapidity charged hadron yield in minimum bias p+p collisions. The relative contribution increases to about 20-30\% in the forward and backward rapidities at 200 GeV. The inclusion of the cold corona is favored by the experimental data in the forward rapidity region.

\section{Initial-state estimator for collision centrality}\label{app:centralityEst}

In event-by-event simulations, it is practical to find some initial-state variables which have a strong correlation with the final-state charged hadron multiplicity. These initial-state variables are useful to speed up simulations by pre-sorting events into the correct centrality bin, and only running the full simulation in the centrality bin of interest.

\begin{figure}[ht!]
  \centering
  \includegraphics[width=0.9\linewidth]{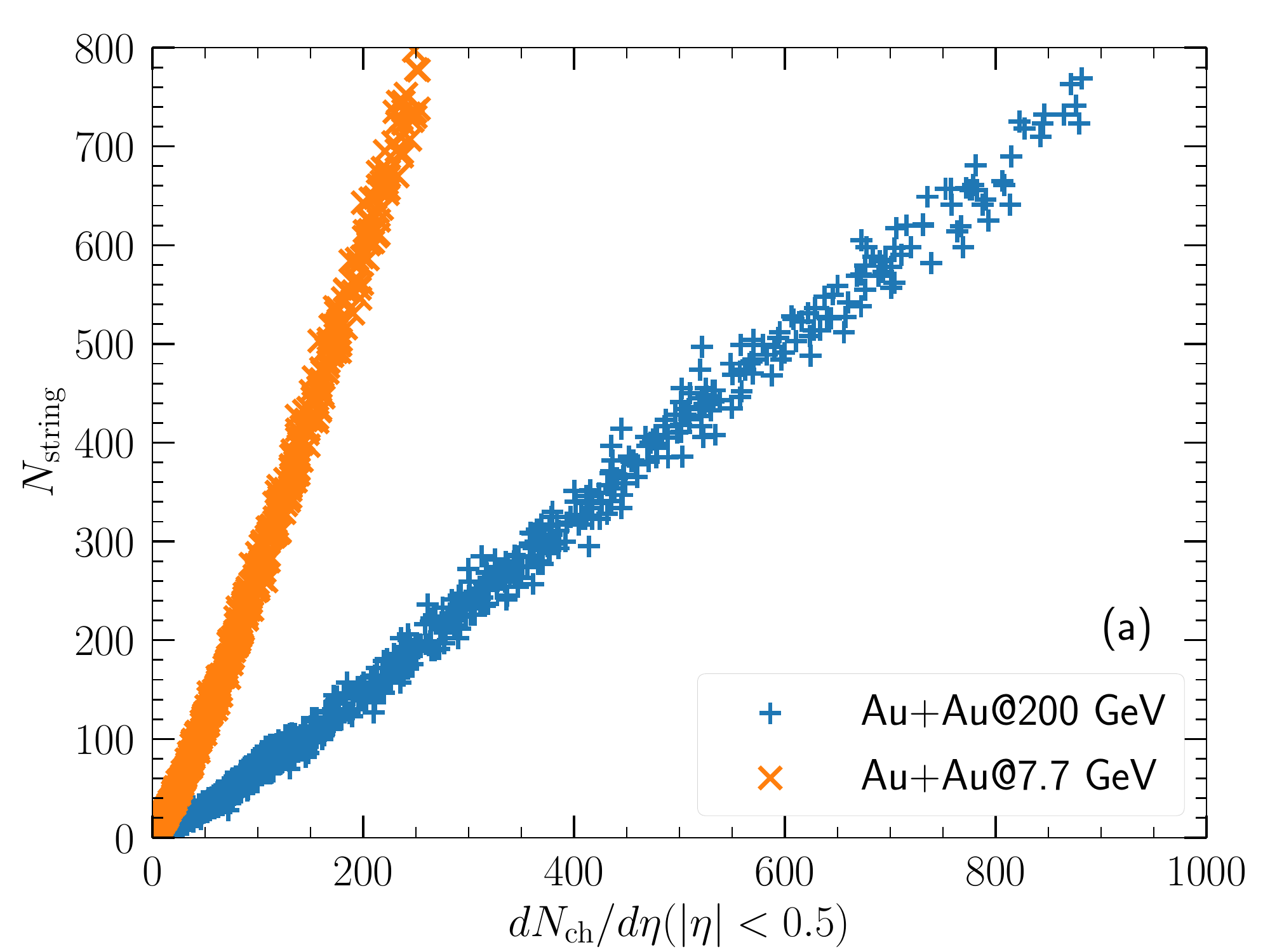}
  \includegraphics[width=0.9\linewidth]{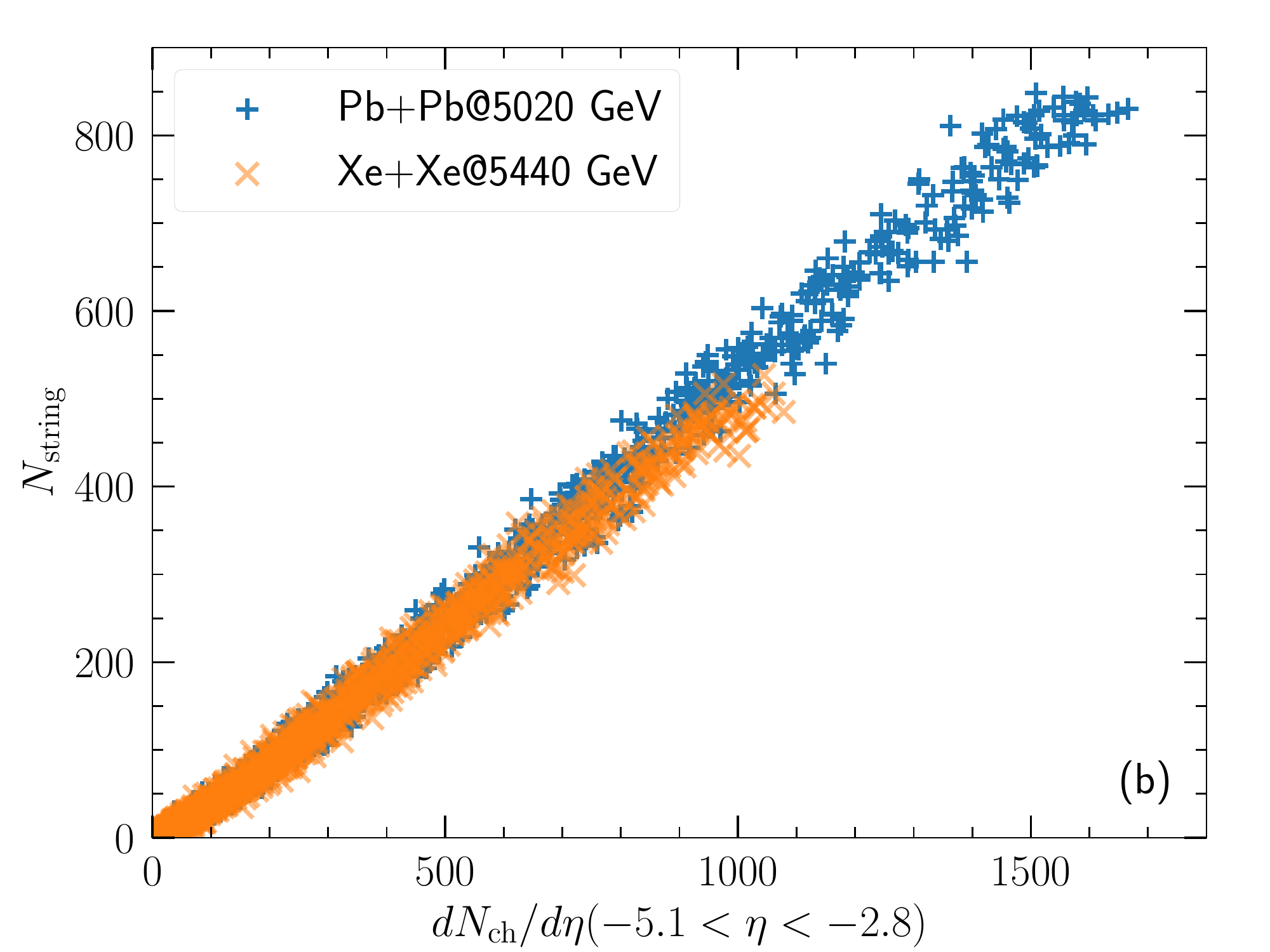}
  \includegraphics[width=0.9\linewidth]{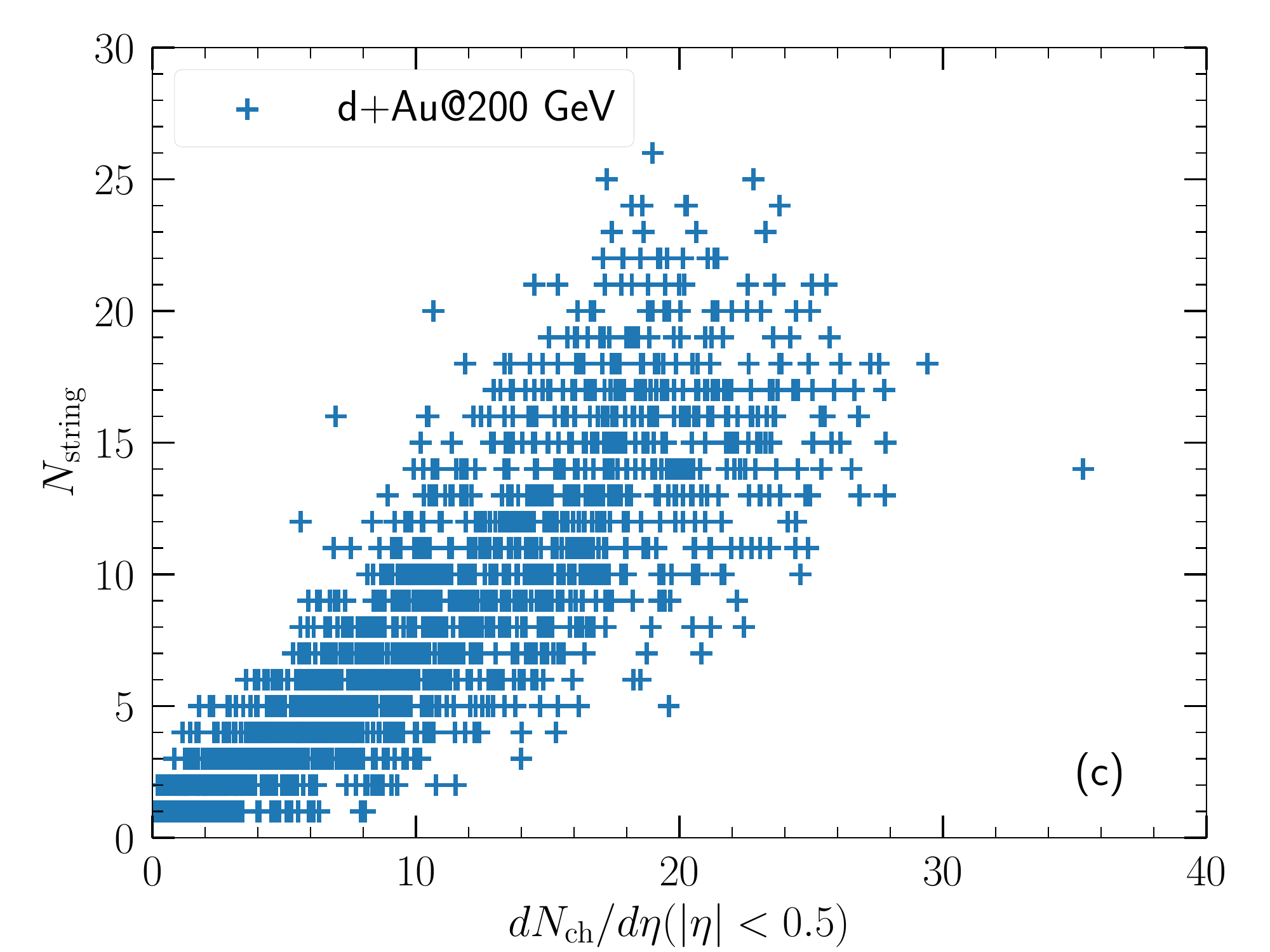}
  \caption{(Color online) Scatter plots for the initial-state number of strings vs.~final-state charged hadron multiplicity in Au+Au collisions at two RHIC energies (a), heavy-ion collisions at LHC energies (b), and d+Au collisions at 200 GeV (c).}
  \label{fig:centralityEst}
\end{figure}

In our \Glauber{} model, the energy near mid-rapidity is supplied by the strings. Therefore, the number of produced hadrons is expected to have a strong correlation with the number of strings in the initial state. Figures~\ref{fig:centralityEst}a and \ref{fig:centralityEst}b show that this correlation is strong for symmetric heavy-ion collisions at the RHIC BES and LHC energies. The Pearson coefficients between $N_\mathrm{string}$ and \dNdeta{} reach 0.997 for these collisions. Therefore, one can pre-select collision events using the number of strings $N_\mathrm{string}$ in the initial state to speed up the event-by-event simulations.

The correlation between $N_\mathrm{string}$ and \dNdeta{} becomes weaker for asymmetric d+Au collisions as shown in Fig.~\ref{fig:centralityEst}c. In contrast to the symmetric heavy-ion collisions, the corresponding Pearson coefficient for d+Au collisions at 200 GeV reduces to 0.957. Hence, we need to simulate minimum bias collisions for asymmetric collisions and determine the centrality class for each collision event using the final-state observables. 

\bibliography{ref}

\begin{thebibliography}{80}%
\makeatletter
\providecommand \@ifxundefined [1]{%
 \@ifx{#1\undefined}
}%
\providecommand \@ifnum [1]{%
 \ifnum #1\expandafter \@firstoftwo
 \else \expandafter \@secondoftwo
 \fi
}%
\providecommand \@ifx [1]{%
 \ifx #1\expandafter \@firstoftwo
 \else \expandafter \@secondoftwo
 \fi
}%
\providecommand \natexlab [1]{#1}%
\providecommand \enquote  [1]{``#1''}%
\providecommand \bibnamefont  [1]{#1}%
\providecommand \bibfnamefont [1]{#1}%
\providecommand \citenamefont [1]{#1}%
\providecommand \href@noop [0]{\@secondoftwo}%
\providecommand \href [0]{\begingroup \@sanitize@url \@href}%
\providecommand \@href[1]{\@@startlink{#1}\@@href}%
\providecommand \@@href[1]{\endgroup#1\@@endlink}%
\providecommand \@sanitize@url [0]{\catcode `\\12\catcode `\$12\catcode
  `\&12\catcode `\#12\catcode `\^12\catcode `\_12\catcode `\%12\relax}%
\providecommand \@@startlink[1]{}%
\providecommand \@@endlink[0]{}%
\providecommand \url  [0]{\begingroup\@sanitize@url \@url }%
\providecommand \@url [1]{\endgroup\@href {#1}{\urlprefix }}%
\providecommand \urlprefix  [0]{URL }%
\providecommand \Eprint [0]{\href }%
\providecommand \doibase [0]{http://dx.doi.org/}%
\providecommand \selectlanguage [0]{\@gobble}%
\providecommand \bibinfo  [0]{\@secondoftwo}%
\providecommand \bibfield  [0]{\@secondoftwo}%
\providecommand \translation [1]{[#1]}%
\providecommand \BibitemOpen [0]{}%
\providecommand \bibitemStop [0]{}%
\providecommand \bibitemNoStop [0]{.\EOS\space}%
\providecommand \EOS [0]{\spacefactor3000\relax}%
\providecommand \BibitemShut  [1]{\csname bibitem#1\endcsname}%
\let\auto@bib@innerbib\@empty
\bibitem [{\citenamefont {Evans}\ and\ \citenamefont
  {Bryant}(2008)}]{Evans:2008zzb}%
  \BibitemOpen
  \bibfield  {author} {\bibinfo {author} {\bibfnamefont {Lyndon}\ \bibnamefont
  {Evans}}\ and\ \bibinfo {author} {\bibfnamefont {Philip}\ \bibnamefont
  {Bryant}},\ }\bibfield  {title} {\enquote {\bibinfo {title} {{LHC
  Machine}},}\ }\href {\doibase 10.1088/1748-0221/3/08/S08001} {\bibfield
  {journal} {\bibinfo  {journal} {JINST}\ }\textbf {\bibinfo {volume} {3}},\
  \bibinfo {pages} {S08001} (\bibinfo {year} {2008})}\BibitemShut {NoStop}%
\bibitem [{\citenamefont {Adams}\ \emph {et~al.}(2005)\citenamefont {Adams}
  \emph {et~al.}}]{STAR:2005gfr}%
  \BibitemOpen
  \bibfield  {author} {\bibinfo {author} {\bibfnamefont {John}\ \bibnamefont
  {Adams}} \emph {et~al.} (\bibinfo {collaboration} {STAR}),\ }\bibfield
  {title} {\enquote {\bibinfo {title} {{Experimental and theoretical challenges
  in the search for the quark gluon plasma: The STAR Collaboration's critical
  assessment of the evidence from RHIC collisions}},}\ }\href {\doibase
  10.1016/j.nuclphysa.2005.03.085} {\bibfield  {journal} {\bibinfo  {journal}
  {Nucl. Phys. A}\ }\textbf {\bibinfo {volume} {757}},\ \bibinfo {pages}
  {102--183} (\bibinfo {year} {2005})},\ \Eprint
  {http://arxiv.org/abs/nucl-ex/0501009} {arXiv:nucl-ex/0501009} \BibitemShut
  {NoStop}%
\bibitem [{\citenamefont {Adcox}\ \emph {et~al.}(2005)\citenamefont {Adcox}
  \emph {et~al.}}]{PHENIX:2004vcz}%
  \BibitemOpen
  \bibfield  {author} {\bibinfo {author} {\bibfnamefont {K.}~\bibnamefont
  {Adcox}} \emph {et~al.} (\bibinfo {collaboration} {PHENIX}),\ }\bibfield
  {title} {\enquote {\bibinfo {title} {{Formation of dense partonic matter in
  relativistic nucleus-nucleus collisions at RHIC: Experimental evaluation by
  the PHENIX collaboration}},}\ }\href {\doibase
  10.1016/j.nuclphysa.2005.03.086} {\bibfield  {journal} {\bibinfo  {journal}
  {Nucl. Phys. A}\ }\textbf {\bibinfo {volume} {757}},\ \bibinfo {pages}
  {184--283} (\bibinfo {year} {2005})},\ \Eprint
  {http://arxiv.org/abs/nucl-ex/0410003} {arXiv:nucl-ex/0410003} \BibitemShut
  {NoStop}%
\bibitem [{\citenamefont {Back}\ \emph {et~al.}(2005)\citenamefont {Back} \emph
  {et~al.}}]{PHOBOS:2004zne}%
  \BibitemOpen
  \bibfield  {author} {\bibinfo {author} {\bibfnamefont {B.~B.}\ \bibnamefont
  {Back}} \emph {et~al.} (\bibinfo {collaboration} {PHOBOS}),\ }\bibfield
  {title} {\enquote {\bibinfo {title} {{The PHOBOS perspective on discoveries
  at RHIC}},}\ }\href {\doibase 10.1016/j.nuclphysa.2005.03.084} {\bibfield
  {journal} {\bibinfo  {journal} {Nucl. Phys. A}\ }\textbf {\bibinfo {volume}
  {757}},\ \bibinfo {pages} {28--101} (\bibinfo {year} {2005})},\ \Eprint
  {http://arxiv.org/abs/nucl-ex/0410022} {arXiv:nucl-ex/0410022} \BibitemShut
  {NoStop}%
\bibitem [{\citenamefont {Arsene}\ \emph {et~al.}(2005)\citenamefont {Arsene}
  \emph {et~al.}}]{BRAHMS:2004adc}%
  \BibitemOpen
  \bibfield  {author} {\bibinfo {author} {\bibfnamefont {I.}~\bibnamefont
  {Arsene}} \emph {et~al.} (\bibinfo {collaboration} {BRAHMS}),\ }\bibfield
  {title} {\enquote {\bibinfo {title} {{Quark gluon plasma and color glass
  condensate at RHIC? The Perspective from the BRAHMS experiment}},}\ }\href
  {\doibase 10.1016/j.nuclphysa.2005.02.130} {\bibfield  {journal} {\bibinfo
  {journal} {Nucl. Phys. A}\ }\textbf {\bibinfo {volume} {757}},\ \bibinfo
  {pages} {1--27} (\bibinfo {year} {2005})},\ \Eprint
  {http://arxiv.org/abs/nucl-ex/0410020} {arXiv:nucl-ex/0410020} \BibitemShut
  {NoStop}%
\bibitem [{\citenamefont {Caines}(2009)}]{Caines:2009yu}%
  \BibitemOpen
  \bibfield  {author} {\bibinfo {author} {\bibfnamefont {Helen}\ \bibnamefont
  {Caines}} (\bibinfo {collaboration} {STAR}),\ }\bibfield  {title} {\enquote
  {\bibinfo {title} {{The RHIC Beam Energy Scan: STAR'S Perspective}},}\ }in\
  \href@noop {} {\emph {\bibinfo {booktitle} {{44th Rencontres de Moriond on
  QCD and High Energy Interactions}}}}\ (\bibinfo {year} {2009})\ pp.\ \bibinfo
  {pages} {375--378},\ \Eprint {http://arxiv.org/abs/0906.0305}
  {arXiv:0906.0305 [nucl-ex]} \BibitemShut {NoStop}%
\bibitem [{\citenamefont {Mohanty}(2011)}]{Mohanty:2011nm}%
  \BibitemOpen
  \bibfield  {author} {\bibinfo {author} {\bibfnamefont {Bedangadas}\
  \bibnamefont {Mohanty}} (\bibinfo {collaboration} {STAR}),\ }\bibfield
  {title} {\enquote {\bibinfo {title} {{STAR experiment results from the beam
  energy scan program at RHIC}},}\ }\href {\doibase
  10.1088/0954-3899/38/12/124023} {\bibfield  {journal} {\bibinfo  {journal}
  {J. Phys. G}\ }\textbf {\bibinfo {volume} {38}},\ \bibinfo {pages} {124023}
  (\bibinfo {year} {2011})},\ \Eprint {http://arxiv.org/abs/1106.5902}
  {arXiv:1106.5902 [nucl-ex]} \BibitemShut {NoStop}%
\bibitem [{\citenamefont {Mitchell}(2013)}]{Mitchell:2012mx}%
  \BibitemOpen
  \bibfield  {author} {\bibinfo {author} {\bibfnamefont {Jeffery~T.}\
  \bibnamefont {Mitchell}} (\bibinfo {collaboration} {PHENIX}),\ }\bibfield
  {title} {\enquote {\bibinfo {title} {{The RHIC Beam Energy Scan Program:
  Results from the PHENIX Experiment}},}\ }\href {\doibase
  10.1016/j.nuclphysa.2013.02.161} {\bibfield  {journal} {\bibinfo  {journal}
  {Nucl. Phys. A}\ }\textbf {\bibinfo {volume} {904-905}},\ \bibinfo {pages}
  {903c--906c} (\bibinfo {year} {2013})},\ \Eprint
  {http://arxiv.org/abs/1211.6139} {arXiv:1211.6139 [nucl-ex]} \BibitemShut
  {NoStop}%
\bibitem [{\citenamefont {Odyniec}(2015)}]{Odyniec:2015iaa}%
  \BibitemOpen
  \bibfield  {author} {\bibinfo {author} {\bibfnamefont {Grazyna}\ \bibnamefont
  {Odyniec}},\ }\bibfield  {title} {\enquote {\bibinfo {title} {{Future of the
  beam energy scan program at RHIC}},}\ }\href {\doibase
  10.1051/epjconf/20149503027} {\bibfield  {journal} {\bibinfo  {journal} {EPJ
  Web Conf.}\ }\textbf {\bibinfo {volume} {95}},\ \bibinfo {pages} {03027}
  (\bibinfo {year} {2015})}\BibitemShut {NoStop}%
\bibitem [{\citenamefont {Gazdzicki}(2009)}]{Gazdzicki:2008kk}%
  \BibitemOpen
  \bibfield  {author} {\bibinfo {author} {\bibfnamefont {Marek}\ \bibnamefont
  {Gazdzicki}} (\bibinfo {collaboration} {NA61/SHINE}),\ }\bibfield  {title}
  {\enquote {\bibinfo {title} {{Ion Program of Na61/Shine at the CERN SPS}},}\
  }\href {\doibase 10.1088/0954-3899/36/6/064039} {\bibfield  {journal}
  {\bibinfo  {journal} {J. Phys. G}\ }\textbf {\bibinfo {volume} {36}},\
  \bibinfo {pages} {064039} (\bibinfo {year} {2009})},\ \Eprint
  {http://arxiv.org/abs/0812.4415} {arXiv:0812.4415 [nucl-ex]} \BibitemShut
  {NoStop}%
\bibitem [{\citenamefont {Abgrall}\ \emph {et~al.}(2014)\citenamefont {Abgrall}
  \emph {et~al.}}]{Abgrall:2014xwa}%
  \BibitemOpen
  \bibfield  {author} {\bibinfo {author} {\bibfnamefont {N.}~\bibnamefont
  {Abgrall}} \emph {et~al.} (\bibinfo {collaboration} {NA61}),\ }\bibfield
  {title} {\enquote {\bibinfo {title} {{NA61/SHINE facility at the CERN SPS:
  beams and detector system}},}\ }\href {\doibase
  10.1088/1748-0221/9/06/P06005} {\bibfield  {journal} {\bibinfo  {journal}
  {JINST}\ }\textbf {\bibinfo {volume} {9}},\ \bibinfo {pages} {P06005}
  (\bibinfo {year} {2014})},\ \Eprint {http://arxiv.org/abs/1401.4699}
  {arXiv:1401.4699 [physics.ins-det]} \BibitemShut {NoStop}%
\bibitem [{\citenamefont {Bzdak}\ \emph {et~al.}(2020)\citenamefont {Bzdak},
  \citenamefont {Esumi}, \citenamefont {Koch}, \citenamefont {Liao},
  \citenamefont {Stephanov},\ and\ \citenamefont {Xu}}]{Bzdak:2019pkr}%
  \BibitemOpen
  \bibfield  {author} {\bibinfo {author} {\bibfnamefont {Adam}\ \bibnamefont
  {Bzdak}}, \bibinfo {author} {\bibfnamefont {Shinichi}\ \bibnamefont {Esumi}},
  \bibinfo {author} {\bibfnamefont {Volker}\ \bibnamefont {Koch}}, \bibinfo
  {author} {\bibfnamefont {Jinfeng}\ \bibnamefont {Liao}}, \bibinfo {author}
  {\bibfnamefont {Mikhail}\ \bibnamefont {Stephanov}}, \ and\ \bibinfo {author}
  {\bibfnamefont {Nu}~\bibnamefont {Xu}},\ }\bibfield  {title} {\enquote
  {\bibinfo {title} {{Mapping the Phases of Quantum Chromodynamics with Beam
  Energy Scan}},}\ }\href {\doibase 10.1016/j.physrep.2020.01.005} {\bibfield
  {journal} {\bibinfo  {journal} {Phys. Rept.}\ }\textbf {\bibinfo {volume}
  {853}},\ \bibinfo {pages} {1--87} (\bibinfo {year} {2020})},\ \Eprint
  {http://arxiv.org/abs/1906.00936} {arXiv:1906.00936 [nucl-th]} \BibitemShut
  {NoStop}%
\bibitem [{\citenamefont {Wu}\ \emph {et~al.}(2021)\citenamefont {Wu},
  \citenamefont {Shen},\ and\ \citenamefont {Song}}]{Wu:2021xgu}%
  \BibitemOpen
  \bibfield  {author} {\bibinfo {author} {\bibfnamefont {Shanjin}\ \bibnamefont
  {Wu}}, \bibinfo {author} {\bibfnamefont {Chun}\ \bibnamefont {Shen}}, \ and\
  \bibinfo {author} {\bibfnamefont {Huichao}\ \bibnamefont {Song}},\ }\bibfield
   {title} {\enquote {\bibinfo {title} {{Dynamically Exploring the QCD Matter
  at Finite Temperatures and Densities: A Short Review}},}\ }\href {\doibase
  10.1088/0256-307X/38/8/081201} {\bibfield  {journal} {\bibinfo  {journal}
  {Chin. Phys. Lett.}\ }\textbf {\bibinfo {volume} {38}},\ \bibinfo {pages}
  {081201} (\bibinfo {year} {2021})},\ \Eprint
  {http://arxiv.org/abs/2104.13250} {arXiv:2104.13250 [nucl-th]} \BibitemShut
  {NoStop}%
\bibitem [{\citenamefont {An}\ \emph {et~al.}(2022)\citenamefont {An} \emph
  {et~al.}}]{An:2021wof}%
  \BibitemOpen
  \bibfield  {author} {\bibinfo {author} {\bibfnamefont {Xin}\ \bibnamefont
  {An}} \emph {et~al.},\ }\bibfield  {title} {\enquote {\bibinfo {title} {{The
  BEST framework for the search for the QCD critical point and the chiral
  magnetic effect}},}\ }\href {\doibase 10.1016/j.nuclphysa.2021.122343}
  {\bibfield  {journal} {\bibinfo  {journal} {Nucl. Phys. A}\ }\textbf
  {\bibinfo {volume} {1017}},\ \bibinfo {pages} {122343} (\bibinfo {year}
  {2022})},\ \Eprint {http://arxiv.org/abs/2108.13867} {arXiv:2108.13867
  [nucl-th]} \BibitemShut {NoStop}%
\bibitem [{\citenamefont {Gale}\ \emph {et~al.}(2013)\citenamefont {Gale},
  \citenamefont {Jeon},\ and\ \citenamefont {Schenke}}]{Gale:2013da}%
  \BibitemOpen
  \bibfield  {author} {\bibinfo {author} {\bibfnamefont {Charles}\ \bibnamefont
  {Gale}}, \bibinfo {author} {\bibfnamefont {Sangyong}\ \bibnamefont {Jeon}}, \
  and\ \bibinfo {author} {\bibfnamefont {Bjoern}\ \bibnamefont {Schenke}},\
  }\bibfield  {title} {\enquote {\bibinfo {title} {{Hydrodynamic Modeling of
  Heavy-Ion Collisions}},}\ }\href {\doibase 10.1142/S0217751X13400113}
  {\bibfield  {journal} {\bibinfo  {journal} {Int. J. Mod. Phys. A}\ }\textbf
  {\bibinfo {volume} {28}},\ \bibinfo {pages} {1340011} (\bibinfo {year}
  {2013})},\ \Eprint {http://arxiv.org/abs/1301.5893} {arXiv:1301.5893
  [nucl-th]} \BibitemShut {NoStop}%
\bibitem [{\citenamefont {Shen}\ and\ \citenamefont
  {Yan}(2020)}]{Shen:2020mgh}%
  \BibitemOpen
  \bibfield  {author} {\bibinfo {author} {\bibfnamefont {Chun}\ \bibnamefont
  {Shen}}\ and\ \bibinfo {author} {\bibfnamefont {Li}~\bibnamefont {Yan}},\
  }\bibfield  {title} {\enquote {\bibinfo {title} {{Recent development of
  hydrodynamic modeling in heavy-ion collisions}},}\ }\href {\doibase
  10.1007/s41365-020-00829-z} {\bibfield  {journal} {\bibinfo  {journal} {Nucl.
  Sci. Tech.}\ }\textbf {\bibinfo {volume} {31}},\ \bibinfo {pages} {122}
  (\bibinfo {year} {2020})},\ \Eprint {http://arxiv.org/abs/2010.12377}
  {arXiv:2010.12377 [nucl-th]} \BibitemShut {NoStop}%
\bibitem [{\citenamefont {Karpenko}\ \emph {et~al.}(2015)\citenamefont
  {Karpenko}, \citenamefont {Huovinen}, \citenamefont {Petersen},\ and\
  \citenamefont {Bleicher}}]{Karpenko:2015xea}%
  \BibitemOpen
  \bibfield  {author} {\bibinfo {author} {\bibfnamefont {Iu.~A.}\ \bibnamefont
  {Karpenko}}, \bibinfo {author} {\bibfnamefont {P.}~\bibnamefont {Huovinen}},
  \bibinfo {author} {\bibfnamefont {H.}~\bibnamefont {Petersen}}, \ and\
  \bibinfo {author} {\bibfnamefont {M.}~\bibnamefont {Bleicher}},\ }\bibfield
  {title} {\enquote {\bibinfo {title} {{Estimation of the shear viscosity at
  finite net-baryon density from $A+A$ collision data at $\sqrt{s_\mathrm{NN}}
  = 7.7-200$ GeV}},}\ }\href {\doibase 10.1103/PhysRevC.91.064901} {\bibfield
  {journal} {\bibinfo  {journal} {Phys. Rev. C}\ }\textbf {\bibinfo {volume}
  {91}},\ \bibinfo {pages} {064901} (\bibinfo {year} {2015})},\ \Eprint
  {http://arxiv.org/abs/1502.01978} {arXiv:1502.01978 [nucl-th]} \BibitemShut
  {NoStop}%
\bibitem [{\citenamefont {Shen}\ and\ \citenamefont
  {Schenke}(2018{\natexlab{a}})}]{Shen:2017bsr}%
  \BibitemOpen
  \bibfield  {author} {\bibinfo {author} {\bibfnamefont {Chun}\ \bibnamefont
  {Shen}}\ and\ \bibinfo {author} {\bibfnamefont {Bj\"orn}\ \bibnamefont
  {Schenke}},\ }\bibfield  {title} {\enquote {\bibinfo {title} {{Dynamical
  initial state model for relativistic heavy-ion collisions}},}\ }\href
  {\doibase 10.1103/PhysRevC.97.024907} {\bibfield  {journal} {\bibinfo
  {journal} {Phys. Rev. C}\ }\textbf {\bibinfo {volume} {97}},\ \bibinfo
  {pages} {024907} (\bibinfo {year} {2018}{\natexlab{a}})},\ \Eprint
  {http://arxiv.org/abs/1710.00881} {arXiv:1710.00881 [nucl-th]} \BibitemShut
  {NoStop}%
\bibitem [{\citenamefont {Sch\"afer}\ \emph {et~al.}(2021)\citenamefont
  {Sch\"afer}, \citenamefont {Karpenko}, \citenamefont {Wu}, \citenamefont
  {Hammelmann},\ and\ \citenamefont {Elfner}}]{Schafer:2021csj}%
  \BibitemOpen
  \bibfield  {author} {\bibinfo {author} {\bibfnamefont {Anna}\ \bibnamefont
  {Sch\"afer}}, \bibinfo {author} {\bibfnamefont {Iurii}\ \bibnamefont
  {Karpenko}}, \bibinfo {author} {\bibfnamefont {Xiang-Yu}\ \bibnamefont {Wu}},
  \bibinfo {author} {\bibfnamefont {Jan}\ \bibnamefont {Hammelmann}}, \ and\
  \bibinfo {author} {\bibfnamefont {Hannah}\ \bibnamefont {Elfner}},\
  }\bibfield  {title} {\enquote {\bibinfo {title} {{Particle production in a
  hybrid approach for a beam energy scan of Au+Au/Pb+Pb collisions between
  $\sqrt{s_\mathrm{NN}}$ = 4.3 GeV and $\sqrt{s_\mathrm{NN}}$ = 200.0 GeV}},}\
  }\href@noop {} {\  (\bibinfo {year} {2021})},\ \Eprint
  {http://arxiv.org/abs/2112.08724} {arXiv:2112.08724 [hep-ph]} \BibitemShut
  {NoStop}%
\bibitem [{\citenamefont {Sun}\ \emph {et~al.}(2017)\citenamefont {Sun},
  \citenamefont {Chen}, \citenamefont {Ko},\ and\ \citenamefont
  {Xu}}]{Sun:2017xrx}%
  \BibitemOpen
  \bibfield  {author} {\bibinfo {author} {\bibfnamefont {Kai-Jia}\ \bibnamefont
  {Sun}}, \bibinfo {author} {\bibfnamefont {Lie-Wen}\ \bibnamefont {Chen}},
  \bibinfo {author} {\bibfnamefont {Che~Ming}\ \bibnamefont {Ko}}, \ and\
  \bibinfo {author} {\bibfnamefont {Zhangbu}\ \bibnamefont {Xu}},\ }\bibfield
  {title} {\enquote {\bibinfo {title} {{Probing QCD critical fluctuations from
  light nuclei production in relativistic heavy-ion collisions}},}\ }\href
  {\doibase 10.1016/j.physletb.2017.09.056} {\bibfield  {journal} {\bibinfo
  {journal} {Phys. Lett. B}\ }\textbf {\bibinfo {volume} {774}},\ \bibinfo
  {pages} {103--107} (\bibinfo {year} {2017})},\ \Eprint
  {http://arxiv.org/abs/1702.07620} {arXiv:1702.07620 [nucl-th]} \BibitemShut
  {NoStop}%
\bibitem [{\citenamefont {Sun}\ \emph {et~al.}(2018)\citenamefont {Sun},
  \citenamefont {Chen}, \citenamefont {Ko}, \citenamefont {Pu},\ and\
  \citenamefont {Xu}}]{Sun:2018jhg}%
  \BibitemOpen
  \bibfield  {author} {\bibinfo {author} {\bibfnamefont {Kai-Jia}\ \bibnamefont
  {Sun}}, \bibinfo {author} {\bibfnamefont {Lie-Wen}\ \bibnamefont {Chen}},
  \bibinfo {author} {\bibfnamefont {Che~Ming}\ \bibnamefont {Ko}}, \bibinfo
  {author} {\bibfnamefont {Jie}\ \bibnamefont {Pu}}, \ and\ \bibinfo {author}
  {\bibfnamefont {Zhangbu}\ \bibnamefont {Xu}},\ }\bibfield  {title} {\enquote
  {\bibinfo {title} {{Light nuclei production as a probe of the QCD phase
  diagram}},}\ }\href {\doibase 10.1016/j.physletb.2018.04.035} {\bibfield
  {journal} {\bibinfo  {journal} {Phys. Lett. B}\ }\textbf {\bibinfo {volume}
  {781}},\ \bibinfo {pages} {499--504} (\bibinfo {year} {2018})},\ \Eprint
  {http://arxiv.org/abs/1801.09382} {arXiv:1801.09382 [nucl-th]} \BibitemShut
  {NoStop}%
\bibitem [{\citenamefont {Oliinychenko}\ \emph {et~al.}(2021)\citenamefont
  {Oliinychenko}, \citenamefont {Shen},\ and\ \citenamefont
  {Koch}}]{Oliinychenko:2020znl}%
  \BibitemOpen
  \bibfield  {author} {\bibinfo {author} {\bibfnamefont {Dmytro}\ \bibnamefont
  {Oliinychenko}}, \bibinfo {author} {\bibfnamefont {Chun}\ \bibnamefont
  {Shen}}, \ and\ \bibinfo {author} {\bibfnamefont {Volker}\ \bibnamefont
  {Koch}},\ }\bibfield  {title} {\enquote {\bibinfo {title} {{Deuteron
  production in AuAu collisions at $\sqrt{s_{NN}}=$7\textendash{}200 GeV via
  pion catalysis}},}\ }\href {\doibase 10.1103/PhysRevC.103.034913} {\bibfield
  {journal} {\bibinfo  {journal} {Phys. Rev. C}\ }\textbf {\bibinfo {volume}
  {103}},\ \bibinfo {pages} {034913} (\bibinfo {year} {2021})},\ \Eprint
  {http://arxiv.org/abs/2009.01915} {arXiv:2009.01915 [hep-ph]} \BibitemShut
  {NoStop}%
\bibitem [{\citenamefont {Zhao}\ \emph {et~al.}(2020)\citenamefont {Zhao},
  \citenamefont {Shen}, \citenamefont {Ko}, \citenamefont {Liu},\ and\
  \citenamefont {Song}}]{Zhao:2020irc}%
  \BibitemOpen
  \bibfield  {author} {\bibinfo {author} {\bibfnamefont {Wenbin}\ \bibnamefont
  {Zhao}}, \bibinfo {author} {\bibfnamefont {Chun}\ \bibnamefont {Shen}},
  \bibinfo {author} {\bibfnamefont {Che~Ming}\ \bibnamefont {Ko}}, \bibinfo
  {author} {\bibfnamefont {Quansheng}\ \bibnamefont {Liu}}, \ and\ \bibinfo
  {author} {\bibfnamefont {Huichao}\ \bibnamefont {Song}},\ }\bibfield  {title}
  {\enquote {\bibinfo {title} {{Beam-energy dependence of the production of
  light nuclei in Au + Au collisions}},}\ }\href {\doibase
  10.1103/PhysRevC.102.044912} {\bibfield  {journal} {\bibinfo  {journal}
  {Phys. Rev. C}\ }\textbf {\bibinfo {volume} {102}},\ \bibinfo {pages}
  {044912} (\bibinfo {year} {2020})},\ \Eprint
  {http://arxiv.org/abs/2009.06959} {arXiv:2009.06959 [nucl-th]} \BibitemShut
  {NoStop}%
\bibitem [{\citenamefont {Sun}\ \emph {et~al.}(2021)\citenamefont {Sun},
  \citenamefont {Wang}, \citenamefont {Ko}, \citenamefont {Ma},\ and\
  \citenamefont {Shen}}]{Sun:2021dlz}%
  \BibitemOpen
  \bibfield  {author} {\bibinfo {author} {\bibfnamefont {Kai-Jia}\ \bibnamefont
  {Sun}}, \bibinfo {author} {\bibfnamefont {Rui}\ \bibnamefont {Wang}},
  \bibinfo {author} {\bibfnamefont {Che~Ming}\ \bibnamefont {Ko}}, \bibinfo
  {author} {\bibfnamefont {Yu-Gang}\ \bibnamefont {Ma}}, \ and\ \bibinfo
  {author} {\bibfnamefont {Chun}\ \bibnamefont {Shen}},\ }\bibfield  {title}
  {\enquote {\bibinfo {title} {{Relativistic kinetic approach to light nuclei
  production in high-energy nuclear collisions}},}\ }\href@noop {} {\
  (\bibinfo {year} {2021})},\ \Eprint {http://arxiv.org/abs/2106.12742}
  {arXiv:2106.12742 [nucl-th]} \BibitemShut {NoStop}%
\bibitem [{\citenamefont {Bearden}\ \emph {et~al.}(2004)\citenamefont {Bearden}
  \emph {et~al.}}]{BRAHMS:2003wwg}%
  \BibitemOpen
  \bibfield  {author} {\bibinfo {author} {\bibfnamefont {I.~G.}\ \bibnamefont
  {Bearden}} \emph {et~al.} (\bibinfo {collaboration} {BRAHMS}),\ }\bibfield
  {title} {\enquote {\bibinfo {title} {{Nuclear stopping in Au + Au collisions
  at s(NN)**(1/2) = 200-GeV}},}\ }\href {\doibase
  10.1103/PhysRevLett.93.102301} {\bibfield  {journal} {\bibinfo  {journal}
  {Phys. Rev. Lett.}\ }\textbf {\bibinfo {volume} {93}},\ \bibinfo {pages}
  {102301} (\bibinfo {year} {2004})},\ \Eprint
  {http://arxiv.org/abs/nucl-ex/0312023} {arXiv:nucl-ex/0312023} \BibitemShut
  {NoStop}%
\bibitem [{\citenamefont {Arsene}\ \emph {et~al.}(2009)\citenamefont {Arsene}
  \emph {et~al.}}]{BRAHMS:2009wlg}%
  \BibitemOpen
  \bibfield  {author} {\bibinfo {author} {\bibfnamefont {I.~C.}\ \bibnamefont
  {Arsene}} \emph {et~al.} (\bibinfo {collaboration} {BRAHMS}),\ }\bibfield
  {title} {\enquote {\bibinfo {title} {{Nuclear stopping and rapidity loss in
  Au+Au collisions at s(NN)**(1/2) = 62.4-GeV}},}\ }\href {\doibase
  10.1016/j.physletb.2009.05.049} {\bibfield  {journal} {\bibinfo  {journal}
  {Phys. Lett. B}\ }\textbf {\bibinfo {volume} {677}},\ \bibinfo {pages}
  {267--271} (\bibinfo {year} {2009})},\ \Eprint
  {http://arxiv.org/abs/0901.0872} {arXiv:0901.0872 [nucl-ex]} \BibitemShut
  {NoStop}%
\bibitem [{\citenamefont {Abelev}\ \emph {et~al.}(2009)\citenamefont {Abelev}
  \emph {et~al.}}]{STAR:2008med}%
  \BibitemOpen
  \bibfield  {author} {\bibinfo {author} {\bibfnamefont {B.~I.}\ \bibnamefont
  {Abelev}} \emph {et~al.} (\bibinfo {collaboration} {STAR}),\ }\bibfield
  {title} {\enquote {\bibinfo {title} {{Systematic Measurements of Identified
  Particle Spectra in $p p, d^+$ Au and Au+Au Collisions from STAR}},}\ }\href
  {\doibase 10.1103/PhysRevC.79.034909} {\bibfield  {journal} {\bibinfo
  {journal} {Phys. Rev. C}\ }\textbf {\bibinfo {volume} {79}},\ \bibinfo
  {pages} {034909} (\bibinfo {year} {2009})},\ \Eprint
  {http://arxiv.org/abs/0808.2041} {arXiv:0808.2041 [nucl-ex]} \BibitemShut
  {NoStop}%
\bibitem [{\citenamefont {Adamczyk}\ \emph {et~al.}(2017)\citenamefont
  {Adamczyk} \emph {et~al.}}]{STAR:2017sal}%
  \BibitemOpen
  \bibfield  {author} {\bibinfo {author} {\bibfnamefont {L.}~\bibnamefont
  {Adamczyk}} \emph {et~al.} (\bibinfo {collaboration} {STAR}),\ }\bibfield
  {title} {\enquote {\bibinfo {title} {{Bulk Properties of the Medium Produced
  in Relativistic Heavy-Ion Collisions from the Beam Energy Scan Program}},}\
  }\href {\doibase 10.1103/PhysRevC.96.044904} {\bibfield  {journal} {\bibinfo
  {journal} {Phys. Rev. C}\ }\textbf {\bibinfo {volume} {96}},\ \bibinfo
  {pages} {044904} (\bibinfo {year} {2017})},\ \Eprint
  {http://arxiv.org/abs/1701.07065} {arXiv:1701.07065 [nucl-ex]} \BibitemShut
  {NoStop}%
\bibitem [{\citenamefont {Woods}\ and\ \citenamefont
  {Saxon}(1954)}]{PhysRev.95.577}%
  \BibitemOpen
  \bibfield  {author} {\bibinfo {author} {\bibfnamefont {Roger~D.}\
  \bibnamefont {Woods}}\ and\ \bibinfo {author} {\bibfnamefont {David~S.}\
  \bibnamefont {Saxon}},\ }\bibfield  {title} {\enquote {\bibinfo {title}
  {Diffuse surface optical model for nucleon-nuclei scattering},}\ }\href
  {\doibase 10.1103/PhysRev.95.577} {\bibfield  {journal} {\bibinfo  {journal}
  {Phys. Rev.}\ }\textbf {\bibinfo {volume} {95}},\ \bibinfo {pages} {577--578}
  (\bibinfo {year} {1954})}\BibitemShut {NoStop}%
\bibitem [{\citenamefont {Eskola}\ \emph {et~al.}(2009)\citenamefont {Eskola},
  \citenamefont {Paukkunen},\ and\ \citenamefont {Salgado}}]{Eskola:2009uj}%
  \BibitemOpen
  \bibfield  {author} {\bibinfo {author} {\bibfnamefont {K.~J.}\ \bibnamefont
  {Eskola}}, \bibinfo {author} {\bibfnamefont {H.}~\bibnamefont {Paukkunen}}, \
  and\ \bibinfo {author} {\bibfnamefont {C.~A.}\ \bibnamefont {Salgado}},\
  }\bibfield  {title} {\enquote {\bibinfo {title} {{EPS09: A New Generation of
  NLO and LO Nuclear Parton Distribution Functions}},}\ }\href {\doibase
  10.1088/1126-6708/2009/04/065} {\bibfield  {journal} {\bibinfo  {journal}
  {JHEP}\ }\textbf {\bibinfo {volume} {04}},\ \bibinfo {pages} {065} (\bibinfo
  {year} {2009})},\ \Eprint {http://arxiv.org/abs/0902.4154} {arXiv:0902.4154
  [hep-ph]} \BibitemShut {NoStop}%
\bibitem [{\citenamefont {Gao}\ \emph {et~al.}(2014)\citenamefont {Gao},
  \citenamefont {Guzzi}, \citenamefont {Huston}, \citenamefont {Lai},
  \citenamefont {Li}, \citenamefont {Nadolsky}, \citenamefont {Pumplin},
  \citenamefont {Stump},\ and\ \citenamefont {Yuan}}]{Gao:2013xoa}%
  \BibitemOpen
  \bibfield  {author} {\bibinfo {author} {\bibfnamefont {Jun}\ \bibnamefont
  {Gao}}, \bibinfo {author} {\bibfnamefont {Marco}\ \bibnamefont {Guzzi}},
  \bibinfo {author} {\bibfnamefont {Joey}\ \bibnamefont {Huston}}, \bibinfo
  {author} {\bibfnamefont {Hung-Liang}\ \bibnamefont {Lai}}, \bibinfo {author}
  {\bibfnamefont {Zhao}\ \bibnamefont {Li}}, \bibinfo {author} {\bibfnamefont
  {Pavel}\ \bibnamefont {Nadolsky}}, \bibinfo {author} {\bibfnamefont {Jon}\
  \bibnamefont {Pumplin}}, \bibinfo {author} {\bibfnamefont {Daniel}\
  \bibnamefont {Stump}}, \ and\ \bibinfo {author} {\bibfnamefont {C.~P.}\
  \bibnamefont {Yuan}},\ }\bibfield  {title} {\enquote {\bibinfo {title} {{CT10
  next-to-next-to-leading order global analysis of QCD}},}\ }\href {\doibase
  10.1103/PhysRevD.89.033009} {\bibfield  {journal} {\bibinfo  {journal} {Phys.
  Rev. D}\ }\textbf {\bibinfo {volume} {89}},\ \bibinfo {pages} {033009}
  (\bibinfo {year} {2014})},\ \Eprint {http://arxiv.org/abs/1302.6246}
  {arXiv:1302.6246 [hep-ph]} \BibitemShut {NoStop}%
\bibitem [{\citenamefont {Li}\ and\ \citenamefont
  {Kapusta}(2019)}]{Li:2018ini}%
  \BibitemOpen
  \bibfield  {author} {\bibinfo {author} {\bibfnamefont {Ming}\ \bibnamefont
  {Li}}\ and\ \bibinfo {author} {\bibfnamefont {Joseph~I.}\ \bibnamefont
  {Kapusta}},\ }\bibfield  {title} {\enquote {\bibinfo {title} {{Large Baryon
  Densities Achievable in High Energy Heavy Ion Collisions Outside the Central
  Rapidity Region}},}\ }\href {\doibase 10.1103/PhysRevC.99.014906} {\bibfield
  {journal} {\bibinfo  {journal} {Phys. Rev. C}\ }\textbf {\bibinfo {volume}
  {99}},\ \bibinfo {pages} {014906} (\bibinfo {year} {2019})},\ \Eprint
  {http://arxiv.org/abs/1808.05751} {arXiv:1808.05751 [nucl-th]} \BibitemShut
  {NoStop}%
\bibitem [{\citenamefont {McLerran}\ \emph {et~al.}(2019)\citenamefont
  {McLerran}, \citenamefont {Schlichting},\ and\ \citenamefont
  {Sen}}]{McLerran:2018avb}%
  \BibitemOpen
  \bibfield  {author} {\bibinfo {author} {\bibfnamefont {Larry~D.}\
  \bibnamefont {McLerran}}, \bibinfo {author} {\bibfnamefont {S\"oren}\
  \bibnamefont {Schlichting}}, \ and\ \bibinfo {author} {\bibfnamefont
  {Srimoyee}\ \bibnamefont {Sen}},\ }\bibfield  {title} {\enquote {\bibinfo
  {title} {{Spacetime picture of baryon stopping in the color-glass
  condensate}},}\ }\href {\doibase 10.1103/PhysRevD.99.074009} {\bibfield
  {journal} {\bibinfo  {journal} {Phys. Rev. D}\ }\textbf {\bibinfo {volume}
  {99}},\ \bibinfo {pages} {074009} (\bibinfo {year} {2019})},\ \Eprint
  {http://arxiv.org/abs/1811.04089} {arXiv:1811.04089 [hep-ph]} \BibitemShut
  {NoStop}%
\bibitem [{\citenamefont {Mishustin}\ and\ \citenamefont
  {Kapusta}(2002)}]{Mishustin:2001ib}%
  \BibitemOpen
  \bibfield  {author} {\bibinfo {author} {\bibfnamefont {I.~N.}\ \bibnamefont
  {Mishustin}}\ and\ \bibinfo {author} {\bibfnamefont {Joseph~I.}\ \bibnamefont
  {Kapusta}},\ }\bibfield  {title} {\enquote {\bibinfo {title} {{Collective
  deceleration of ultrarelativistic nuclei and creation of quark gluon
  plasma}},}\ }\href {\doibase 10.1103/PhysRevLett.88.112501} {\bibfield
  {journal} {\bibinfo  {journal} {Phys. Rev. Lett.}\ }\textbf {\bibinfo
  {volume} {88}},\ \bibinfo {pages} {112501} (\bibinfo {year} {2002})},\
  \Eprint {http://arxiv.org/abs/hep-ph/0110321} {arXiv:hep-ph/0110321}
  \BibitemShut {NoStop}%
\bibitem [{\citenamefont {Bialas}\ \emph {et~al.}(2018)\citenamefont {Bialas},
  \citenamefont {Bzdak},\ and\ \citenamefont {Koch}}]{Bialas:2016epd}%
  \BibitemOpen
  \bibfield  {author} {\bibinfo {author} {\bibfnamefont {Andrzej}\ \bibnamefont
  {Bialas}}, \bibinfo {author} {\bibfnamefont {Adam}\ \bibnamefont {Bzdak}}, \
  and\ \bibinfo {author} {\bibfnamefont {Volker}\ \bibnamefont {Koch}},\
  }\bibfield  {title} {\enquote {\bibinfo {title} {{Stopped nucleons in
  configuration space}},}\ }\href {\doibase 10.5506/APhysPolB.49.103}
  {\bibfield  {journal} {\bibinfo  {journal} {Acta Phys. Polon. B}\ }\textbf
  {\bibinfo {volume} {49}},\ \bibinfo {pages} {103} (\bibinfo {year} {2018})},\
  \Eprint {http://arxiv.org/abs/1608.07041} {arXiv:1608.07041 [hep-ph]}
  \BibitemShut {NoStop}%
\bibitem [{\citenamefont {Shen}\ and\ \citenamefont
  {Schenke}(2018{\natexlab{b}})}]{Shen:2017fnn}%
  \BibitemOpen
  \bibfield  {author} {\bibinfo {author} {\bibfnamefont {Chun}\ \bibnamefont
  {Shen}}\ and\ \bibinfo {author} {\bibfnamefont {Bj\"orn}\ \bibnamefont
  {Schenke}},\ }\bibfield  {title} {\enquote {\bibinfo {title} {{Initial state
  and hydrodynamic modeling of heavy-ion collisions at RHIC BES energies}},}\
  }\href {\doibase 10.22323/1.311.0006} {\bibfield  {journal} {\bibinfo
  {journal} {PoS}\ }\textbf {\bibinfo {volume} {CPOD2017}},\ \bibinfo {pages}
  {006} (\bibinfo {year} {2018}{\natexlab{b}})},\ \Eprint
  {http://arxiv.org/abs/1711.10544} {arXiv:1711.10544 [nucl-th]} \BibitemShut
  {NoStop}%
\bibitem [{\citenamefont {Kharzeev}(1996)}]{Kharzeev:1996sq}%
  \BibitemOpen
  \bibfield  {author} {\bibinfo {author} {\bibfnamefont {D.}~\bibnamefont
  {Kharzeev}},\ }\bibfield  {title} {\enquote {\bibinfo {title} {{Can gluons
  trace baryon number?}}}\ }\href {\doibase 10.1016/0370-2693(96)00435-2}
  {\bibfield  {journal} {\bibinfo  {journal} {Phys. Lett. B}\ }\textbf
  {\bibinfo {volume} {378}},\ \bibinfo {pages} {238--246} (\bibinfo {year}
  {1996})},\ \Eprint {http://arxiv.org/abs/nucl-th/9602027}
  {arXiv:nucl-th/9602027} \BibitemShut {NoStop}%
\bibitem [{\citenamefont {Adamczyk}\ \emph {et~al.}(2018)\citenamefont
  {Adamczyk} \emph {et~al.}}]{STAR:2017ieb}%
  \BibitemOpen
  \bibfield  {author} {\bibinfo {author} {\bibfnamefont {L.}~\bibnamefont
  {Adamczyk}} \emph {et~al.} (\bibinfo {collaboration} {STAR}),\ }\bibfield
  {title} {\enquote {\bibinfo {title} {{Beam Energy Dependence of Jet-Quenching
  Effects in Au+Au Collisions at $\sqrt{s_{_{ \mathrm{NN}}}}$ = 7.7, 11.5,
  14.5, 19.6, 27, 39, and 62.4 GeV}},}\ }\href {\doibase
  10.1103/PhysRevLett.121.032301} {\bibfield  {journal} {\bibinfo  {journal}
  {Phys. Rev. Lett.}\ }\textbf {\bibinfo {volume} {121}},\ \bibinfo {pages}
  {032301} (\bibinfo {year} {2018})},\ \Eprint
  {http://arxiv.org/abs/1707.01988} {arXiv:1707.01988 [nucl-ex]} \BibitemShut
  {NoStop}%
\bibitem [{\citenamefont {Monnai}\ \emph {et~al.}(2019)\citenamefont {Monnai},
  \citenamefont {Schenke},\ and\ \citenamefont {Shen}}]{Monnai:2019hkn}%
  \BibitemOpen
  \bibfield  {author} {\bibinfo {author} {\bibfnamefont {Akihiko}\ \bibnamefont
  {Monnai}}, \bibinfo {author} {\bibfnamefont {Bj\"orn}\ \bibnamefont
  {Schenke}}, \ and\ \bibinfo {author} {\bibfnamefont {Chun}\ \bibnamefont
  {Shen}},\ }\bibfield  {title} {\enquote {\bibinfo {title} {{Equation of state
  at finite densities for QCD matter in nuclear collisions}},}\ }\href
  {\doibase 10.1103/PhysRevC.100.024907} {\bibfield  {journal} {\bibinfo
  {journal} {Phys. Rev. C}\ }\textbf {\bibinfo {volume} {100}},\ \bibinfo
  {pages} {024907} (\bibinfo {year} {2019})},\ \Eprint
  {http://arxiv.org/abs/1902.05095} {arXiv:1902.05095 [nucl-th]} \BibitemShut
  {NoStop}%
\bibitem [{\citenamefont {Schenke}\ \emph {et~al.}(2010)\citenamefont
  {Schenke}, \citenamefont {Jeon},\ and\ \citenamefont
  {Gale}}]{Schenke:2010nt}%
  \BibitemOpen
  \bibfield  {author} {\bibinfo {author} {\bibfnamefont {Bjoern}\ \bibnamefont
  {Schenke}}, \bibinfo {author} {\bibfnamefont {Sangyong}\ \bibnamefont
  {Jeon}}, \ and\ \bibinfo {author} {\bibfnamefont {Charles}\ \bibnamefont
  {Gale}},\ }\bibfield  {title} {\enquote {\bibinfo {title} {{(3+1)D
  hydrodynamic simulation of relativistic heavy-ion collisions}},}\ }\href
  {\doibase 10.1103/PhysRevC.82.014903} {\bibfield  {journal} {\bibinfo
  {journal} {Phys. Rev. C}\ }\textbf {\bibinfo {volume} {82}},\ \bibinfo
  {pages} {014903} (\bibinfo {year} {2010})},\ \Eprint
  {http://arxiv.org/abs/1004.1408} {arXiv:1004.1408 [hep-ph]} \BibitemShut
  {NoStop}%
\bibitem [{\citenamefont {Schenke}\ \emph {et~al.}(2012)\citenamefont
  {Schenke}, \citenamefont {Jeon},\ and\ \citenamefont
  {Gale}}]{Schenke:2011bn}%
  \BibitemOpen
  \bibfield  {author} {\bibinfo {author} {\bibfnamefont {Bjorn}\ \bibnamefont
  {Schenke}}, \bibinfo {author} {\bibfnamefont {Sangyong}\ \bibnamefont
  {Jeon}}, \ and\ \bibinfo {author} {\bibfnamefont {Charles}\ \bibnamefont
  {Gale}},\ }\bibfield  {title} {\enquote {\bibinfo {title} {{Higher flow
  harmonics from (3+1)D event-by-event viscous hydrodynamics}},}\ }\href
  {\doibase 10.1103/PhysRevC.85.024901} {\bibfield  {journal} {\bibinfo
  {journal} {Phys. Rev. C}\ }\textbf {\bibinfo {volume} {85}},\ \bibinfo
  {pages} {024901} (\bibinfo {year} {2012})},\ \Eprint
  {http://arxiv.org/abs/1109.6289} {arXiv:1109.6289 [hep-ph]} \BibitemShut
  {NoStop}%
\bibitem [{\citenamefont {Paquet}\ \emph {et~al.}(2016)\citenamefont {Paquet},
  \citenamefont {Shen}, \citenamefont {Denicol}, \citenamefont {Luzum},
  \citenamefont {Schenke}, \citenamefont {Jeon},\ and\ \citenamefont
  {Gale}}]{Paquet:2015lta}%
  \BibitemOpen
  \bibfield  {author} {\bibinfo {author} {\bibfnamefont {Jean-Fran\c{c}ois}\
  \bibnamefont {Paquet}}, \bibinfo {author} {\bibfnamefont {Chun}\ \bibnamefont
  {Shen}}, \bibinfo {author} {\bibfnamefont {Gabriel~S.}\ \bibnamefont
  {Denicol}}, \bibinfo {author} {\bibfnamefont {Matthew}\ \bibnamefont
  {Luzum}}, \bibinfo {author} {\bibfnamefont {Bj\"orn}\ \bibnamefont
  {Schenke}}, \bibinfo {author} {\bibfnamefont {Sangyong}\ \bibnamefont
  {Jeon}}, \ and\ \bibinfo {author} {\bibfnamefont {Charles}\ \bibnamefont
  {Gale}},\ }\bibfield  {title} {\enquote {\bibinfo {title} {{Production of
  photons in relativistic heavy-ion collisions}},}\ }\href {\doibase
  10.1103/PhysRevC.93.044906} {\bibfield  {journal} {\bibinfo  {journal} {Phys.
  Rev. C}\ }\textbf {\bibinfo {volume} {93}},\ \bibinfo {pages} {044906}
  (\bibinfo {year} {2016})},\ \Eprint {http://arxiv.org/abs/1509.06738}
  {arXiv:1509.06738 [hep-ph]} \BibitemShut {NoStop}%
\bibitem [{\citenamefont {Denicol}\ \emph {et~al.}(2018)\citenamefont
  {Denicol}, \citenamefont {Gale}, \citenamefont {Jeon}, \citenamefont
  {Monnai}, \citenamefont {Schenke},\ and\ \citenamefont
  {Shen}}]{Denicol:2018wdp}%
  \BibitemOpen
  \bibfield  {author} {\bibinfo {author} {\bibfnamefont {Gabriel~S.}\
  \bibnamefont {Denicol}}, \bibinfo {author} {\bibfnamefont {Charles}\
  \bibnamefont {Gale}}, \bibinfo {author} {\bibfnamefont {Sangyong}\
  \bibnamefont {Jeon}}, \bibinfo {author} {\bibfnamefont {Akihiko}\
  \bibnamefont {Monnai}}, \bibinfo {author} {\bibfnamefont {Bj\"orn}\
  \bibnamefont {Schenke}}, \ and\ \bibinfo {author} {\bibfnamefont {Chun}\
  \bibnamefont {Shen}},\ }\bibfield  {title} {\enquote {\bibinfo {title} {{Net
  baryon diffusion in fluid dynamic simulations of relativistic heavy-ion
  collisions}},}\ }\href {\doibase 10.1103/PhysRevC.98.034916} {\bibfield
  {journal} {\bibinfo  {journal} {Phys. Rev. C}\ }\textbf {\bibinfo {volume}
  {98}},\ \bibinfo {pages} {034916} (\bibinfo {year} {2018})},\ \Eprint
  {http://arxiv.org/abs/1804.10557} {arXiv:1804.10557 [nucl-th]} \BibitemShut
  {NoStop}%
\bibitem [{\citenamefont {Cooper}\ and\ \citenamefont
  {Frye}(1974)}]{Cooper:1974mv}%
  \BibitemOpen
  \bibfield  {author} {\bibinfo {author} {\bibfnamefont {Fred}\ \bibnamefont
  {Cooper}}\ and\ \bibinfo {author} {\bibfnamefont {Graham}\ \bibnamefont
  {Frye}},\ }\bibfield  {title} {\enquote {\bibinfo {title} {{Comment on the
  Single Particle Distribution in the Hydrodynamic and Statistical
  Thermodynamic Models of Multiparticle Production}},}\ }\href {\doibase
  10.1103/PhysRevD.10.186} {\bibfield  {journal} {\bibinfo  {journal} {Phys.
  Rev. D}\ }\textbf {\bibinfo {volume} {10}},\ \bibinfo {pages} {186} (\bibinfo
  {year} {1974})}\BibitemShut {NoStop}%
\bibitem [{\citenamefont {Huovinen}\ and\ \citenamefont
  {Petersen}(2012)}]{Huovinen:2012is}%
  \BibitemOpen
  \bibfield  {author} {\bibinfo {author} {\bibfnamefont {Pasi}\ \bibnamefont
  {Huovinen}}\ and\ \bibinfo {author} {\bibfnamefont {Hannah}\ \bibnamefont
  {Petersen}},\ }\bibfield  {title} {\enquote {\bibinfo {title}
  {{Particlization in hybrid models}},}\ }\href {\doibase
  10.1140/epja/i2012-12171-9} {\bibfield  {journal} {\bibinfo  {journal} {Eur.
  Phys. J. A}\ }\textbf {\bibinfo {volume} {48}},\ \bibinfo {pages} {171}
  (\bibinfo {year} {2012})},\ \Eprint {http://arxiv.org/abs/1206.3371}
  {arXiv:1206.3371 [nucl-th]} \BibitemShut {NoStop}%
\bibitem [{\citenamefont {Bass}\ \emph {et~al.}(1998)\citenamefont {Bass} \emph
  {et~al.}}]{Bass:1998ca}%
  \BibitemOpen
  \bibfield  {author} {\bibinfo {author} {\bibfnamefont {S.~A.}\ \bibnamefont
  {Bass}} \emph {et~al.},\ }\bibfield  {title} {\enquote {\bibinfo {title}
  {{Microscopic models for ultrarelativistic heavy ion collisions}},}\ }\href
  {\doibase 10.1016/S0146-6410(98)00058-1} {\bibfield  {journal} {\bibinfo
  {journal} {Prog. Part. Nucl. Phys.}\ }\textbf {\bibinfo {volume} {41}},\
  \bibinfo {pages} {255--369} (\bibinfo {year} {1998})},\ \Eprint
  {http://arxiv.org/abs/nucl-th/9803035} {arXiv:nucl-th/9803035} \BibitemShut
  {NoStop}%
\bibitem [{\citenamefont {Bleicher}\ \emph {et~al.}(1999)\citenamefont
  {Bleicher} \emph {et~al.}}]{Bleicher:1999xi}%
  \BibitemOpen
  \bibfield  {author} {\bibinfo {author} {\bibfnamefont {M.}~\bibnamefont
  {Bleicher}} \emph {et~al.},\ }\bibfield  {title} {\enquote {\bibinfo {title}
  {{Relativistic hadron hadron collisions in the ultrarelativistic quantum
  molecular dynamics model}},}\ }\href {\doibase 10.1088/0954-3899/25/9/308}
  {\bibfield  {journal} {\bibinfo  {journal} {J. Phys. G}\ }\textbf {\bibinfo
  {volume} {25}},\ \bibinfo {pages} {1859--1896} (\bibinfo {year} {1999})},\
  \Eprint {http://arxiv.org/abs/hep-ph/9909407} {arXiv:hep-ph/9909407}
  \BibitemShut {NoStop}%
\bibitem [{iEB()}]{iEBEMUSIC}%
  \BibitemOpen
  \href@noop {} {}\bibinfo {note} {The open-source \textsc{iebe-music}
  overarching framework can be download from
  \url{https://github.com/chunshen1987/iEBE-MUSIC}}\BibitemShut {NoStop}%
\bibitem [{\citenamefont {Thome}\ \emph {et~al.}(1977)\citenamefont {Thome}
  \emph {et~al.}}]{Aachen-CERN-Heidelberg-Munich:1977izz}%
  \BibitemOpen
  \bibfield  {author} {\bibinfo {author} {\bibfnamefont {W.}~\bibnamefont
  {Thome}} \emph {et~al.} (\bibinfo {collaboration}
  {Aachen-CERN-Heidelberg-Munich}),\ }\bibfield  {title} {\enquote {\bibinfo
  {title} {{Charged Particle Multiplicity Distributions in p p Collisions at
  ISR Energies}},}\ }\href {\doibase 10.1016/0550-3213(77)90122-5} {\bibfield
  {journal} {\bibinfo  {journal} {Nucl. Phys. B}\ }\textbf {\bibinfo {volume}
  {129}},\ \bibinfo {pages} {365} (\bibinfo {year} {1977})}\BibitemShut
  {NoStop}%
\bibitem [{\citenamefont {Alner}\ \emph {et~al.}(1986)\citenamefont {Alner}
  \emph {et~al.}}]{UA5:1986yef}%
  \BibitemOpen
  \bibfield  {author} {\bibinfo {author} {\bibfnamefont {G.~J.}\ \bibnamefont
  {Alner}} \emph {et~al.} (\bibinfo {collaboration} {UA5}),\ }\bibfield
  {title} {\enquote {\bibinfo {title} {{Scaling of Pseudorapidity Distributions
  at c.m. Energies Up to 0.9-TeV}},}\ }\href {\doibase 10.1007/BF01410446}
  {\bibfield  {journal} {\bibinfo  {journal} {Z. Phys. C}\ }\textbf {\bibinfo
  {volume} {33}},\ \bibinfo {pages} {1--6} (\bibinfo {year}
  {1986})}\BibitemShut {NoStop}%
\bibitem [{\citenamefont {Alver}\ \emph {et~al.}(2011)\citenamefont {Alver}
  \emph {et~al.}}]{PHOBOS:2010eyu}%
  \BibitemOpen
  \bibfield  {author} {\bibinfo {author} {\bibfnamefont {B.}~\bibnamefont
  {Alver}} \emph {et~al.} (\bibinfo {collaboration} {PHOBOS}),\ }\bibfield
  {title} {\enquote {\bibinfo {title} {{Phobos results on charged particle
  multiplicity and pseudorapidity distributions in Au+Au, Cu+Cu, d+Au, and p+p
  collisions at ultra-relativistic energies}},}\ }\href {\doibase
  10.1103/PhysRevC.83.024913} {\bibfield  {journal} {\bibinfo  {journal} {Phys.
  Rev. C}\ }\textbf {\bibinfo {volume} {83}},\ \bibinfo {pages} {024913}
  (\bibinfo {year} {2011})},\ \Eprint {http://arxiv.org/abs/1011.1940}
  {arXiv:1011.1940 [nucl-ex]} \BibitemShut {NoStop}%
\bibitem [{\citenamefont {Ansorge}\ \emph {et~al.}(1989)\citenamefont {Ansorge}
  \emph {et~al.}}]{UA5:1988gup}%
  \BibitemOpen
  \bibfield  {author} {\bibinfo {author} {\bibfnamefont {R.~E.}\ \bibnamefont
  {Ansorge}} \emph {et~al.} (\bibinfo {collaboration} {UA5}),\ }\bibfield
  {title} {\enquote {\bibinfo {title} {{Charged Particle Multiplicity
  Distributions at 200-GeV and 900-GeV Center-Of-Mass Energy}},}\ }\href
  {\doibase 10.1007/BF01506531} {\bibfield  {journal} {\bibinfo  {journal} {Z.
  Phys. C}\ }\textbf {\bibinfo {volume} {43}},\ \bibinfo {pages} {357}
  (\bibinfo {year} {1989})}\BibitemShut {NoStop}%
\bibitem [{\citenamefont {Adare}\ \emph {et~al.}(2018)\citenamefont {Adare}
  \emph {et~al.}}]{PHENIX:2018hho}%
  \BibitemOpen
  \bibfield  {author} {\bibinfo {author} {\bibfnamefont {A.}~\bibnamefont
  {Adare}} \emph {et~al.} (\bibinfo {collaboration} {PHENIX}),\ }\bibfield
  {title} {\enquote {\bibinfo {title} {{Pseudorapidity Dependence of Particle
  Production and Elliptic Flow in Asymmetric Nuclear Collisions of $p+$Al,
  $p+$Au, $d+$Au, and $^{3}$He$+$Au at $\sqrt{s_{_{NN}}}=200$ GeV}},}\ }\href
  {\doibase 10.1103/PhysRevLett.121.222301} {\bibfield  {journal} {\bibinfo
  {journal} {Phys. Rev. Lett.}\ }\textbf {\bibinfo {volume} {121}},\ \bibinfo
  {pages} {222301} (\bibinfo {year} {2018})},\ \Eprint
  {http://arxiv.org/abs/1807.11928} {arXiv:1807.11928 [nucl-ex]} \BibitemShut
  {NoStop}%
\bibitem [{\citenamefont {Aidala}\ \emph {et~al.}(2017)\citenamefont {Aidala}
  \emph {et~al.}}]{PHENIX:2017nae}%
  \BibitemOpen
  \bibfield  {author} {\bibinfo {author} {\bibfnamefont {C.}~\bibnamefont
  {Aidala}} \emph {et~al.} (\bibinfo {collaboration} {PHENIX}),\ }\bibfield
  {title} {\enquote {\bibinfo {title} {{Measurements of azimuthal anisotropy
  and charged-particle multiplicity in $d+$Au collisions at
  $\sqrt{s_{_{NN}}}=$200, 62.4, 39, and 19.6 GeV}},}\ }\href {\doibase
  10.1103/PhysRevC.96.064905} {\bibfield  {journal} {\bibinfo  {journal} {Phys.
  Rev. C}\ }\textbf {\bibinfo {volume} {96}},\ \bibinfo {pages} {064905}
  (\bibinfo {year} {2017})},\ \Eprint {http://arxiv.org/abs/1708.06983}
  {arXiv:1708.06983 [nucl-ex]} \BibitemShut {NoStop}%
\bibitem [{\citenamefont {Acharya}\ \emph {et~al.}(2021)\citenamefont {Acharya}
  \emph {et~al.}}]{ALICE:2020swj}%
  \BibitemOpen
  \bibfield  {author} {\bibinfo {author} {\bibfnamefont {Shreyasi}\
  \bibnamefont {Acharya}} \emph {et~al.} (\bibinfo {collaboration} {ALICE}),\
  }\bibfield  {title} {\enquote {\bibinfo {title} {{Pseudorapidity
  distributions of charged particles as a function of mid- and forward rapidity
  multiplicities in pp collisions at $\sqrt{s}$~=~5.02, 7 and 13 TeV}},}\
  }\href {\doibase 10.1140/epjc/s10052-021-09349-5} {\bibfield  {journal}
  {\bibinfo  {journal} {Eur. Phys. J. C}\ }\textbf {\bibinfo {volume} {81}},\
  \bibinfo {pages} {630} (\bibinfo {year} {2021})},\ \Eprint
  {http://arxiv.org/abs/2009.09434} {arXiv:2009.09434 [nucl-ex]} \BibitemShut
  {NoStop}%
\bibitem [{\citenamefont {Adam}\ \emph
  {et~al.}(2017{\natexlab{a}})\citenamefont {Adam} \emph
  {et~al.}}]{ALICE:2015olq}%
  \BibitemOpen
  \bibfield  {author} {\bibinfo {author} {\bibfnamefont {Jaroslav}\
  \bibnamefont {Adam}} \emph {et~al.} (\bibinfo {collaboration} {ALICE}),\
  }\bibfield  {title} {\enquote {\bibinfo {title} {{Charged-particle
  multiplicities in proton\textendash{}proton collisions at $\sqrt{s} = 0.9$ to
  8 TeV}},}\ }\href {\doibase 10.1140/epjc/s10052-016-4571-1} {\bibfield
  {journal} {\bibinfo  {journal} {Eur. Phys. J. C}\ }\textbf {\bibinfo {volume}
  {77}},\ \bibinfo {pages} {33} (\bibinfo {year} {2017}{\natexlab{a}})},\
  \Eprint {http://arxiv.org/abs/1509.07541} {arXiv:1509.07541 [nucl-ex]}
  \BibitemShut {NoStop}%
\bibitem [{\citenamefont {Adam}\ \emph {et~al.}(2015)\citenamefont {Adam} \emph
  {et~al.}}]{ALICE:2014xsp}%
  \BibitemOpen
  \bibfield  {author} {\bibinfo {author} {\bibfnamefont {Jaroslav}\
  \bibnamefont {Adam}} \emph {et~al.} (\bibinfo {collaboration} {ALICE}),\
  }\bibfield  {title} {\enquote {\bibinfo {title} {{Centrality dependence of
  particle production in p-Pb collisions at $\sqrt{s_{\rm NN} }$= 5.02 TeV}},}\
  }\href {\doibase 10.1103/PhysRevC.91.064905} {\bibfield  {journal} {\bibinfo
  {journal} {Phys. Rev. C}\ }\textbf {\bibinfo {volume} {91}},\ \bibinfo
  {pages} {064905} (\bibinfo {year} {2015})},\ \Eprint
  {http://arxiv.org/abs/1412.6828} {arXiv:1412.6828 [nucl-ex]} \BibitemShut
  {NoStop}%
\bibitem [{\citenamefont {Acharya}\ \emph
  {et~al.}(2019{\natexlab{a}})\citenamefont {Acharya} \emph
  {et~al.}}]{ALICE:2018wma}%
  \BibitemOpen
  \bibfield  {author} {\bibinfo {author} {\bibfnamefont {Shreyasi}\
  \bibnamefont {Acharya}} \emph {et~al.} (\bibinfo {collaboration} {ALICE}),\
  }\bibfield  {title} {\enquote {\bibinfo {title} {{Charged-particle
  pseudorapidity density at mid-rapidity in p-Pb collisions at
  $\sqrt{s_{\rm{NN}}}$ = 8.16 TeV}},}\ }\href {\doibase
  10.1140/epjc/s10052-019-6801-9} {\bibfield  {journal} {\bibinfo  {journal}
  {Eur. Phys. J. C}\ }\textbf {\bibinfo {volume} {79}},\ \bibinfo {pages} {307}
  (\bibinfo {year} {2019}{\natexlab{a}})},\ \Eprint
  {http://arxiv.org/abs/1812.01312} {arXiv:1812.01312 [nucl-ex]} \BibitemShut
  {NoStop}%
\bibitem [{\citenamefont {Aad}\ \emph {et~al.}(2016)\citenamefont {Aad} \emph
  {et~al.}}]{ATLAS:2015hkr}%
  \BibitemOpen
  \bibfield  {author} {\bibinfo {author} {\bibfnamefont {Georges}\ \bibnamefont
  {Aad}} \emph {et~al.} (\bibinfo {collaboration} {ATLAS}),\ }\bibfield
  {title} {\enquote {\bibinfo {title} {{Measurement of the centrality
  dependence of the charged-particle pseudorapidity distribution in
  proton\textendash{}lead collisions at $\sqrt{s_{_\text {NN}}} = 5.02$ TeV
  with the ATLAS detector}},}\ }\href {\doibase 10.1140/epjc/s10052-016-4002-3}
  {\bibfield  {journal} {\bibinfo  {journal} {Eur. Phys. J. C}\ }\textbf
  {\bibinfo {volume} {76}},\ \bibinfo {pages} {199} (\bibinfo {year} {2016})},\
  \Eprint {http://arxiv.org/abs/1508.00848} {arXiv:1508.00848 [hep-ex]}
  \BibitemShut {NoStop}%
\bibitem [{\citenamefont {Christensen}(2017)}]{Christensen:2017zqh}%
  \BibitemOpen
  \bibfield  {author} {\bibinfo {author} {\bibfnamefont {Christian~Holm}\
  \bibnamefont {Christensen}} (\bibinfo {collaboration} {ALICE}),\ }\bibfield
  {title} {\enquote {\bibinfo {title} {{System-size dependence of the
  charged-particle pseudorapidity density at $\sqrt {s_{NN}}$ = 5.02 TeV with
  ALICE}},}\ }\href {\doibase 10.1016/j.nuclphysa.2017.05.066} {\bibfield
  {journal} {\bibinfo  {journal} {Nucl. Phys. A}\ }\textbf {\bibinfo {volume}
  {967}},\ \bibinfo {pages} {301--304} (\bibinfo {year} {2017})}\BibitemShut
  {NoStop}%
\bibitem [{\citenamefont {Back}\ \emph {et~al.}(2006)\citenamefont {Back} \emph
  {et~al.}}]{PHOBOS:2005zhy}%
  \BibitemOpen
  \bibfield  {author} {\bibinfo {author} {\bibfnamefont {B.~B.}\ \bibnamefont
  {Back}} \emph {et~al.} (\bibinfo {collaboration} {PHOBOS}),\ }\bibfield
  {title} {\enquote {\bibinfo {title} {{Charged-particle pseudorapidity
  distributions in Au+Au collisions at $s(NN) ^{1/2}$ = 62.4-GeV}},}\ }\href
  {\doibase 10.1103/PhysRevC.74.021901} {\bibfield  {journal} {\bibinfo
  {journal} {Phys. Rev. C}\ }\textbf {\bibinfo {volume} {74}},\ \bibinfo
  {pages} {021901} (\bibinfo {year} {2006})},\ \Eprint
  {http://arxiv.org/abs/nucl-ex/0509034} {arXiv:nucl-ex/0509034} \BibitemShut
  {NoStop}%
\bibitem [{\citenamefont {Vovchenko}\ \emph {et~al.}(2022)\citenamefont
  {Vovchenko}, \citenamefont {Koch},\ and\ \citenamefont
  {Shen}}]{Vovchenko:2021kxx}%
  \BibitemOpen
  \bibfield  {author} {\bibinfo {author} {\bibfnamefont {Volodymyr}\
  \bibnamefont {Vovchenko}}, \bibinfo {author} {\bibfnamefont {Volker}\
  \bibnamefont {Koch}}, \ and\ \bibinfo {author} {\bibfnamefont {Chun}\
  \bibnamefont {Shen}},\ }\bibfield  {title} {\enquote {\bibinfo {title}
  {{Proton number cumulants and correlation functions in Au-Au collisions at
  sNN=7.7\textendash{}200 GeV from hydrodynamics}},}\ }\href {\doibase
  10.1103/PhysRevC.105.014904} {\bibfield  {journal} {\bibinfo  {journal}
  {Phys. Rev. C}\ }\textbf {\bibinfo {volume} {105}},\ \bibinfo {pages}
  {014904} (\bibinfo {year} {2022})},\ \Eprint
  {http://arxiv.org/abs/2107.00163} {arXiv:2107.00163 [hep-ph]} \BibitemShut
  {NoStop}%
\bibitem [{\citenamefont {Shen}(2022)}]{Shen:2021nbe}%
  \BibitemOpen
  \bibfield  {author} {\bibinfo {author} {\bibfnamefont {Chun}\ \bibnamefont
  {Shen}},\ }\bibfield  {title} {\enquote {\bibinfo {title} {{Dynamic modeling
  for heavy-ion collisions}},}\ }\href {\doibase 10.1051/epjconf/202225902001}
  {\bibfield  {journal} {\bibinfo  {journal} {EPJ Web Conf.}\ }\textbf
  {\bibinfo {volume} {259}},\ \bibinfo {pages} {02001} (\bibinfo {year}
  {2022})},\ \Eprint {http://arxiv.org/abs/2108.04987} {arXiv:2108.04987
  [nucl-th]} \BibitemShut {NoStop}%
\bibitem [{\citenamefont {Klay}\ \emph {et~al.}(2002)\citenamefont {Klay} \emph
  {et~al.}}]{E895:2001zms}%
  \BibitemOpen
  \bibfield  {author} {\bibinfo {author} {\bibfnamefont {J.~L.}\ \bibnamefont
  {Klay}} \emph {et~al.} (\bibinfo {collaboration} {E895}),\ }\bibfield
  {title} {\enquote {\bibinfo {title} {{Longitudinal flow from 2-A-GeV to
  8-A-GeV Au+Au collisions at the Brookhaven AGS}},}\ }\href {\doibase
  10.1103/PhysRevLett.88.102301} {\bibfield  {journal} {\bibinfo  {journal}
  {Phys. Rev. Lett.}\ }\textbf {\bibinfo {volume} {88}},\ \bibinfo {pages}
  {102301} (\bibinfo {year} {2002})},\ \Eprint
  {http://arxiv.org/abs/nucl-ex/0111006} {arXiv:nucl-ex/0111006} \BibitemShut
  {NoStop}%
\bibitem [{\citenamefont {Afanasiev}\ \emph {et~al.}(2002)\citenamefont
  {Afanasiev} \emph {et~al.}}]{NA49:2002pzu}%
  \BibitemOpen
  \bibfield  {author} {\bibinfo {author} {\bibfnamefont {S.~V.}\ \bibnamefont
  {Afanasiev}} \emph {et~al.} (\bibinfo {collaboration} {NA49}),\ }\bibfield
  {title} {\enquote {\bibinfo {title} {{Energy dependence of pion and kaon
  production in central Pb + Pb collisions}},}\ }\href {\doibase
  10.1103/PhysRevC.66.054902} {\bibfield  {journal} {\bibinfo  {journal} {Phys.
  Rev. C}\ }\textbf {\bibinfo {volume} {66}},\ \bibinfo {pages} {054902}
  (\bibinfo {year} {2002})},\ \Eprint {http://arxiv.org/abs/nucl-ex/0205002}
  {arXiv:nucl-ex/0205002} \BibitemShut {NoStop}%
\bibitem [{\citenamefont {Alt}\ \emph {et~al.}(2005)\citenamefont {Alt} \emph
  {et~al.}}]{NA49:2004irs}%
  \BibitemOpen
  \bibfield  {author} {\bibinfo {author} {\bibfnamefont {C.}~\bibnamefont
  {Alt}} \emph {et~al.} (\bibinfo {collaboration} {NA49}),\ }\bibfield  {title}
  {\enquote {\bibinfo {title} {{Omega- and anti-Omega+ production in central Pb
  + Pb collisions at 40-AGeV and 158-AGeV}},}\ }\href {\doibase
  10.1103/PhysRevLett.94.192301} {\bibfield  {journal} {\bibinfo  {journal}
  {Phys. Rev. Lett.}\ }\textbf {\bibinfo {volume} {94}},\ \bibinfo {pages}
  {192301} (\bibinfo {year} {2005})},\ \Eprint
  {http://arxiv.org/abs/nucl-ex/0409004} {arXiv:nucl-ex/0409004} \BibitemShut
  {NoStop}%
\bibitem [{\citenamefont {Alt}\ \emph {et~al.}(2008)\citenamefont {Alt} \emph
  {et~al.}}]{NA49:2008ysv}%
  \BibitemOpen
  \bibfield  {author} {\bibinfo {author} {\bibfnamefont {C.}~\bibnamefont
  {Alt}} \emph {et~al.} (\bibinfo {collaboration} {NA49}),\ }\bibfield  {title}
  {\enquote {\bibinfo {title} {{Energy dependence of Lambda and Xi production
  in central Pb+Pb collisions at A-20, A-30, A-40, A-80, and A-158 GeV measured
  at the CERN Super Proton Synchrotron}},}\ }\href {\doibase
  10.1103/PhysRevC.78.034918} {\bibfield  {journal} {\bibinfo  {journal} {Phys.
  Rev. C}\ }\textbf {\bibinfo {volume} {78}},\ \bibinfo {pages} {034918}
  (\bibinfo {year} {2008})},\ \Eprint {http://arxiv.org/abs/0804.3770}
  {arXiv:0804.3770 [nucl-ex]} \BibitemShut {NoStop}%
\bibitem [{\citenamefont {Adam}\ \emph {et~al.}(2020)\citenamefont {Adam} \emph
  {et~al.}}]{STAR:2019bjj}%
  \BibitemOpen
  \bibfield  {author} {\bibinfo {author} {\bibfnamefont {Jaroslav}\
  \bibnamefont {Adam}} \emph {et~al.} (\bibinfo {collaboration} {STAR}),\
  }\bibfield  {title} {\enquote {\bibinfo {title} {{Strange hadron production
  in Au+Au collisions at $\sqrt{s_{NN}}=$7.7 , 11.5, 19.6, 27, and 39 GeV}},}\
  }\href {\doibase 10.1103/PhysRevC.102.034909} {\bibfield  {journal} {\bibinfo
   {journal} {Phys. Rev. C}\ }\textbf {\bibinfo {volume} {102}},\ \bibinfo
  {pages} {034909} (\bibinfo {year} {2020})},\ \Eprint
  {http://arxiv.org/abs/1906.03732} {arXiv:1906.03732 [nucl-ex]} \BibitemShut
  {NoStop}%
\bibitem [{\citenamefont {Monnai}\ \emph {et~al.}(2021)\citenamefont {Monnai},
  \citenamefont {Schenke},\ and\ \citenamefont {Shen}}]{Monnai:2021kgu}%
  \BibitemOpen
  \bibfield  {author} {\bibinfo {author} {\bibfnamefont {Akihiko}\ \bibnamefont
  {Monnai}}, \bibinfo {author} {\bibfnamefont {Bj\"orn}\ \bibnamefont
  {Schenke}}, \ and\ \bibinfo {author} {\bibfnamefont {Chun}\ \bibnamefont
  {Shen}},\ }\bibfield  {title} {\enquote {\bibinfo {title} {{QCD Equation of
  State at Finite Chemical Potentials for Relativistic Nuclear Collisions}},}\
  }\href {\doibase 10.1142/S0217751X21300076} {\bibfield  {journal} {\bibinfo
  {journal} {Int. J. Mod. Phys. A}\ }\textbf {\bibinfo {volume} {36}},\
  \bibinfo {pages} {2130007} (\bibinfo {year} {2021})},\ \Eprint
  {http://arxiv.org/abs/2101.11591} {arXiv:2101.11591 [nucl-th]} \BibitemShut
  {NoStop}%
\bibitem [{\citenamefont {Anticic}\ \emph {et~al.}(2012)\citenamefont {Anticic}
  \emph {et~al.}}]{NA49:2012rsi}%
  \BibitemOpen
  \bibfield  {author} {\bibinfo {author} {\bibfnamefont {T.}~\bibnamefont
  {Anticic}} \emph {et~al.} (\bibinfo {collaboration} {NA49}),\ }\bibfield
  {title} {\enquote {\bibinfo {title} {{System-size and centrality dependence
  of charged kaon and pion production in nucleus-nucleus collisions at 40A GeV
  and158A GeV beam energy}},}\ }\href {\doibase 10.1103/PhysRevC.86.054903}
  {\bibfield  {journal} {\bibinfo  {journal} {Phys. Rev. C}\ }\textbf {\bibinfo
  {volume} {86}},\ \bibinfo {pages} {054903} (\bibinfo {year} {2012})},\
  \Eprint {http://arxiv.org/abs/1207.0348} {arXiv:1207.0348 [nucl-ex]}
  \BibitemShut {NoStop}%
\bibitem [{\citenamefont {Abbas}\ \emph {et~al.}(2013)\citenamefont {Abbas}
  \emph {et~al.}}]{ALICE:2013jfw}%
  \BibitemOpen
  \bibfield  {author} {\bibinfo {author} {\bibfnamefont {Ehab}\ \bibnamefont
  {Abbas}} \emph {et~al.} (\bibinfo {collaboration} {ALICE}),\ }\bibfield
  {title} {\enquote {\bibinfo {title} {{Centrality dependence of the
  pseudorapidity density distribution for charged particles in Pb-Pb collisions
  at $\sqrt{s_{\rm NN}}$ = 2.76 TeV}},}\ }\href {\doibase
  10.1016/j.physletb.2013.09.022} {\bibfield  {journal} {\bibinfo  {journal}
  {Phys. Lett. B}\ }\textbf {\bibinfo {volume} {726}},\ \bibinfo {pages}
  {610--622} (\bibinfo {year} {2013})},\ \Eprint
  {http://arxiv.org/abs/1304.0347} {arXiv:1304.0347 [nucl-ex]} \BibitemShut
  {NoStop}%
\bibitem [{\citenamefont {Adam}\ \emph
  {et~al.}(2017{\natexlab{b}})\citenamefont {Adam} \emph
  {et~al.}}]{ALICE:2016fbt}%
  \BibitemOpen
  \bibfield  {author} {\bibinfo {author} {\bibfnamefont {Jaroslav}\
  \bibnamefont {Adam}} \emph {et~al.} (\bibinfo {collaboration} {ALICE}),\
  }\bibfield  {title} {\enquote {\bibinfo {title} {{Centrality dependence of
  the pseudorapidity density distribution for charged particles in Pb-Pb
  collisions at $\sqrt{s_{\rm NN}}=5.02$ TeV}},}\ }\href {\doibase
  10.1016/j.physletb.2017.07.017} {\bibfield  {journal} {\bibinfo  {journal}
  {Phys. Lett. B}\ }\textbf {\bibinfo {volume} {772}},\ \bibinfo {pages}
  {567--577} (\bibinfo {year} {2017}{\natexlab{b}})},\ \Eprint
  {http://arxiv.org/abs/1612.08966} {arXiv:1612.08966 [nucl-ex]} \BibitemShut
  {NoStop}%
\bibitem [{\citenamefont {Acharya}\ \emph
  {et~al.}(2019{\natexlab{b}})\citenamefont {Acharya} \emph
  {et~al.}}]{ALICE:2018cpu}%
  \BibitemOpen
  \bibfield  {author} {\bibinfo {author} {\bibfnamefont {Shreyasi}\
  \bibnamefont {Acharya}} \emph {et~al.} (\bibinfo {collaboration} {ALICE}),\
  }\bibfield  {title} {\enquote {\bibinfo {title} {{Centrality and
  pseudorapidity dependence of the charged-particle multiplicity density in
  Xe\textendash{}Xe collisions at $\sqrt{s_{\rm NN}}$ =5.44TeV}},}\ }\href
  {\doibase 10.1016/j.physletb.2018.12.048} {\bibfield  {journal} {\bibinfo
  {journal} {Phys. Lett. B}\ }\textbf {\bibinfo {volume} {790}},\ \bibinfo
  {pages} {35--48} (\bibinfo {year} {2019}{\natexlab{b}})},\ \Eprint
  {http://arxiv.org/abs/1805.04432} {arXiv:1805.04432 [nucl-ex]} \BibitemShut
  {NoStop}%
\bibitem [{\citenamefont {Aamodt}\ \emph {et~al.}(2010)\citenamefont {Aamodt}
  \emph {et~al.}}]{ALICE:2010khr}%
  \BibitemOpen
  \bibfield  {author} {\bibinfo {author} {\bibfnamefont {K}~\bibnamefont
  {Aamodt}} \emph {et~al.} (\bibinfo {collaboration} {ALICE}),\ }\bibfield
  {title} {\enquote {\bibinfo {title} {{Charged-particle multiplicity density
  at mid-rapidity in central Pb-Pb collisions at $\sqrt{s_{NN}} = 2.76$
  TeV}},}\ }\href {\doibase 10.1103/PhysRevLett.105.252301} {\bibfield
  {journal} {\bibinfo  {journal} {Phys. Rev. Lett.}\ }\textbf {\bibinfo
  {volume} {105}},\ \bibinfo {pages} {252301} (\bibinfo {year} {2010})},\
  \Eprint {http://arxiv.org/abs/1011.3916} {arXiv:1011.3916 [nucl-ex]}
  \BibitemShut {NoStop}%
\bibitem [{\citenamefont {Shen}\ \emph {et~al.}(2016)\citenamefont {Shen},
  \citenamefont {Paquet}, \citenamefont {Denicol}, \citenamefont {Jeon},\ and\
  \citenamefont {Gale}}]{Shen:2015qba}%
  \BibitemOpen
  \bibfield  {author} {\bibinfo {author} {\bibfnamefont {C.}~\bibnamefont
  {Shen}}, \bibinfo {author} {\bibfnamefont {J.~F.}\ \bibnamefont {Paquet}},
  \bibinfo {author} {\bibfnamefont {G.~S.}\ \bibnamefont {Denicol}}, \bibinfo
  {author} {\bibfnamefont {S.}~\bibnamefont {Jeon}}, \ and\ \bibinfo {author}
  {\bibfnamefont {C.}~\bibnamefont {Gale}},\ }\bibfield  {title} {\enquote
  {\bibinfo {title} {{Thermal photon radiation in high multiplicity p+Pb
  collisions at the Large Hadron Collider}},}\ }\href {\doibase
  10.1103/PhysRevLett.116.072301} {\bibfield  {journal} {\bibinfo  {journal}
  {Phys. Rev. Lett.}\ }\textbf {\bibinfo {volume} {116}},\ \bibinfo {pages}
  {072301} (\bibinfo {year} {2016})},\ \Eprint
  {http://arxiv.org/abs/1504.07989} {arXiv:1504.07989 [nucl-th]} \BibitemShut
  {NoStop}%
\bibitem [{\citenamefont {Shen}\ \emph {et~al.}(2017)\citenamefont {Shen},
  \citenamefont {Paquet}, \citenamefont {Denicol}, \citenamefont {Jeon},\ and\
  \citenamefont {Gale}}]{Shen:2016zpp}%
  \BibitemOpen
  \bibfield  {author} {\bibinfo {author} {\bibfnamefont {Chun}\ \bibnamefont
  {Shen}}, \bibinfo {author} {\bibfnamefont {Jean-Fran\c{c}ois}\ \bibnamefont
  {Paquet}}, \bibinfo {author} {\bibfnamefont {Gabriel~S.}\ \bibnamefont
  {Denicol}}, \bibinfo {author} {\bibfnamefont {Sangyong}\ \bibnamefont
  {Jeon}}, \ and\ \bibinfo {author} {\bibfnamefont {Charles}\ \bibnamefont
  {Gale}},\ }\bibfield  {title} {\enquote {\bibinfo {title} {{Collectivity and
  electromagnetic radiation in small systems}},}\ }\href {\doibase
  10.1103/PhysRevC.95.014906} {\bibfield  {journal} {\bibinfo  {journal} {Phys.
  Rev. C}\ }\textbf {\bibinfo {volume} {95}},\ \bibinfo {pages} {014906}
  (\bibinfo {year} {2017})},\ \Eprint {http://arxiv.org/abs/1609.02590}
  {arXiv:1609.02590 [nucl-th]} \BibitemShut {NoStop}%
\bibitem [{\citenamefont {Gale}\ \emph {et~al.}(2022)\citenamefont {Gale},
  \citenamefont {Paquet}, \citenamefont {Schenke},\ and\ \citenamefont
  {Shen}}]{Gale:2021emg}%
  \BibitemOpen
  \bibfield  {author} {\bibinfo {author} {\bibfnamefont {Charles}\ \bibnamefont
  {Gale}}, \bibinfo {author} {\bibfnamefont {Jean-Fran\c{c}ois}\ \bibnamefont
  {Paquet}}, \bibinfo {author} {\bibfnamefont {Bj\"orn}\ \bibnamefont
  {Schenke}}, \ and\ \bibinfo {author} {\bibfnamefont {Chun}\ \bibnamefont
  {Shen}},\ }\bibfield  {title} {\enquote {\bibinfo {title} {{Multimessenger
  heavy-ion collision physics}},}\ }\href {\doibase
  10.1103/PhysRevC.105.014909} {\bibfield  {journal} {\bibinfo  {journal}
  {Phys. Rev. C}\ }\textbf {\bibinfo {volume} {105}},\ \bibinfo {pages}
  {014909} (\bibinfo {year} {2022})},\ \Eprint
  {http://arxiv.org/abs/2106.11216} {arXiv:2106.11216 [nucl-th]} \BibitemShut
  {NoStop}%
\bibitem [{\citenamefont {Park}\ \emph {et~al.}(2017)\citenamefont {Park},
  \citenamefont {Shen}, \citenamefont {Jeon},\ and\ \citenamefont
  {Gale}}]{Park:2016jap}%
  \BibitemOpen
  \bibfield  {author} {\bibinfo {author} {\bibfnamefont {Chanwook}\
  \bibnamefont {Park}}, \bibinfo {author} {\bibfnamefont {Chun}\ \bibnamefont
  {Shen}}, \bibinfo {author} {\bibfnamefont {Sangyong}\ \bibnamefont {Jeon}}, \
  and\ \bibinfo {author} {\bibfnamefont {Charles}\ \bibnamefont {Gale}},\
  }\bibfield  {title} {\enquote {\bibinfo {title} {{Rapidity-dependent jet
  energy loss in small systems with finite-size effects and running
  coupling}},}\ }\href {\doibase 10.1016/j.nuclphysbps.2017.05.066} {\bibfield
  {journal} {\bibinfo  {journal} {Nucl. Part. Phys. Proc.}\ }\textbf {\bibinfo
  {volume} {289-290}},\ \bibinfo {pages} {289--292} (\bibinfo {year} {2017})},\
  \Eprint {http://arxiv.org/abs/1612.06754} {arXiv:1612.06754 [nucl-th]}
  \BibitemShut {NoStop}%
\bibitem [{\citenamefont {Pordes}\ \emph {et~al.}(2007)\citenamefont {Pordes}
  \emph {et~al.}}]{Pordes:2007zzb}%
  \BibitemOpen
  \bibfield  {author} {\bibinfo {author} {\bibfnamefont {Ruth}\ \bibnamefont
  {Pordes}} \emph {et~al.},\ }\bibfield  {title} {\enquote {\bibinfo {title}
  {{The Open Science Grid}},}\ }\href {\doibase 10.1088/1742-6596/78/1/012057}
  {\bibfield  {journal} {\bibinfo  {journal} {J. Phys. Conf. Ser.}\ }\textbf
  {\bibinfo {volume} {78}},\ \bibinfo {pages} {012057} (\bibinfo {year}
  {2007})}\BibitemShut {NoStop}%
\bibitem [{\citenamefont {Sfiligoi}\ \emph {et~al.}(2009)\citenamefont
  {Sfiligoi}, \citenamefont {Bradley}, \citenamefont {Holzman}, \citenamefont
  {Mhashilkar}, \citenamefont {Padhi},\ and\ \citenamefont
  {Wurthwrin}}]{Sfiligoi:2009cct}%
  \BibitemOpen
  \bibfield  {author} {\bibinfo {author} {\bibfnamefont {Igor}\ \bibnamefont
  {Sfiligoi}}, \bibinfo {author} {\bibfnamefont {Daniel~C.}\ \bibnamefont
  {Bradley}}, \bibinfo {author} {\bibfnamefont {Burt}\ \bibnamefont {Holzman}},
  \bibinfo {author} {\bibfnamefont {Parag}\ \bibnamefont {Mhashilkar}},
  \bibinfo {author} {\bibfnamefont {Sanjay}\ \bibnamefont {Padhi}}, \ and\
  \bibinfo {author} {\bibfnamefont {Frank}\ \bibnamefont {Wurthwrin}},\
  }\bibfield  {title} {\enquote {\bibinfo {title} {{The pilot way to Grid
  resources using glideinWMS}},}\ }\href {\doibase 10.1109/CSIE.2009.950}
  {\bibfield  {journal} {\bibinfo  {journal} {WRI World Congress}\ }\textbf
  {\bibinfo {volume} {2}},\ \bibinfo {pages} {428--432} (\bibinfo {year}
  {2009})}\BibitemShut {NoStop}%
\end{thebibliography}%

\end{document}